  \documentclass[
preprint,
prd,
amsmath, amssymb, floatfix,
nofootinbib,wrapfloat byrevtex, tightenlines,11pt]{revtex4}

 \usepackage{amsmath,mathrsfs,graphicx,epstopdf,placeins,float}
\usepackage{multirow}
\usepackage[utf8]{inputenc}

\usepackage[colorlinks]{hyperref} 
\hypersetup{
	citecolor=blue,
	linkcolor=blue,
	urlcolor=blue
}

 \newcommand{\beq}[1]{\begin{equation}\label{#1}}
 \newcommand{\eeq}{\end{equation}}
 \newcommand{\bea}[1]{\begin{eqnarray}\label{#1}}
 \newcommand{\eea}{\end{eqnarray}}
 \newcommand\figcaption{\def\@captype{figure}\caption}
 \newcommand\tabcaption{\def\@captype{table}\caption}

 \newcommand{\PYTHIA}{{\tt PYTHIA~8.2}}
\newcommand{\EPOSLHC}{{\tt EPOS-LHC}}
\newcommand{\DPMJET}{{\tt DPMJET-}III}
\newcommand{\GALPROP}{{\tt GALPROP-v54}}

\newcommand{\pHebar}{p_0^{\scriptscriptstyle\overline{\mathrm{He}}}}

\newcommand{\pDbar}{p_0^{\scriptscriptstyle\overline{\mathrm{D}}}}
\newcommand{\Hebar}{^3\overline{\text{He}}}
\newcommand{\Dbar}{\overline{\text{D}}}

\begin{document}

 \title{The high energy window of probing dark matter with cosmic-ray antideuterium and antihelium}
\author{
 Yu-Chen Ding$^{1}$\footnote{dingyuchen@itp.ac.cn},
 Nan Li$^{1,3}$\footnote{linan2016@itp.ac.cn},
 Chun-Cheng Wei$^{1}$\footnote{ccwei@itp.ac.cn}
and
 Yu-Feng Zhou$^{1,2}$\footnote{yfzhou@itp.ac.cn}}
 \affiliation{$^{1}$CAS Key Laboratory of Theoretical Physics,
Institute of Theoretical Physics, Chinese Academy of Sciences, Beijing 100190,
China.\\
University of Chinese Academy of Sciences, Beijing 100049, China.\\
$^{2}$School of Fundamental Physics and Mathematical Sciences, Hangzhou Institute for Advanced Study, UCAS, Hangzhou 310024, China.\\
International Centre for Theoretical Physics Asia-Pacific, Beijing/Hangzhou, China.\\
$^{3}$Key Laboratory of Particle Acceleration Physics and Technology, Institute of High Energy Physics,\\
Chinese Academy of Sciences, Beijing 100049, China.}
 \begin{abstract}
Cosmic-ray (CR) anti-nuclei are often considered as important observables
for dark matter (DM) indirect detections at
low kinetic energies below GeV per nucleon. Since the primary CR fluxes
drop quickly towards high energies, the secondary
anti-nuclei in CR are expected to be significantly suppressed in high energy regions ($\gtrsim 100$ GeV per nucleon).
If DM particles are heavy, the annihilation productions of DM can be highly boosted,
thus the fluxes of anti-nuclei produced by DM annihilations may exceed the secondary background at high energies,
which opens a high energy window for DM indirect detections.
We investigate the possibility of detecting heavy DM particles which
annihilate into high energy anti-nuclei. We use Monte-Carlo generators
{\tt PYTHIA, EPOS-LHC} and {\tt DPMJET} and the coalescence model to simulate the production
of anti-nuclei, and constrain the DM annihilation cross sections
by using the AMS-02 and HAWC antiproton data and the HESS galactic center gamma-ray data.
We find that the conclusion depends on the choice of DM density profiles.
For the ``Cored'' type profile with a DM particle mass $\gtrsim 10$ TeV, the contributions
from DM annihilations can exceed the secondary background in
high energy regions, which opens the high energy window.
While for the ``Cuspy'' type profile, the excess disappears. \\[0.8em]
\textbf{\textsf{ArXiv ePrint:}} 2006.14681
 \end{abstract}

 \maketitle

\section{Introduction}
The existence of dark matter (DM) is supported by various astronomic
observations at different scales, but the particle nature
of DM is still mysterious. As an important probe in DM indirect
detections, the antiparticles in cosmic rays (CR) may shed light
on the properties of DM. In recent years, a number of experiments have
shown an unexpected structure in the CR positron data~\cite{Beatty:2004cy,Adriani:2008zr,FermiLAT:2011ab,Accardo:2014lma},
which could be related to the DM annihilation or decay~\cite{Kopp:2013eka,%
Bergstrom:2013jra,Ibarra:2013zia,Jin:2013nta}. Unlike CR positrons, the
CR antiproton flux data from PAMELA~\cite{Adriani:2012paa}, BESS-polar II~\cite{Abe:2011nx} and
AMS-02~\cite{Aguilar:2016kjl} do not show significant discrepancies with the secondary production of antiproton,
and these null results can be used to
place stringent constraints on the DM annihilation cross sections~\cite{Giesen:2015ufa,%
Jin:2015sqa,%
Lin:2016ezz,%
Reinert:2017aga}.

CR heavy anti-nuclei such as antideuterium ($\Dbar$) and antihelium-3 ($\Hebar$)
are supposed to be important probes for
the DM~\cite{Donato:1999gy,Carlson:2014ssa,Cirelli:2014qia}. The $\Dbar$ and $\Hebar$ in CR can
be generated as secondary productions by the collisions between the primary CR
particles and the interstellar gas, or they can be produced by the DM annihilation or decay.
However, the secondary $\Dbar$ and $\Hebar$ are boosted to high kinetic energies because of
the high production threshold in $pp$-collisions (17$m_{p}$ for
$\overline{\textrm{D}}$ and 31$m_{p}$ for $^{3}\overline{\text{He}}$, where
$m_{p}$ is the proton mass), thus the signal from DM can be distinguished in low energy
regions (below GeV per nucleon). Although the fluxes of anti-nuclei decrease rapidly with
the increase of the atom mass number $A$, the high signal-to-background
ratio at low energies and the experiments with high sensitivities (such as AMS-02~\cite{Giovacchini:2007dwa,Kounine:2010js}
and GAPS~\cite{Aramaki:2015laa}) make
it possible to distinguish the contributions originated from DM interactions.
Furthermore, an advantage for considering $\Dbar$ and $\Hebar$ is that their productions
are highly correlated with CR antiprotons, the uncertainties of the $\Dbar$ and $\Hebar$ fluxes
can be greatly reduced by the CR $\bar{p}$ data~\cite{Li:2018dxj}.

In the literature (for a recent review, see Ref.~\cite{vonDoetinchem:2020vbj}),
the analysis of the DM produced $\Dbar$ and $\Hebar$ are focused on
the low kinetic energy regions, which we refer as the low energy window.
In our previous analysis~\cite{Li:2018dxj}, we have studied the prospects of detecting DM through
the low energy antihelium. We systematically analysed the uncertainties from propagation models,
DM density profiles and MC generators, and reduced the uncertainties by constraining the DM annihilation
cross sections with the AMS-02 $\bar{p}/p$ data.
However, the low energy window suffers from the uncertainties
of solar activities (solar modulations).
In this work, we investigate the possibility of
probing DM with high energy CR $\Dbar$ and $\Hebar$ particles. In
high energy regions (typically above 100 GeV per nucleon), the flux of primary CR particles
drops quickly (the flux of CR proton is proportional to $E^{-2.75}$), which leads to a
suppression on the high energy secondary CR particles.
As a result, the fluxes of anti-nuclei produced by DM annihilations may exceed the secondary background
and open a high energy window for probing the DM.

We use the Monte-Carlo (MC) event generators \PYTHIA~\cite{Sjostrand:2006za,Sjostrand:2014zea},
\EPOSLHC~\cite{Werner:2005jf,Pierog:2013ria} and \DPMJET~\cite{Roesler:2001mn}
to fit the coalescence momenta for anti-nuclei with the experiments including
ALEPH~\cite{Schael:2006fd}, CERN ISR~\cite{Henning:1977mt} and ALICE~\cite{Acharya:2017fvb},
and generate the energy spectra of anti-nuclei.
The propagation of CR particles are calculated by using the GALPROP code.
We use the AMS-02~\cite{Aguilar:2016kjl} and HAWC $\bar{p}/p$ data~\cite{Abeysekara:2018syp} and the
HESS galactic center (GC) $\gamma$-ray data~\cite{Abramowski:2011hc,Abdallah:2016ygi} to constrain the DM
annihilation cross sections. We find that the conclusion is depend on the choice
of DM density profiles. For a large DM mass ($\gtrsim 10$ TeV) with
the relatively flat ``Cored'' type DM profile, the high energy window exist.
While for a typical steep DM profile like ``Cuspy'' type, the high energy window closes.

This paper is organized as follows:
In section~\ref{sec:model}, we briefly review
the coalescence model and determine the coalescence
momenta for anti-nuclei by fitting the ALEPH, ALICE and CERN-ISR data. In section~\ref{sec:propagation},
we review the theory of CR propagation. In section~\ref{sec:limit}, we constrain the DM
annihilation cross section by using the $\bar{p}/p$ data from AMS-02 and HAWC and $\gamma-$ray data
from HESS. The fluxes of $\Dbar$ and $\Hebar$ for DM direct annihilation and
annihilation through mediator channels are presented in section~\ref{sec:flux}.
The conclusions are summarized in section~\ref{sec:conclusion}.

\section{The coalescence model and coalescence momenta}\label{sec:model}
The formation of anti-nuclei can be described by the coalescence model~\cite{Butler:1963pp,Schwarzschild:1963zz,Csernai:1986qf},
which uses a single parameter, the coalescence momentum
$p_0^{\scriptscriptstyle\bar{\mathrm{A}}}$ to quantify the
probability of anti-nucleons merge into an anti-nucleus $\bar{\mathrm{A}}$.
The basic idea of this model is that anti-nucleons combine into an anti-nucleus if the relative four-momenta of
a proper set of nucleons is less than the coalescence momentum.
For example, the coalescence criterion for $\Dbar$ is written as:
\beq{Dbar1}
||k_{\bar{p}}-k_{\bar{n}}|| = \sqrt{(\Delta \vec{k})^2-(\Delta E)^2} <\pDbar,
\eeq
where $k_{\bar{p}}$ and $k_{\bar{n}}$ are the four-momenta of antiproton and
antineutron respectively, and $p_0^{\bar{\scriptscriptstyle \mathrm{D}}}$ is the
coalescence momentum of $\Dbar$. If we assume that the momenta distribution of
the $\bar{p}$ and $\bar{n}$ in one collision event are uncorrelated and isotropic,
the spectrum of $\Dbar$ can be derived by the phase-space analysis:
\begin{align}\label{eq:p0-isotropic}
\gamma_{\bar{\scriptscriptstyle \mathrm{D}}}
\frac{d^{3}N_{\bar{\scriptscriptstyle \mathrm{D}}}}{d^{3}\vec{k}_{\bar{\scriptscriptstyle \mathrm{D}}}}
(\vec{k}_{\bar{\scriptscriptstyle \mathrm{D}}})
=
\frac{\pi}{6}
\left(p_{0}^{\bar{\scriptscriptstyle \mathrm{D}}}\right)^{3}
\cdot
\gamma_{\bar p} \frac{d^{3}N_{\bar p}}{d^{3}\vec{k}_{\bar p}}(\vec{k}_{\bar p})
\cdot
\gamma_{\bar n} \frac{d^{3}N_{\bar n}}{d^{3}\vec{k}_{\bar n}}(\vec{k}_{\bar n}) ,
\end{align}
where $\gamma_{\bar{\scriptscriptstyle \mathrm{D}}, \bar p, \bar n}$ are the
Lorentz factors, and $\vec{k}_{\bar p}\approx \vec{k}_{\bar n}\approx
\vec{k}_{\bar{\scriptscriptstyle \mathrm{D}}}/2$.

For $\Hebar$, we adopt the same coalescence criterion as in our previous analysis~\cite{Li:2018dxj}.
We compose a triangle using the norms of the three relative four-momenta
$ l_1 = ||k_1-k_2||,~~ l_2 = ||k_2-k_3||$ and $ l_3 =||k_1-k_3||$,  where $k_1, k_2,
k_3$ are the four-momenta of the three anti-nucleons respectively.
And then, making a circle with minimal diameter to envelop the triangle,
if the diameter of this circle is smaller than $\pHebar$, an $\Hebar$
is generated. If the triangle is acute, the
minimal circle is just the circumcircle of this triangle, and the
coalescence criterion can be expressed as follows:
\begin{equation}
\label{eq:pHebar-def1}
d_{\mathrm{circ}}=\frac{l_1l_2l_3}{\sqrt{(l_1+l_2+l_3)(-l_1+l_2+l_3)(l_1-l_2+l_3)(l_1+l_2-l_3)}}<
p_0^{\scriptscriptstyle \overline{\mathrm{He}}}.
\end{equation}
Otherwise, the minimal diameter is equal to the longest side of the triangle, and the
criterion can be simply written as
$\text{max}\{l_1,l_2,l_3\} < p_0^{\scriptscriptstyle
  \overline{\mathrm{He}}}$.
See Ref.~\cite{Li:2018dxj} for more details.

We use the MC generators \PYTHIA,
\EPOSLHC~and \DPMJET~to simulate the hadronization after DM
annihilations and $pp-$collisions, and then adopt the coalescence model
to produce anti-nuclei form the final state $\bar{p}$ or $\bar{n}$.
The spatial distance between each pair of anti-nucleons also needs to be
considered, because anti-nucleons should be close enough to under go
the nuclear reactions and then merge into anti-nuclei.
We set all the particles with
lifetime $\tau \gtrsim 2\,\, \mathrm{fm}/c$ to be
stable to ensure that every pair of anti-nucleons are located in a
short enough distance~\cite{Carlson:2014ssa}, where 2 fm is approximately
the size of the $^3\overline{\textrm{He}}$ nucleus.

The value of coalescence momenta should be determined by the experimental
data, which are often released in the form of coalescence parameters $B_A$.
The definition of the coalescence parameter $B_A$ is expressed by the formula:
\begin{align}\label{BA}
E_A\frac{d^3N_A}{dp_A^3}
=
B_A  \left(E_p\frac{d^3N_p}{dp_p^3}\right)^Z
\left(E_n\frac{d^3N_n}{dp_n^3}\right)^N
, \ \ \ \vec{p}_p=\vec{p}_n=\vec{p}_A/A ,
\end{align}
where $A=Z+N$ is the mass number of the nucleus, $Z$ and $N$ are
the proton number and the neutron number respectively. Under the assumption
that the momenta distribution of the $\bar{p}$ and $\bar{n}$
are uncorrelated and isotropic, the relation $B_A\propto p_0^{3(A-1)}$
is expected by comparing Eq.~\eqref{eq:p0-isotropic} and Eq.~\eqref{BA}. However, the jet structure and
the correlation between $\bar{p}$ and $\bar{n}$ play important roles in
the formation of anti-nuclei. So we derive the coalescence momenta by
using the MC generators to fit the experimental data, thus the effect of jet structures and
correlations are included.

\begin{figure}[ht]
\includegraphics[width=0.43\textwidth]{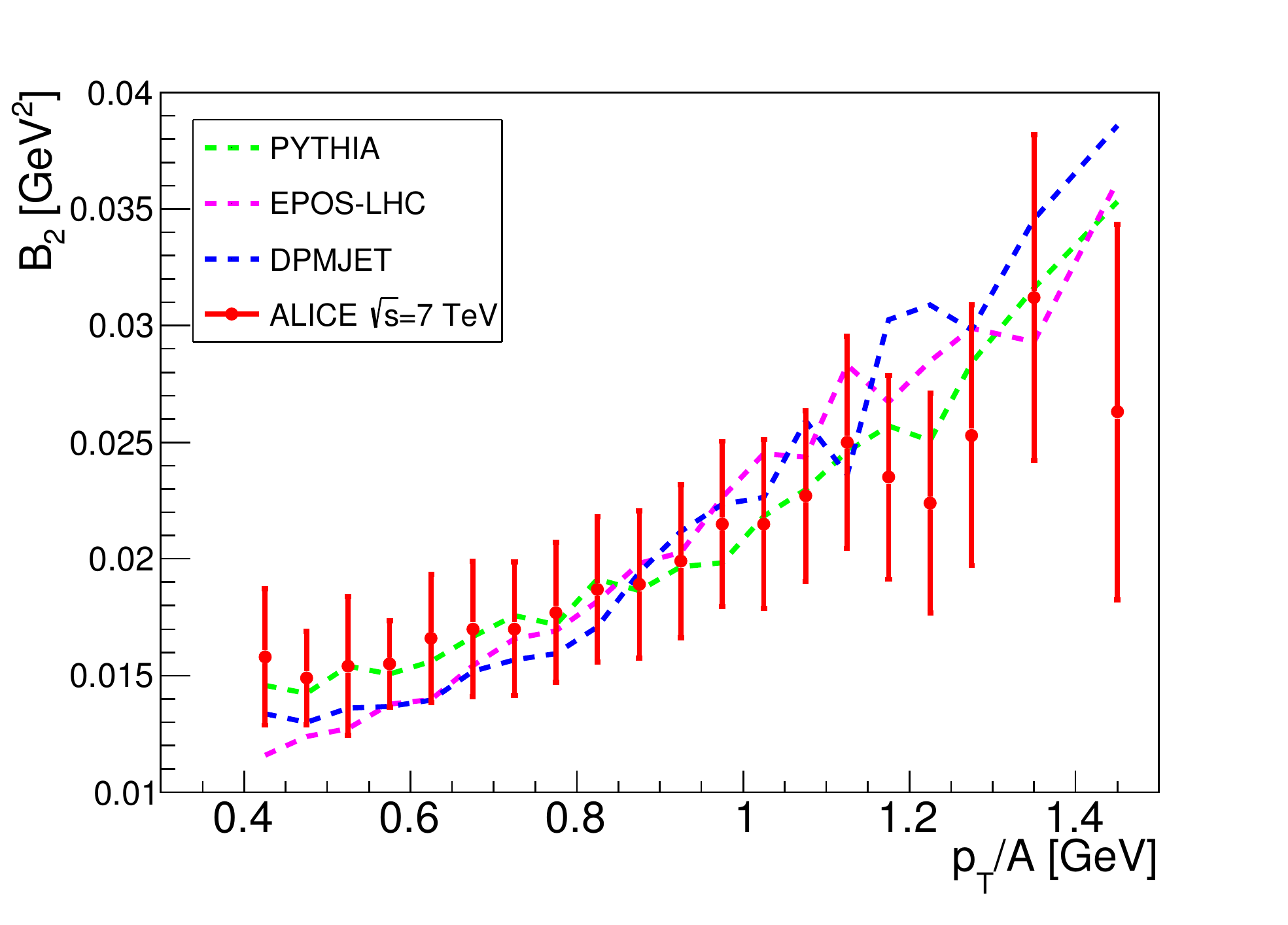}
\includegraphics[width=0.43\textwidth]{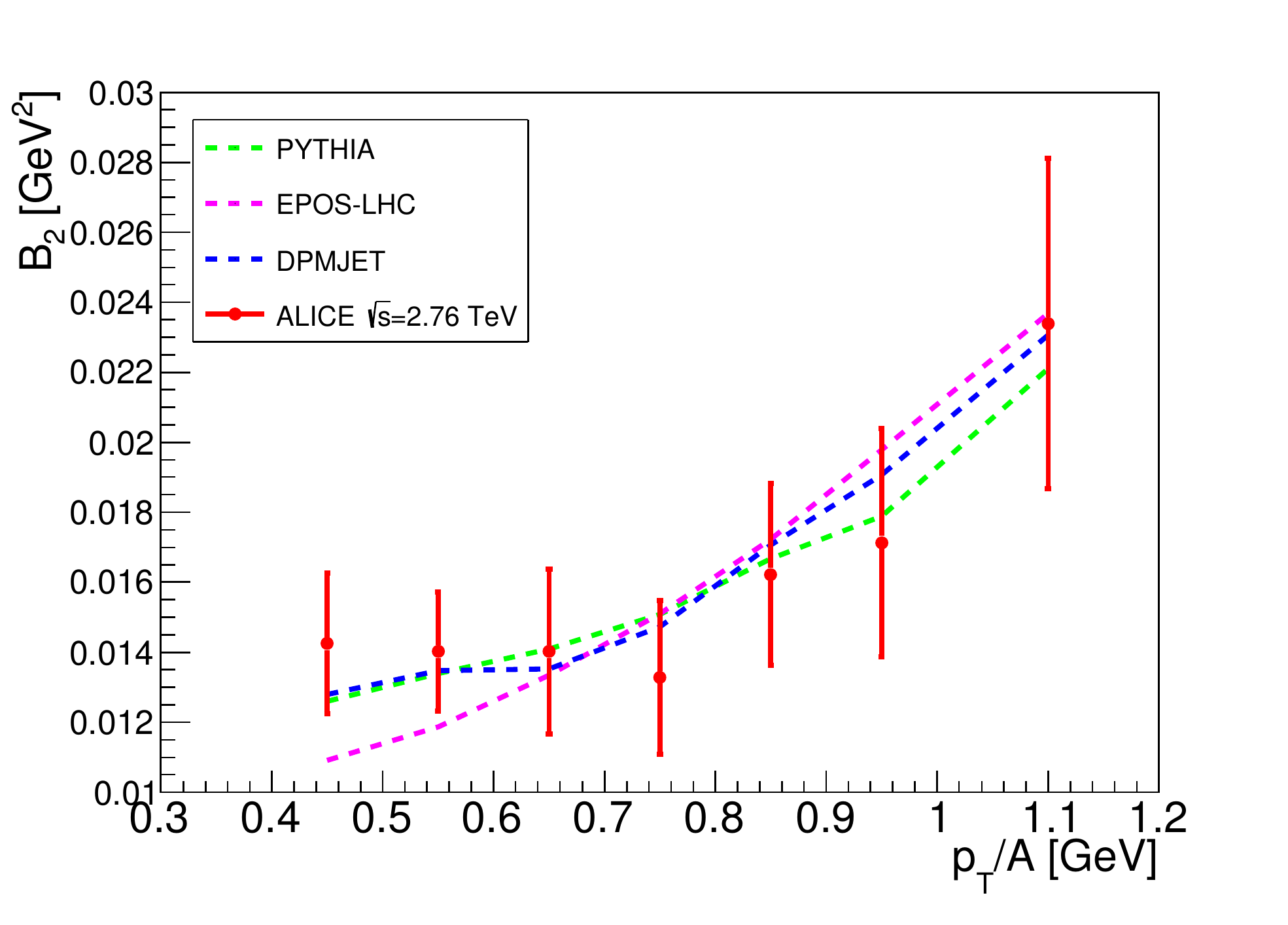}\\
\includegraphics[width=0.43\textwidth]{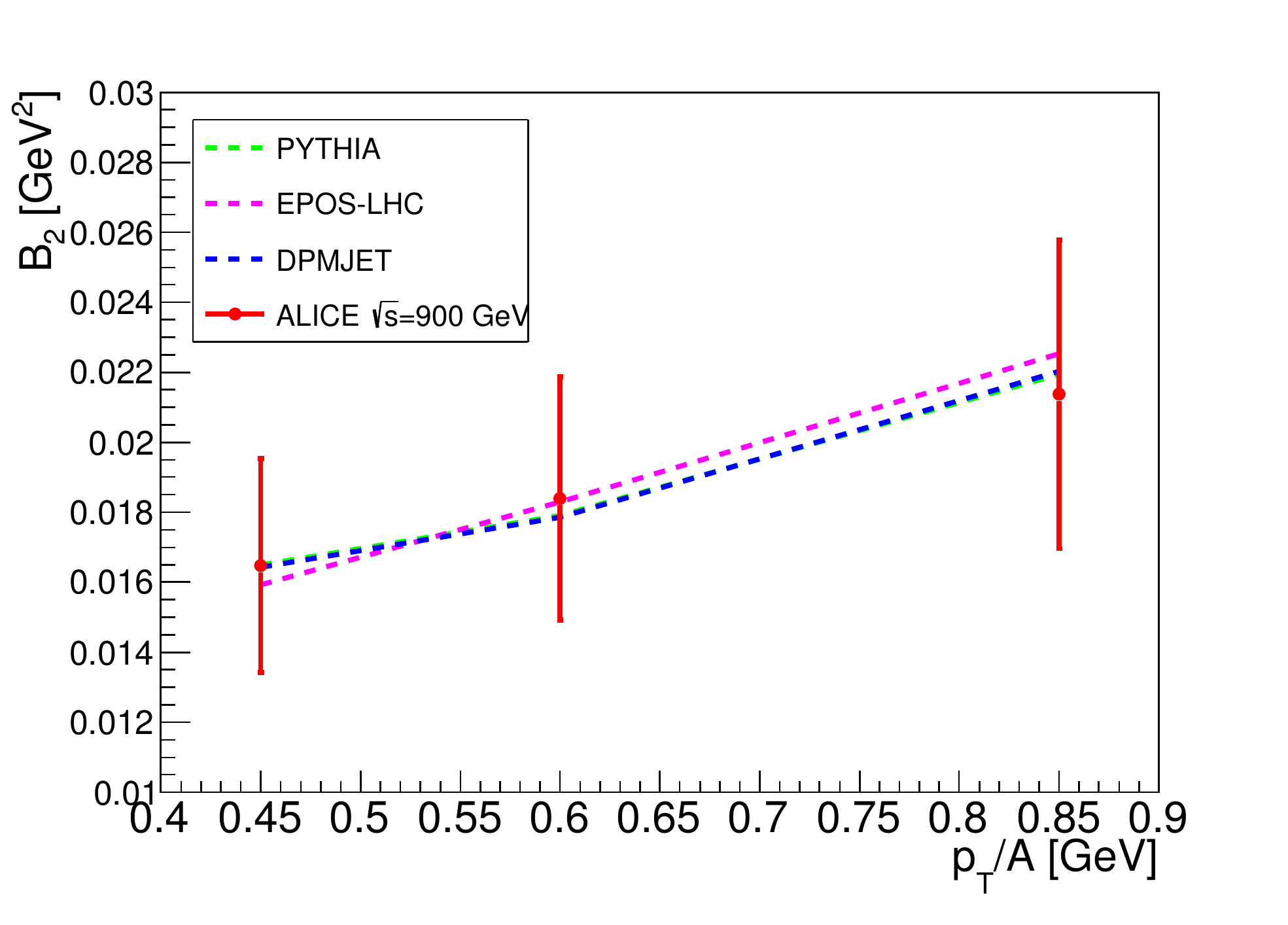}
\includegraphics[width=0.43\textwidth]{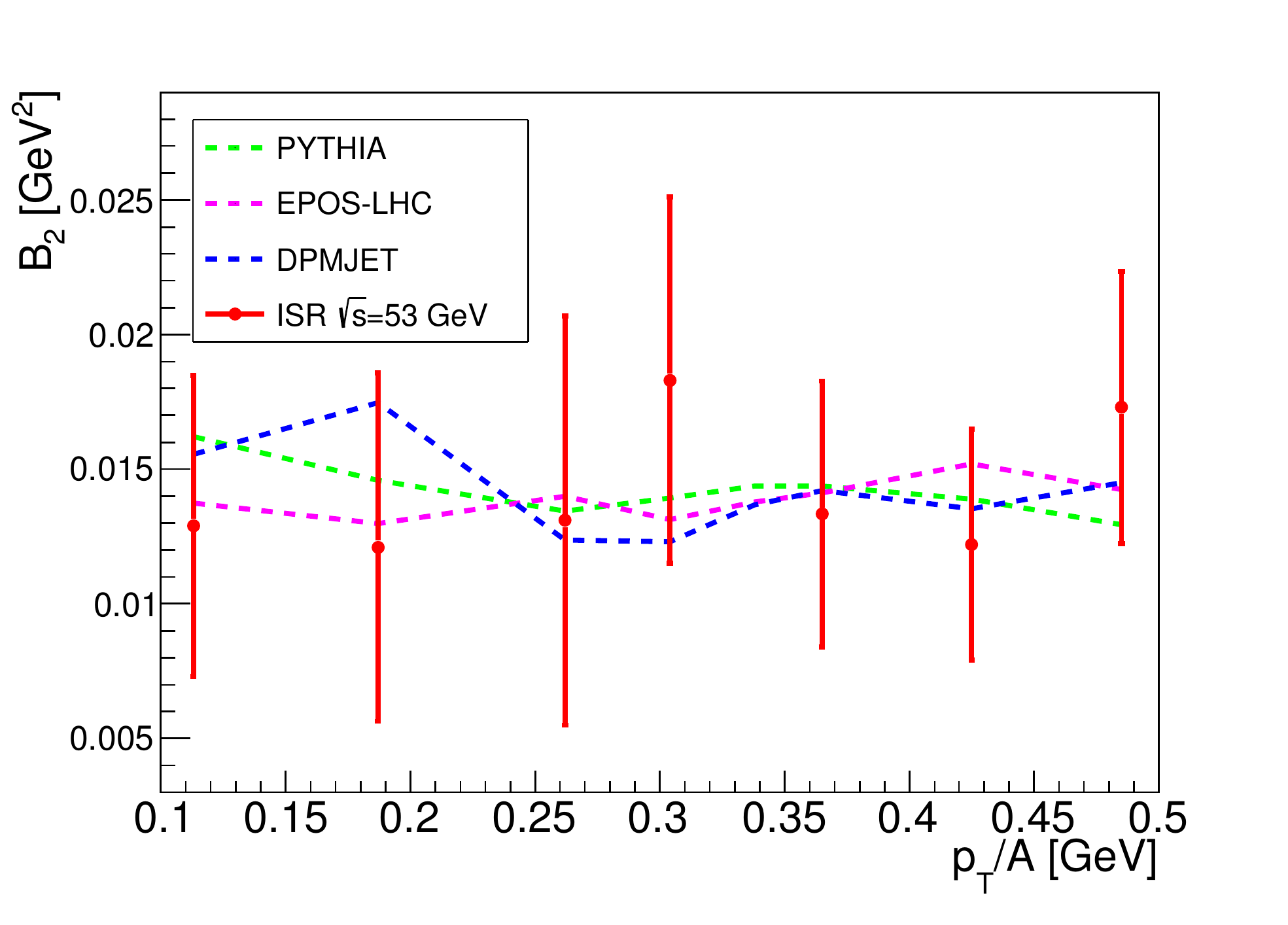}
\caption{The best-fit $B_2$ values from MC generators compare with the ALICE~\cite{Acharya:2017fvb}
and CERN-ISR~\cite{Henning:1977mt} $pp$-collision data.
$topleft)$ ALICE-7 TeV, $topright)$ ALICE-2.76 TeV, $bottomleft)$ ALICE-900 GeV, $bottomright)$
ISR-53 GeV.
}
\label{Dbar_B2}
\end{figure}

To derive the coalescence momentum of $\overline{\mathrm{D}}$ in $pp$ collisions,
we follow the procedures described in Ref.~\cite{Li:2018dxj}
to fit the coalescence parameter $B_2$ data from the ALICE-7 TeV, 2.76 TeV, 900 GeV~\cite{Acharya:2017fvb} and
ISR-53 GeV~\cite{Henning:1977mt} experiments.
We use MC generators to simulate these experiments, and record the
momenta information of $\bar{p}$, $\bar{n}$ that are possible to form a $\overline{\mathrm{D}}$.
We select the $\bar{p}$ and $\bar{n}$ according to a sufficiently large coalescence momentum
$p_{0,\text{max}}$ (600 MeV for $\overline{\mathrm{D}}$, 1 GeV for $\overline{\mathrm{He}}$),
and then calculate the spectra of $\Dbar$ for different $\pDbar$ values that are smaller than $p_{0,\text{max}}$.
The $B_2$ values are calculated by using the Eq.~\eqref{BA}, and we make a $\chi^2$ analysis
to find the value of $\pDbar$.
The best-fit $B_2$ values are shown in Fig.~\ref{Dbar_B2}. We find $\chi^2_{\text{min}}/\text{d.o.f}< 1$
for all these fits, which means the fitting results are in good agreement with
the experimental data. As shown by the figure, the $p_T$-dependence are well
reproduced by the coalescence model and the MC generators.

The fitting result of $\pDbar$ are listed in Tab.~\ref{tab:Dbar_p0}, and the values in brackets
are the results given by ALICE group, which are only available for ALICE 7 TeV and ISR-53 GeV experiments
with \PYTHIA~and \EPOSLHC~generators. We can see that for ALICE 7 TeV, our best-fit values are in
good agreements with the ones from ALICE group, but for ISR-53 GeV, our results are larger.
This is partially because ALICE group have only generated the energy spectra of $\Dbar$ for six $\pDbar$ values~\cite{Acharya:2017fvb},
and used the isotropic approximation $B_A\approx p_0^{3(A-1)}$
to interpolate the spectra for other $\pDbar$ values, while we generate
the $\Dbar$ spectra for every integer number MeV and do not need the approximation.
Moreover, in fitting the ISR-53 GeV experiment, ALICE group have manually rescaled
the spectra of $\bar{p}$ generated by MC to better reproduce the experimental data,
while we do not make this correction for self-consistencies.

\begin{figure}[ht]
\includegraphics[width=0.68\textwidth]{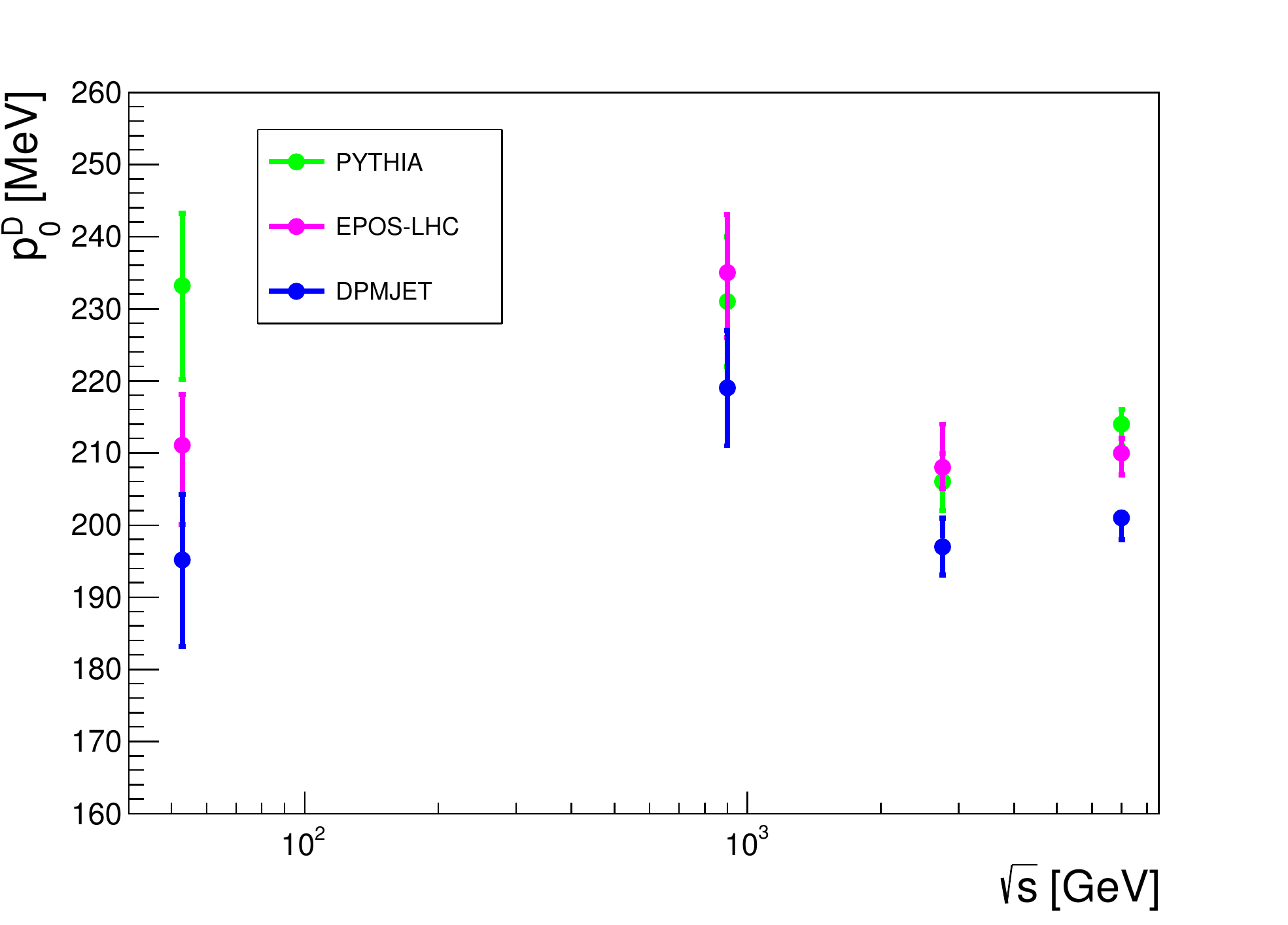}
\caption{The $p_0$ values of ${\overline{\mathrm{D}}}$ from fitting the $pp$-collision data with three MC generators.
}
\label{Dbar_p0}
\end{figure}

\begin{table}[h]
	\begin{tabular}{ccccc}
		\hline\hline
		MC generators    &  ALICE 7 TeV  & ALICE 2.76 TeV  & ALICE 900 GeV  & ISR 53 GeV             \\ \hline
		{\tt PYTHIA 8.2} & $214^{+2}_{-3}~(216\pm 8)$  & $206^{+4}_{-4}$     &  $231^{+9}_{-9}$    &  $233^{+10}_{-13}~(188\pm8)$       \\ \hline
		{\tt EPOS-LHC}   & $210^{+2}_{-3}~(200\pm10)$   & $208^{+6}_{-3}$   &  $235^{+8}_{-9}$  &  $211^{+7}_{-11}~(190\pm8)$         \\ \hline
		{\tt DPMJET}-III   & $201^{+1}_{-3}$   & $197^{+4}_{-4}$     &  $219^{+8}_{-8}$   &  $195^{+9}_{-12}$        \\ \hline\hline
	\end{tabular}
	\caption{
	Best-fit values of $p_0^{\overline{\mathrm{D}}}$ in unit of MeV,
	derived by fitting the $pp$-collision data with three MC
	generators {\tt PYTHIA} 8.2~\cite{Sjostrand:2006za,Sjostrand:2014zea},
	{\tt EPOS-LHC}~\cite{Werner:2005jf,Pierog:2013ria} and
	{\tt DPMJET}-III~\cite{Roesler:2001mn}. The $p_0^{\overline{\mathrm{D}}}$ values do not show a clear
    relation with the $\sqrt{s}$ value of the experiments. The values in brackets are fitting results
    given by ALICE group.
	}
	\label{tab:Dbar_p0}
\end{table}

In Fig.~\ref{Dbar_p0}, we show the $\pDbar$ values for different $pp$ collision experiments with
various MC generators. As can be seen, $\pDbar$ do not show a clear relation with the $\sqrt{s}$ values of the experiments.
Since the fitting results for different center-of-mass energies are similar, we
assume that the $p_0^{\overline{\mathrm{D}}}$ value does not vary with the $\sqrt{s}$ of the experiment.
By fitting these $p_0^{\overline{\mathrm{D}}}$, we get $p_0^{\overline{\mathrm{D}}}(\texttt{PYTHIA)}=213\pm2$ MeV,
$p_0^{\overline{\mathrm{D}}}(\texttt{EPOS-LHC)}=211\pm2$ MeV and $p_0^{\overline{\mathrm{D}}}(\texttt{DPMJET)}=201\pm2$ MeV
for $pp$-collisions.

By using \PYTHIA~ to fit the $B_2=3.3\pm1.0\pm0.8\times10^{-3}$ data from the ALEPH $e^+e^-\rightarrow Z^0 \rightarrow \overline{\mathrm{D}}$
experiment~\cite{Schael:2006fd}, we find $p_0^{\overline{\mathrm{D}}}(Z^0)=190^{+23}_{-27}$ MeV.
Considering the similarity between the dynamics of the
$Z^0$ decay and the DM annihilation, we set $p_0^{\overline{\mathrm{D}}}(Z^0)$ to be
the coalescence momentum for the DM annihilations process $\chi\chi \rightarrow \overline{\mathrm{D}} + X$.

For $\Hebar$, we adopt the $p_0^{\overline{\mathrm{He}}}$ value obtained in our previous work~\cite{Li:2018dxj},
which determining $p_0^{\overline{\mathrm{He}}}$ by fitting the ALICE $\sqrt{s}=7$ TeV $pp$-collision data,
the results are listed in Tab.~\ref{tab:Hebar_p0}. Note that, $\Hebar$ particles can be produced from two
channels: direct formation from the coalescence of $\bar{p}\bar{p}\bar{n}$, or through the $\beta$-decay of an antitriton
$\overline{\text{T}}$ ($\bar{p}\bar{n}\bar{n}$).
The direct formation channel are suppressed by the Coulomb-repulsion between the two antiprotons, thus
some previous works only considered the antitriton channel. However, our calculation shows that the
direct formation channel are not negligible. The coalescence momentum of $^3\overline{\mathrm{He}}$ are
only slightly smaller than that of $\overline{\text{T}}$, and the Gamow factor $\mathcal{G}\sim \exp(-2\pi\alpha m_p/p_0^{\overline{\mathrm{He}}}) \approx 0.8$~\cite{Blum:2017qnn}, these indicate that the Coulomb-repulsion may
not be significant. By these facts, it is reasonable to ignore the Coulomb-repulsion in the MC simulation,
and the contributions from both channels are included in this work. Our MC calculation shows that about $30\%$ of $^3\overline{\mathrm{He}}$ are produced through the direct formation channel.
For the lack of the $e^+e^-\rightarrow \Hebar$ experiments data, we set $p_0^{\overline{\mathrm{He}}/\overline{\mathrm{T}}}(\texttt{PYTHIA)}$ to be
the coalescence momentum for the DM annihilation process $\chi\chi \rightarrow \overline{\mathrm{He}}/\overline{\mathrm{T}} + X$.
It is known that the coalescence momentum varies for different processes and center of mass energies~\cite{Aramaki:2015pii}, and the 
relation $B_A\approx p_0^{3(A-1)}$ makes $p_0$ a crutial factor for the uncertainties of the final fluxes. In some previous analyses, 
the value of $p_0^{\overline{\mathrm{He}}}$ is estimated using various approaches, for example, by using the binding energy relation between
$p_0^{\overline{\mathrm{He}}}$ and  $p_0^{\overline{\mathrm{D}}}$ or assuming the ratio $p_0^{\overline{\mathrm{He}}}/p_0^{\overline{\mathrm{D}}}
=p_0^{\mathrm{He}}/p_0^{\mathrm{D}}$~\cite{Carlson:2014ssa}, and some works just set $p_0^{\overline{\mathrm{He}}}=p_0^{\overline{\mathrm{D}}}$ ~\cite{Cirelli:2014qia}. The $p_0^{\overline{\mathrm{He}}}$ results from these approaches are different. It is worth mention that our $p_0^{\overline{\mathrm{He}}}$ value is relatively small comparing to previous works, which 
leads to conservative DM contributions.

\begin{table}[h]
	\begin{tabular}{cccc}
		\hline\hline
		~MC generators: ~                       & ~~~~ \PYTHIA~~~~~  & ~~~~~\EPOSLHC ~~~~ & ~~~~\DPMJET ~~~~              \\ \hline
		$p_0^{\overline{\mathrm{He}}}$~(MeV) & $224^{+12}_{-16}$   & $227^{+11}_{-16}$     &  $212^{+10}_{-13}$           \\ \hline
		$p_0^{\bar{\mathrm{T}}}$~(MeV)    & $234^{+17}_{-29}$   & $245^{+17}_{-30}$     &  $222^{+16}_{-26}$           \\ \hline\hline
	\end{tabular}
	\caption{
	Best-fit values of $p_0^{\overline{\mathrm{He}}}$ and $p_0^{\bar{\mathrm{T}}}$
	from Ref.~\cite{Li:2018dxj}, which is obtained by
fitting the ALICE $pp$-collision data at $\sqrt{s}=7$~TeV for three MC
	generators \PYTHIA,
	\EPOSLHC~and
	\DPMJET.
	}
	\label{tab:Hebar_p0}
\end{table}

For the primary anti-particles originated from DM annihilations, the
injection spectra are calculated using \PYTHIA. We simulate the annihilation of
Majorana DM particles by a positron-electron annihilation process
$e^+e^- \rightarrow \phi^* \rightarrow f\bar{f}$, where $\phi^*$ is
a fictitious scaler singlet and $f$ is a standard model final state.
We set $\sqrt{s}=2m_{\chi}$ and switch off all initial-state-radiations
in \PYTHIA~to mimic the dynamics of DM annihilation. Three
kinds of final states are considered: $q\bar{q}$ ($q$ stands for $u$ or $d$ quark), $b\bar{b}$ and $W^+W^-$.
In \EPOSLHC~and \DPMJET~generators, only hadrons can be set as
the initial states, which do not resemble the properties of DM annihilations.

For the secondary anti-particles produced in $pp-$collisions,
\EPOSLHC~and \DPMJET~are used to generate the energy spectra.
The default parameters in \PYTHIA~(the Monash tune~\cite{Skands:2014pea})
are focused to reproduce the experimental results at high center-of-mass energies
(like ATLAS at $\sqrt{s}=7$ TeV~\cite{Aad:2011gj}), but are not
optimized for the energy regions around a few tens of GeV, which give the dominating contributions for
the secondary CR anti-particles. To evaluate the performance of MC generators
at low center-of-mass energies, we make a comparison between the
$\bar{p}$ differential invariant cross section
obtained by MC generators and the NA49 data at $\sqrt{s}=17.3$ GeV~\cite{Anticic:2009wd}.
This comparison shows that \EPOSLHC~has the best performance,
\DPMJET~are also in relatively good agreements with the experiment,
while the production cross section of $\bar{p}$ given by \PYTHIA~are larger than the NA49 data roughly by a factor
of two. In this paper, we will draw our conclusion based on the results from
\EPOSLHC, and the difference between \EPOSLHC~and \DPMJET~can be used as a rough
estimation of the uncertainties from different MC generators.

\section{The propagation of cosmic-rays}\label{sec:propagation}
The propagation of charged CR particles are assumed to be random diffusions in
a cylindrical diffusion halo with radius $r_h \approx 20$ kpc and half-height
$z_h = 1 \sim 10$ kpc. The diffusion equation is
written as~\cite{GINZBURG:1990SK,STRONG:2007NH}:
\beq{CR}
\frac{\partial f}{\partial t}= q(\vec{r},p)+\vec{\nabla}\cdot(D_{xx}\vec{\nabla} f-\vec{V}_c  f)+
\frac{\partial}{\partial p}p^2 D_{pp}\frac{\partial}{\partial p}\frac{1}{p^2} f - \frac{\partial}{\partial p}
\left[ \dot{p} f-\frac{p}{3}(\vec{\nabla}\cdot\vec{V}_c) f \right] -\frac{1}{\tau_f} f - \frac{1}{\tau_r} f ~,
\eeq
where $ f(\vec{r},p,t)$ is the number density in phase spaces at the particle momentum $p$ and
position $\vec{r}$, and $q(\vec{r},p)$ is the source term. $D_{xx}$ is the
spatial diffusion coefficient, which is parameterized as $D_{xx} = \beta D_0 (R/R_0)^{\delta}$, where $R = p/(Ze)$ is the
rigidity of the CR particle with electric charge $Ze$, $\delta$ is the spectral power
index which takes two different values $\delta = \delta_{1(2)}$ when $R$ is below (above)
a reference rigidity $R_0$, $D_0$ is a constant normalization coefficient, and $\beta = v/c$
is the velocity of CR particles. $\vec{V}_c$ quantifies the velocity of
the galactic wind convection. The diffusive re-acceleration is described as diffusions in the momentum
space, which is described by the parameter $D_{pp} = p^2V_a^2/(9D_{xx})$,
where $V_a$ is the Alfv\`{e}n velocity that characterizes the propagation of weak disturbances
in a magnetic field. $\dot{p} \equiv dp/dt$ is the momentum loss rate, and $\tau_f$ and $\tau_r$ are
the time scales of particle fragmentation and radioactive decay respectively.
For boundary conditions, we assume that the number densities
of CR particles vanish at the boundary of the halo: $ f(r_h,z,p)= f(r,\pm z_h, p)= 0$.
The steady-state diffusion condition is achieved by setting
$\partial f/\partial t = 0$. We numerically solve the diffusion
equation Eq.~\eqref{CR} by using the \texttt{GALPROP
v54}~code~\cite{Strong:1998pw,MOSKALENKO:2001YA,%
STRONG:2001FU,MOSKALENKO:2002YX,PTUSKIN:2005AX}.
The primary CR nucleus injection spectra are assumed to have a broken power
law behavior $f_p(\vec{r},p)\propto p^{\gamma_p}$, with the injection index
$\gamma_p = \gamma_{p1} (\gamma_{p2})$ for the nucleus rigidity $R_p$ below
(above) a reference value $R_{ps}$. The spatial distribution of the interstellar gas
and the primary sources of CR nuclei are taken from Ref.~\cite{Strong:1998pw}.

The injection of CR particles are described by the source term in the diffusion equation.
For the primary CR antiparticles $\bar{A}~ (\bar{A}=\bar{p}, \Dbar, \Hebar)$  originated from the
annihilation of Majorana DM particles, the source term is given by:
\beq{Hesource}
q_{\bar{A}}(\vec{r},p) =
\frac{\rho_{_{\mathrm{DM}}}^2(\vec{r})}{2m^2_{\chi}}\langle\sigma v\rangle
\frac{dN_{\bar{A}}}{dp}~,
\eeq
where $\rho_{_{\mathrm{DM}}}(\vec{r})$ is the energy density of DM, $\langle\sigma v\rangle$ is the
thermally-averaged DM annihilation cross section and $dN_{\bar{A}}/dp$ is the
energy spectrum of $\bar{A}$ discussed in the previous section.
The spatial distribution of DM are described by DM profiles,
in this work, we consider four commonly used DM profiles to
represent the uncertainties: the Navarfro-Frenk-White (NFW) profile \cite{NAVARRO:1996GJ},
the Isothermal profile \cite{Bergstrom:1997fj},
the Moore profile \cite{Moore:1999nt,Diemand:2004wh} and
the Einasto  profile \cite{Einasto:2009zd}.

For the secondary $\bar{A}$ produced in collisions between the primary CR
and the interstellar gas, the source term can be written as follows:
\beq{secsource}
q_{\bar A}(\vec{r},p) =
\sum_{ij} n_j(\vec{r})
\int \beta_i \,c\, \sigma_{ij\to \bar A}^{\mathrm{inel}}(p^{\prime})
\frac{dN_{\bar{A}}(p,p^{\prime})}{d p}\,n_i(\vec{r},p^{\prime})\,dp^{\prime}~,
\eeq
where
$n_i$ is the number density of CR components (proton, Helium or antiproton) per unit momentum,
$n_j$ is the number density of interstellar gases (hydrogen or Helium), and
$\sigma_{ij}^{\mathrm{inel}}(p^{\prime})$ is the inelastic cross section for
the process $ij\to \bar A+X$, which is provided by the MC generators.
$dN_{\bar{A}}(p,p^{\prime})/d p$ is the energy spectrum of $\bar{A}$
in the collisions, with $p^{\prime}$ stands for the momentum of incident CR particles.
For $\bar p$, we include the contributions from
the collisions of $pp$, $p\text{He}$,
$\text{He}p$, $\text{He}\text{He}$, $\bar p p$ and $\bar p \text{He}$.
For the secondary $\Dbar$ and $^3\overline{\mathrm{He}}$, since the experimental data are
only available in $pp$-collisions, we consider the contribution from collisions
between CR protons and the interstellar hydrogen,
which dominates the secondary background of $\Dbar$ and $^3\overline{\mathrm{He}}$.
The tertiary contributions of $\Dbar$ and $^3\overline{\mathrm{He}}$ are not
included, for they are only important at low kinetic energy regions below 1 GeV$/A$~\cite{Korsmeier:2017xzj},
which do not relevant to our conclusions.

The fragmentation time scale $\tau_f$ in Eq. \eqref{CR}
is inversely proportional to
the inelastic interaction rate between
the nucleus $\bar{A}$ and the interstellar gas,
which is estimated
as~\cite{Carlson:2014ssa,Cirelli:2014qia}
\beq{interact_rate}
\Gamma_{\mathrm{int}} = (n_{_{\mathrm{H}}} + 4^{2/3}
n_{_{\mathrm{He}}})~v~\sigma_{\bar{A}p}~,
\eeq
where
$n_{_{\mathrm{H}}}$ and $n_{_{\mathrm{He}}}$ are the number densities of
interstellar hydrogen and helium, respectively, $4^{2/3}$ is the geometrical factor of helium,
$v$ is the velocity of $\bar{A}$ relative to interstellar gases, and
$\sigma_{\bar{A}p}$ is the total inelastic cross section of
the collisions between  $\bar{A}$ and the interstellar gas.
The number density  ratio $n_{_{\mathrm{He}}}/n_{_{\mathrm{H}}}$ in the interstellar gas is taken to be
0.11~\cite{Strong:1998pw}, which is the default value in {\tt GALPROP}.

Since the experimental data of the inelastic cross sections $\sigma_{\overline{\textrm{D}}p}$ and
$\sigma_{\overline{\textrm{He}}p}$ are currently not available, we
assume the relation $\sigma_{\bar{A} p} = \sigma_{A \bar{p}}$, which is guaranteed by
CP-invariance. For an incident nucleus with atomic mass number $A$, charge
number $Z$ and kinetic energy $T$, the total inelastic cross section
for $A\bar p$ collisions is parameterized by the following
formula~\cite{MOSKALENKO:2001YA}:
\beq{Apbar}
\sigma_{A\bar{p}}^{\mathrm{tot}} = A^{2/3}\left[48.2+19\, x^{-0.55} + (0.1-0.18\, x^{-1.2}) Z +
0.0012\,x^{-1.5}Z^2\right]~\mathrm{mb},
\eeq
where $x=T/(A \cdot \mathrm{GeV})$. For example, by substituting
$A=2$ and $Z=1$, one obtains the cross section
$\sigma_{\overline{\textrm{D}} p}$.

Finally, when anti-nuclei propagate into the heliosphere,
the spectra of charged CR particles are distorted by
the magnetic fields of the solar system and the solar wind.
The effects of solar modulation are quantified by the force-field
approximation~\cite{Gleeson:1968zza}:
\beq{force-field}
\Phi^{\mathrm{TOA}}_{A,Z}(T_{\mathrm{TOA}})=
\left(\frac{2m_A  T_{\mathrm{TOA}}+T^2_{\mathrm{TOA}}}{2m_A T_{\mathrm{IS}}+T^2_{\mathrm{IS}}}\right)
\Phi^{\mathrm{IS}}_{A,Z}(T_{\mathrm{IS}}),
\eeq
where $\Phi$ is the flux of the CR particles, which is related
to the density function $f$ by $\Phi = v f/(4\pi)$, ``TOA'' denotes
the value at the top of the atmosphere of the earth, ``IS'' denotes
the value at the boundary between the interstellar and the heliosphere
and $m$ is the mass of the nucleus.  $T_{\mathrm{IS}}$ is related to
$T_{\mathrm{TOA}}$ as $T_{\mathrm{IS}}=T_{\mathrm{TOA}}+e \phi_F|Z|$.
In this work, we set the value of the Fisk potential fixed at $\phi_F=550$
MV.

\section{The upper limit of DM annihilation cross sections}\label{sec:limit}
\subsection{Constrains from the AMS-02 and HAWC $\bar{p}/p$ data}
The experimental CR $\bar{p}$ data show good agreements with the
scenario of the secondary $\bar{p}$ productions, thus it is expected
to place stringent constraints on the DM annihilation cross sections.
Since the production of anti-nuclei are strongly correlated with
the antiproton, these constraints can greatly reduce the uncertainties
of the maximal flux of $^3\overline{\textrm{D}}$ and $^3\overline{\textrm{He}}$
originated from DM. In the year 2016, AMS-02 group~\cite{Aguilar:2016kjl} released the currently most accurate
$\bar{p}/p$ ratio data in the rigidity range from 1 to 450 GV.
Recently, the HAWC group~\cite{Abeysekara:2018syp}
published the upper limit of the $\bar{p}/p$ ratio
in very high energy regions, which is obtained by
using observations of the moon shadow. In this paper,
we use these two $\bar{p}/p$ ratio data to constrain
the upper limit of the DM annihilation cross sections.

To quantify the uncertainties from the CR propagation,
we consider three different propagation models, i.e.
the ``MIN'', ``MED'' and ``MAX'' models~\cite{JIN:2014ICA}.
The parameters of these models are obtained by making a
global fit to the AMS-02 proton flux and B/C ratio data
using the \GALPROP~ code, and the names of these models
represent the typically minimal, median and
maximal antiproton fluxes due to the uncertainties of propagation.
The parameters of these three models are listed in Tab. \ref{tab:prop_models}.
In our calculations, we adopt the default normalization scheme
in {\tt GALPROP}, which normalize the primary nuclei source term
to reproduce the AMS-02 proton flux at the reference kinetic energy $T =
100$ GeV.

\begin{table}[ht]
\begin{tabular}{ccccccccc}
\hline\hline
 Model         &~$r_h$(kpc)~&$~z_h$(kpc)~&~~$D_0$~~&~~$R_0$(GV)~~&~~~~~$\delta_1/\delta_2$~~~~~&$~V_a$(km/s)~&~$R_{ps}$(GV)~&~~~~$\gamma_{p1}/\gamma_{p2}$~~~~\\ \hline
 MIN           & 20       & 1.8      &  3.53   & 4.0   &  0.3/0.3          &  42.7     & 10.0   &  1.75/2.44               \\ \hline
 MED           & 20       & 3.2      &  6.50   & 4.0   &  0.29/0.29        &  44.8     & 10.0   &  1.79/2.45               \\ \hline
 MAX           & 20       & 6.0      &  10.6   & 4.0   &  0.29/0.29        &  43.4     & 10.0   &  1.81/2.46               \\ \hline\hline
\end{tabular}
\caption{
Values of the main parameters in
the  ``MIN'', ``MED'' and ``MAX'' models derived from fitting to the AMS-02 $B/C$ and
proton data based on the {\tt GALPROP} code~\cite{JIN:2014ICA}. The parameter $D_0$ is in units of
$10^{28}~\mathrm{cm}^2\cdot\mathrm{s}^{-1}$.
}
\label{tab:prop_models}
\end{table}

\begin{figure}
\includegraphics[width=0.82\textwidth]{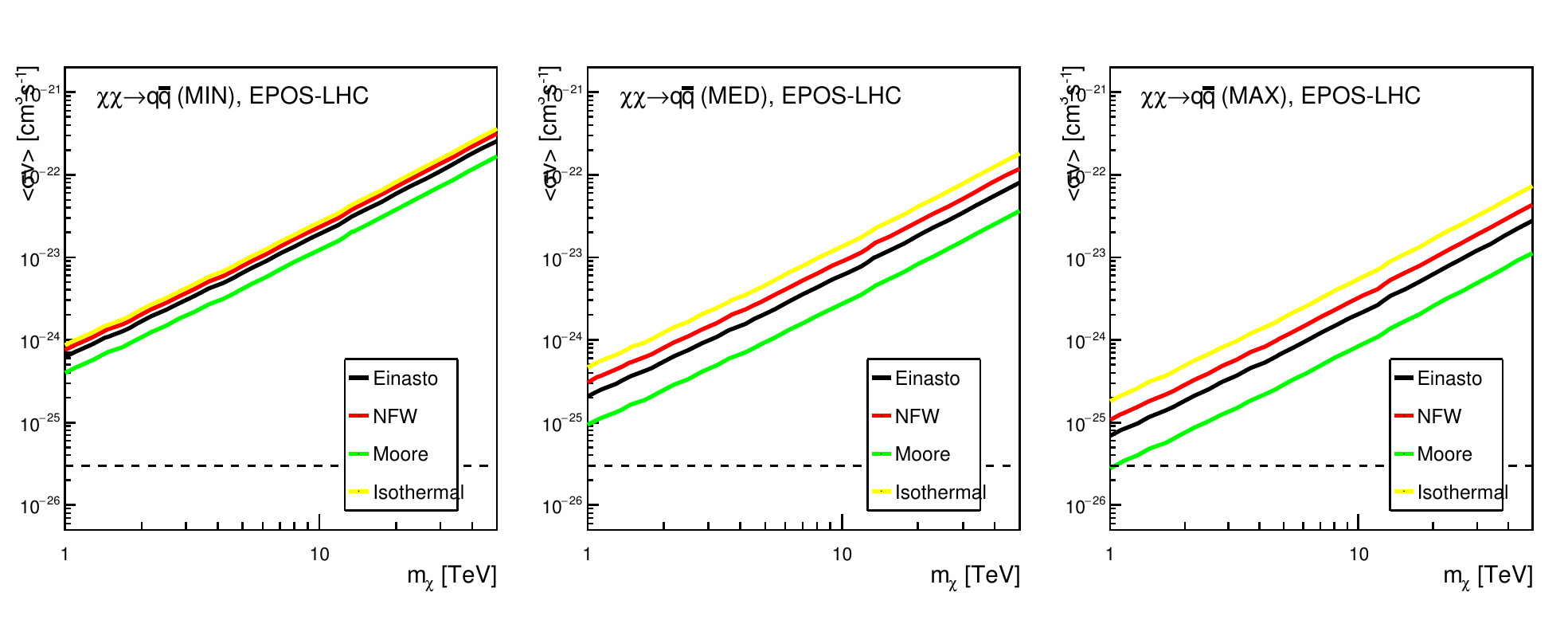}\\
\includegraphics[width=0.82\textwidth]{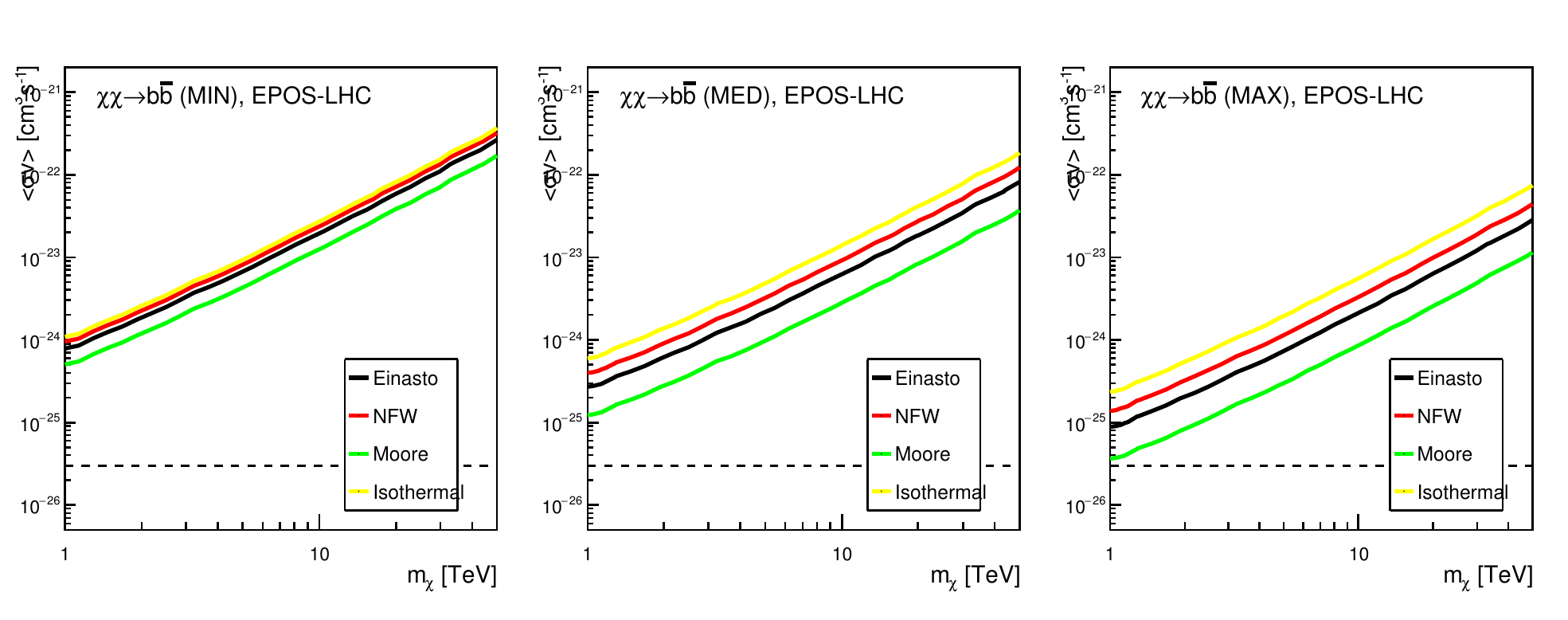}\\
\includegraphics[width=0.82\textwidth]{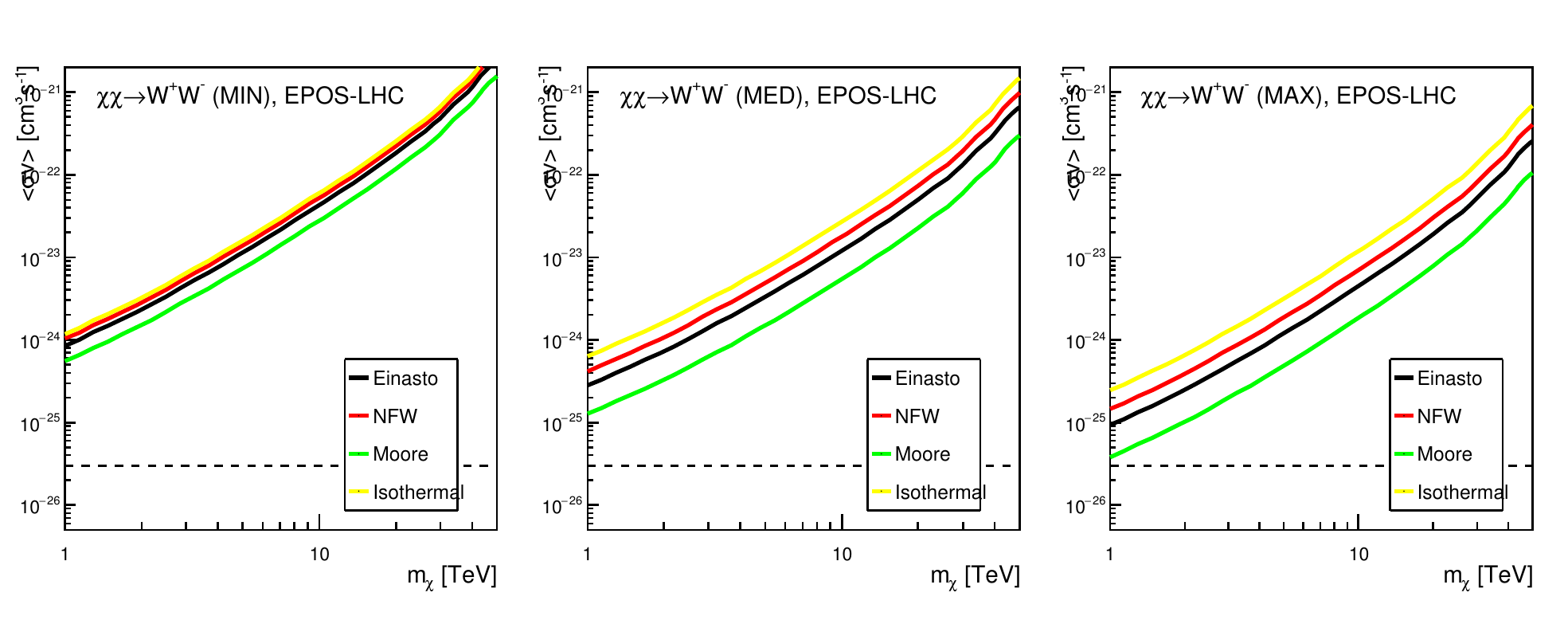}
\caption{The 95$\%$ CL upper limit of DM annihilation cross sections
as functions of DM particle masses, for different decay channels, propagation models
and DM profiles, obtained by using the AMS-02 $\bar{p}/p$ data~\cite{Aguilar:2016kjl}
and the HAWC $\bar{p}/p$ upper limit~\cite{Abeysekara:2018syp}.
The energy spectrum of the secondary $\bar{p}$ are calculated by the {\tt EPOS-LHC} MC generator.
}
\label{fig:limits-EPOS}
\end{figure}

\begin{figure}
\includegraphics[width=0.42\textwidth]{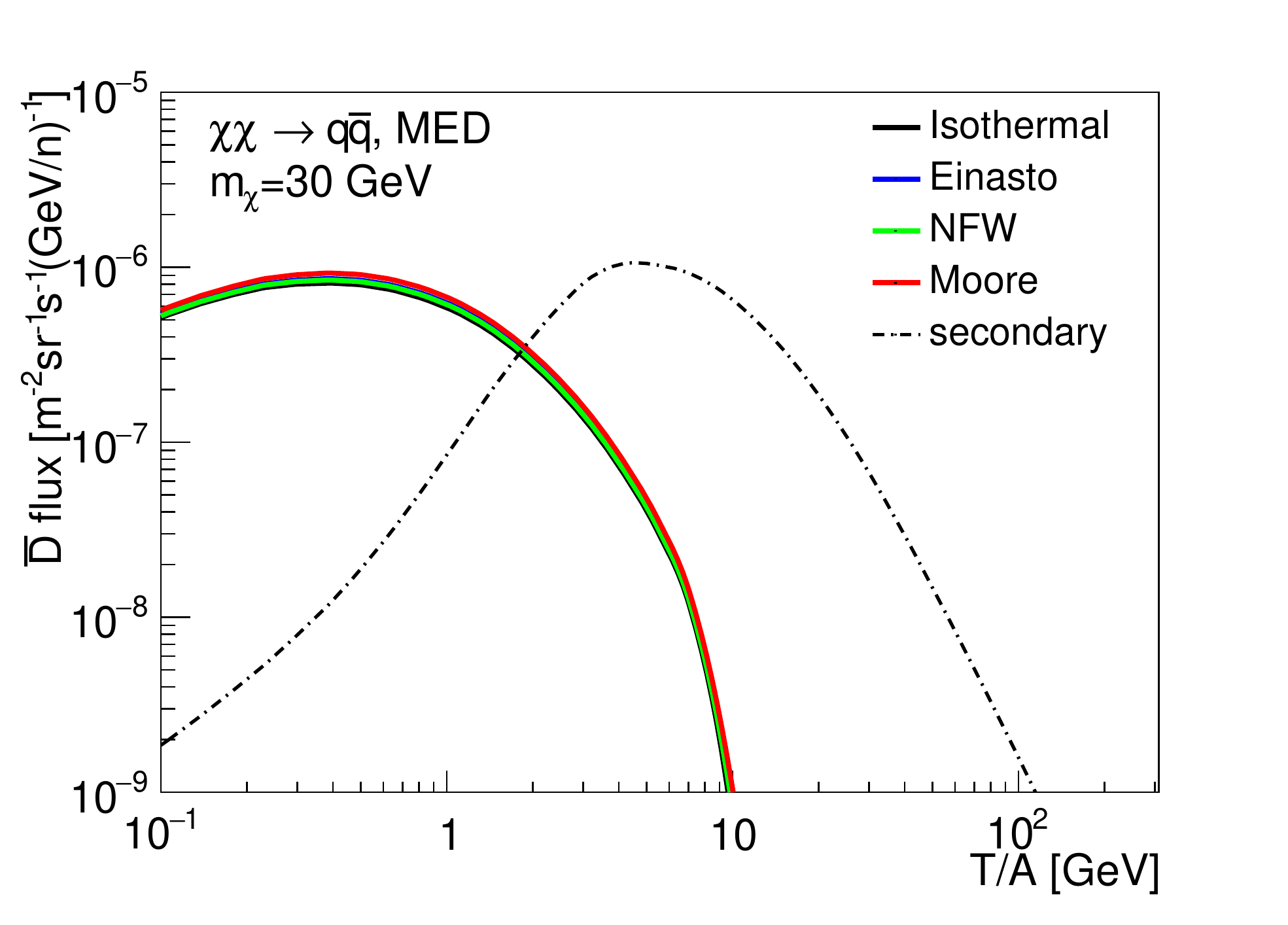}
\includegraphics[width=0.42\textwidth]{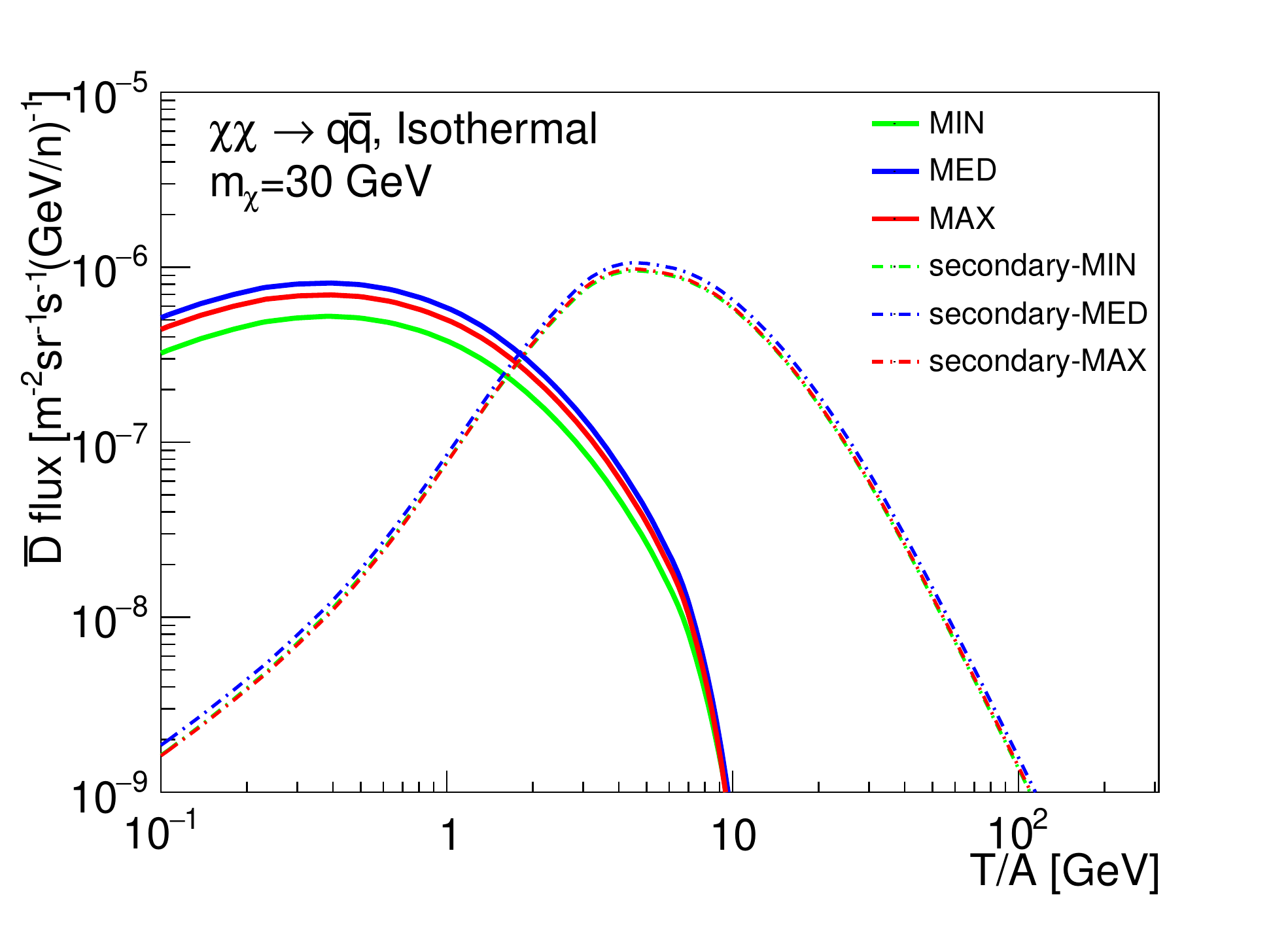}
\caption{
A comparison of the maximal $\Dbar$ fluxes in different propagation models and DM profiles,
with the DM annihilation into $q\bar q $ final states and DM mass $m_{\chi}=30$ GeV.
The DM annihilation cross section are constrained by the AMS-02 and HAWC $\bar{p}/p$ data,
with the spectrum of secondary $\bar{p}$ are calculated by using EPOS-LHC.
\textit{Left)} The flux results in the MED propagation model
with four different DM profiles.
\textit{Right)} The flux results in three different propagation models,
with DM profile fixed to  ``Isothermal''.
The secondary background of the $\overline{\textrm{D}}$ flux are generated
by \EPOSLHC.
}
\label{Dbar_compare}
\end{figure}

The $95\%$ CL upper limits of DM annihilation cross sections
are derived by making a frequentist $\chi^2-$analysis, with
$\chi^2$ defined as:
\beq{chi2_pbar}
\chi^2=\sum_{i}\frac{(f^{\mathrm{th}}_i-f^{\mathrm{exp}}_i)^2}{\sigma^2_i},
\eeq
where $f^{\mathrm{exp}}_i$ is the central value of the experimental $\bar{p}/p$ ratio,
$\sigma_i$ is the data error, $f^{\mathrm{exp}}_i$ is theoretical prediction of $\bar{p}/p$
and $i$ denotes the $i-$th data point. For the $95\%$ CL upper limits of $\bar{p}/p$
given by HAWC experiment, we set $f^{\mathrm{exp}}_i=0$ and the value of upper limit corresponds
to $1.96\sigma_i$. For a specific DM mass, we first calculate the minimal value of $\chi^2$,
and then the $95\%$ CL upper limits on DM annihilation cross sections correspond to
$\Delta\chi^2=3.84$ for one parameter. See Ref.~\cite{Jin:2015sqa} for more details.
The upper limits for different annihilation channels are shown in Fig. \ref{fig:limits-EPOS},
with the production cross section of secondary $\bar{p}$ generated by \EPOSLHC.
We can see that the differences between the upper limits for various propagation models
and DM profiles can reach one to two orders of magnitude.
As shown in previous works \cite{Carlson:2014ssa,Lin:2018avl}, the final flux uncertainties
from propagation models and DM profiles can be larger than one order of magnitude. However, if we use these
cross section upper limits to constrain the maximal flux of $\Dbar$ and $\Hebar$, the final uncertainties from
the propagation model and the DM profile can be reduced to merely $30\%$~\cite{Li:2018dxj}.
A comparison of the maximal $\Dbar$ fluxes in different propagation models and DM profiles are demonstrated
in Fig.~\ref{Dbar_compare}, it can be seen that the uncertainties are small.

\subsection{Constrains from the HESS galactic center $\gamma$-ray data}
The galactic center is a promising place for detecting DM interactions, for
the expected high DM density.
The 10-year HESS $\gamma$-ray data~\cite{Abdallah:2016ygi}
focus on a small area around the galactic center, and
well constraints the DM annihilation cross sections for
``Cuspy'' type DM profiles which have large gradient at the inner galactic halo, such as the
``Einasto'' profile and the ``NFW'' profile. For DM mass $m_{\chi} \gtrsim 1$ TeV,
the galactic center $\gamma$-ray can place more stringent upper limits
than the $\bar{p}/p$ data. However, for ``Cored'' type DM profiles which are flat near
the galactic center, such as the ``Isothermal'' profile, the constraints from the $\gamma$-ray data are relatively weak.

\begin{figure}
\includegraphics[width=0.32\textwidth]{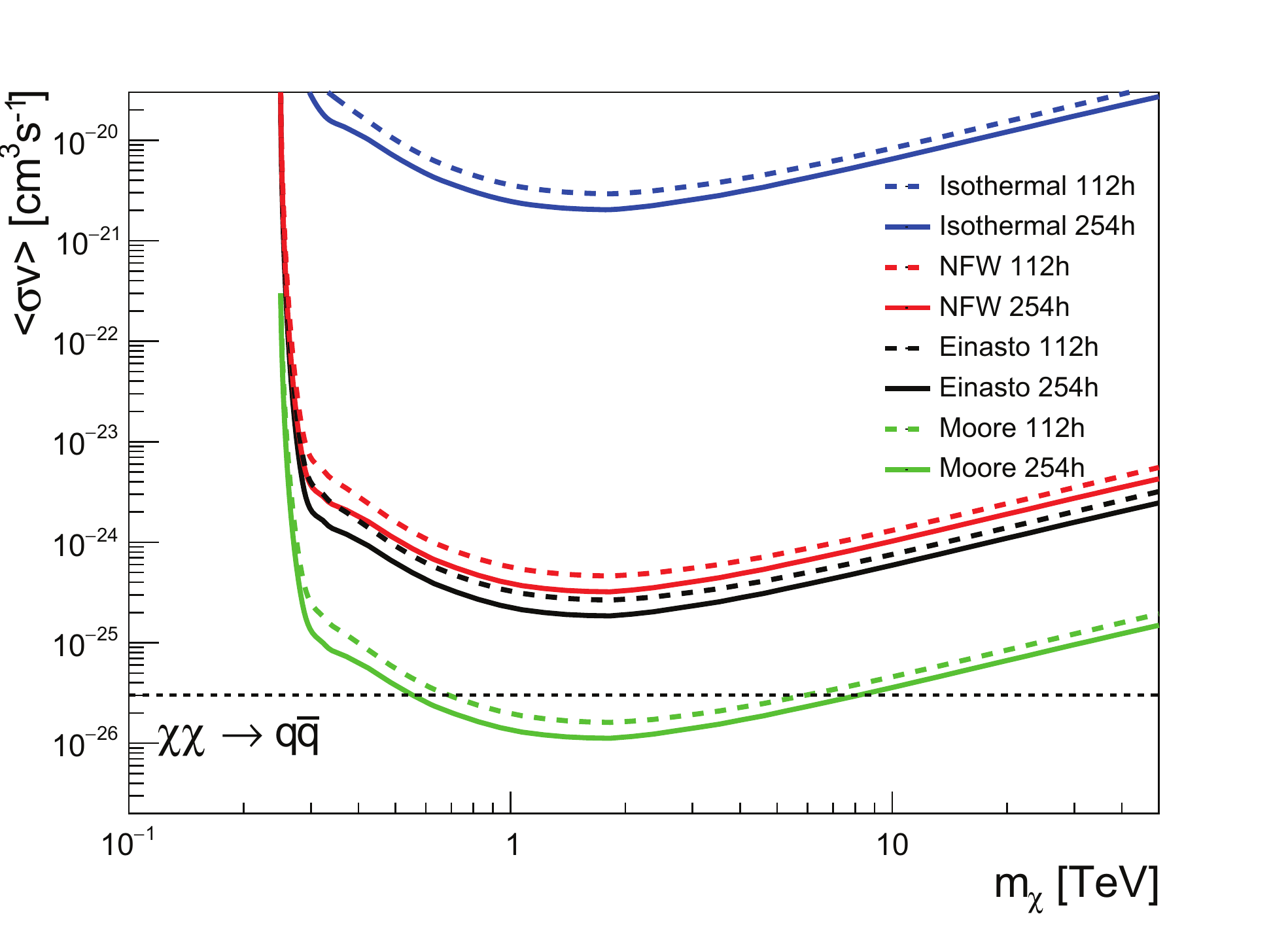}
\includegraphics[width=0.32\textwidth]{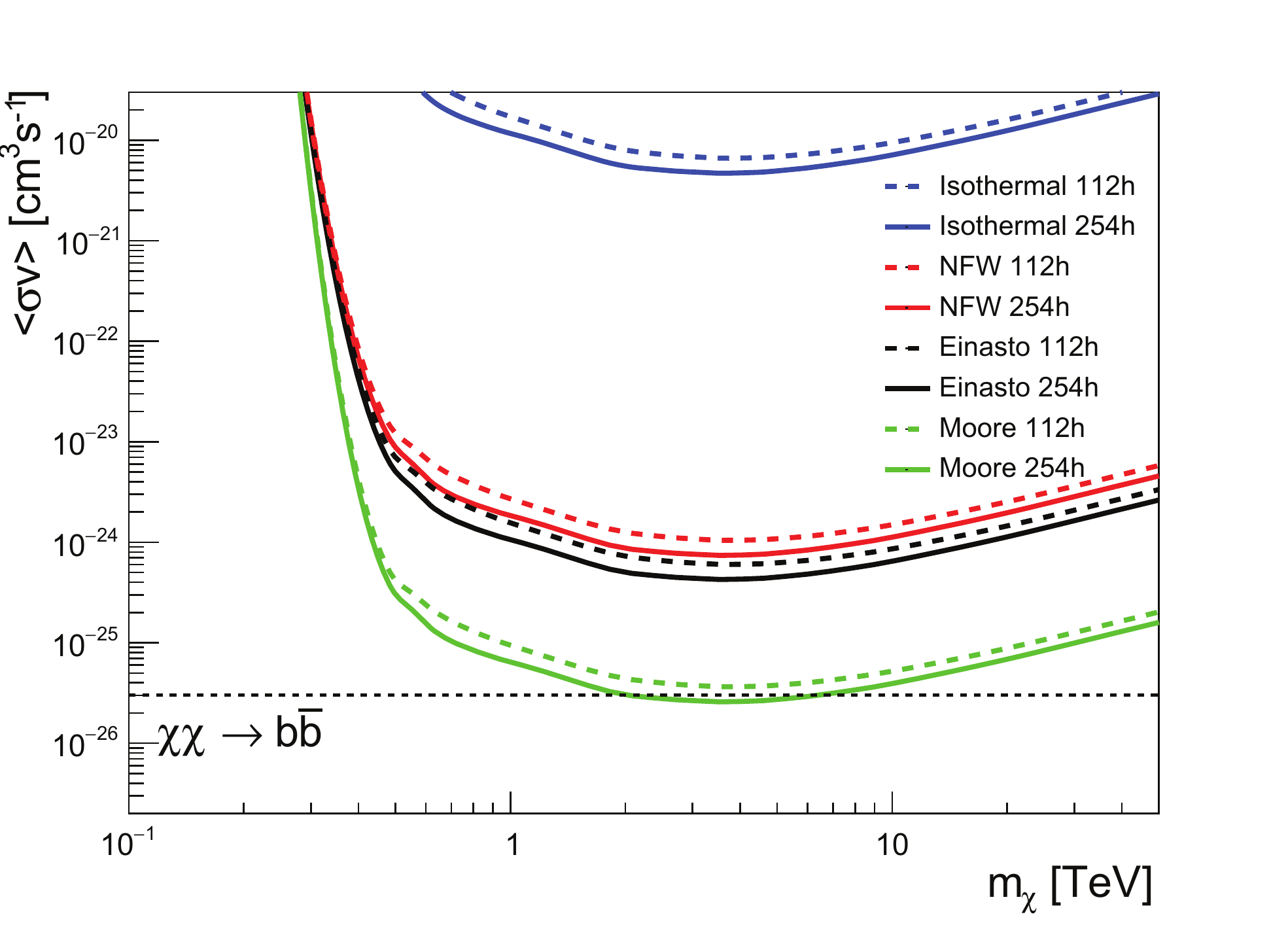}
\includegraphics[width=0.32\textwidth]{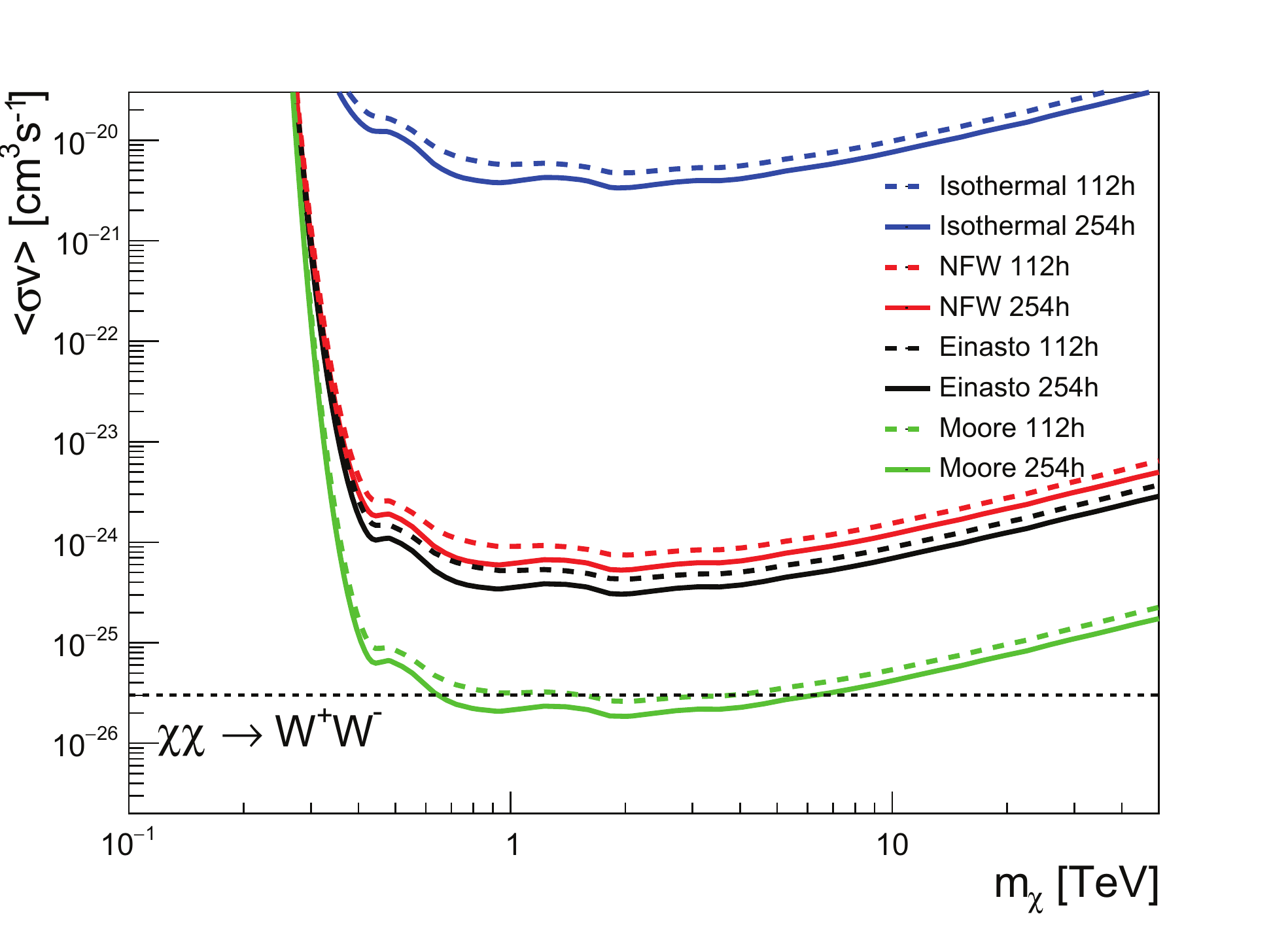}
\caption{The 95$\%$ CL upper limit of DM annihilation cross sections for different annihilation channels and DM profiles, obtained by using the HESS GC $\gamma$-ray data with 112 hours exposure~\cite{Abramowski:2011hc}. The 254 hours results are derived
by rescaling the data errors.
}
\label{our_gamma_limit}
\end{figure}

The latest galactic center $\gamma$-ray analysis with 254 hours exposure was
published in 2016 by HESS~\cite{Abdallah:2016ygi}, the upper limits are calculated for
several DM profiles and DM annihilation channels,
but the $\gamma$-ray flux data are not released in public.
Since the relative statistical error of a data point is inversely proportional to
the square root of the number of counts in the bins, we can approximately
estimate the 254 hours results by rescaling a previous HESS $\gamma$-ray data~\cite{Abramowski:2011hc},
which was published in 2011 and the exposure time are 112 hours.
To obtain the upper limits for other DM profiles and channels, we perform a $\chi^2$ analysis to
the 112 hours $\gamma$-ray flux data and estimate the 254 hours results
by rescaling the data errors by a factor $\sqrt{112/254}$.
The flux residual is defined as
$R^{\mathrm{exp}}=F^{\mathrm{exp}}_{\mathrm{Sr}}-F^{\mathrm{exp}}_{\mathrm{Bg}}$,
where $F^{\mathrm{exp}}_{\mathrm{Sr}}$ and $F^{\mathrm{exp}}_{\mathrm{Bg}}$ are experimental
$\gamma$-ray flux data from the source region and from the background region respectively,
and the error of $R^{\mathrm{exp}}$ is provided in Ref.~\cite{Abramowski:2011hc}.
$R^{\mathrm{th}}=F^{\mathrm{th}}_{\mathrm{Sr}}-F^{\mathrm{th}}_{\mathrm{Bg}}$ is the
theoretical value of the flux residual, and the differential flux
$F^{\mathrm{th}}$ is calculated by the following formula:

\beq{gamma_flux}
F^{\mathrm{th}}=\frac{\mathrm{d}\Phi_{\gamma}}{\Omega\mathrm{d}E_{\gamma}}=\frac{\langle\sigma v\rangle}{8\pi m_{\chi}^2\Omega}\frac{\mathrm{d} N_{\gamma}}{\mathrm{d}E_{\gamma}}
\int_{\Omega}\int_{\mathrm{l.o.s}}\rho^2(r(s,\theta))\mathrm{d}s \mathrm{d}\Omega,
\eeq
where $\mathrm{d} N_{\gamma}/\mathrm{d}E_{\gamma}$ is the energy spectrum of photon produced in one
DM annihilation, $\Omega$ is the total spherical angle of the source or background regions. We adopt
the reflected background technique described in Ref.~\cite{Abramowski:2011hc} to determine the source region and background
regions, and calculate the flux residual to make the $\chi^2$ analysis. We randomly choose 540
pointing positions near the GC, with the maximal distance between the pointing position and the GC is $1.5^{\circ}$.
We first calculate the
minimal $\chi^2$ value, and then the $95\%$ CL upper limit corresponds to $\Delta\chi^2=3.84$.
The results are shown in Fig.~\ref{our_gamma_limit}. We can see that there are large gaps between the
upper limits for different DM profiles. As expected, the constraints are
stringent for DM profiles that are cuspy at GC, but for a cored one
like ``Isothermal'' profile, the limits are rather weak. 
For various DM profiles, the constraints from $\bar{p}/p$ data are are similar, 
because the DM densities in the diffusion halo are similar in different DM profiles,
except for the GC region. However, for $\gamma$-ray data, the GC region provides
the most strigent constraints for heavy DM particles, which leads to the large gaps between DM profiles.

\begin{figure}
\includegraphics[width=0.32\textwidth]{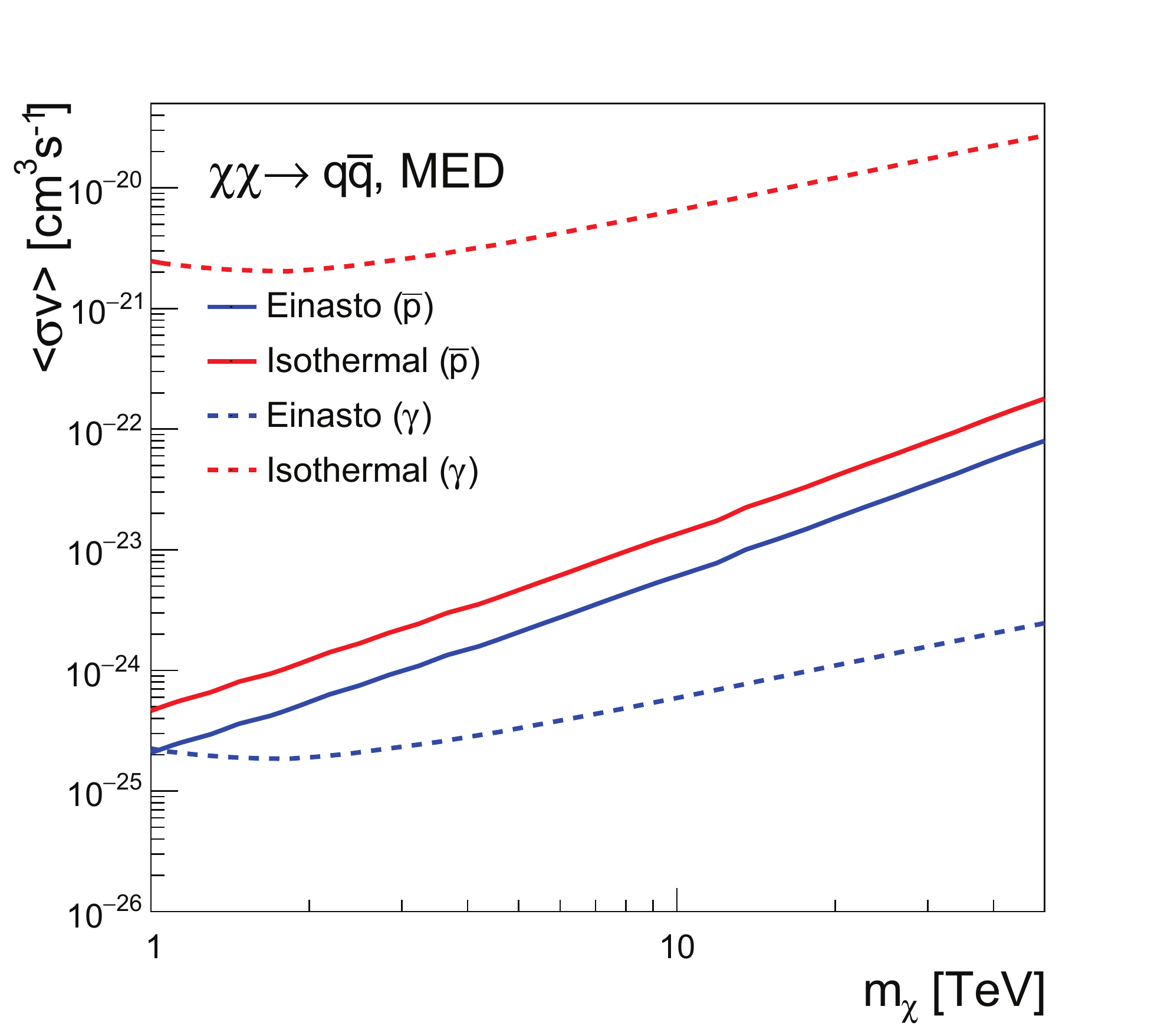}
\includegraphics[width=0.32\textwidth]{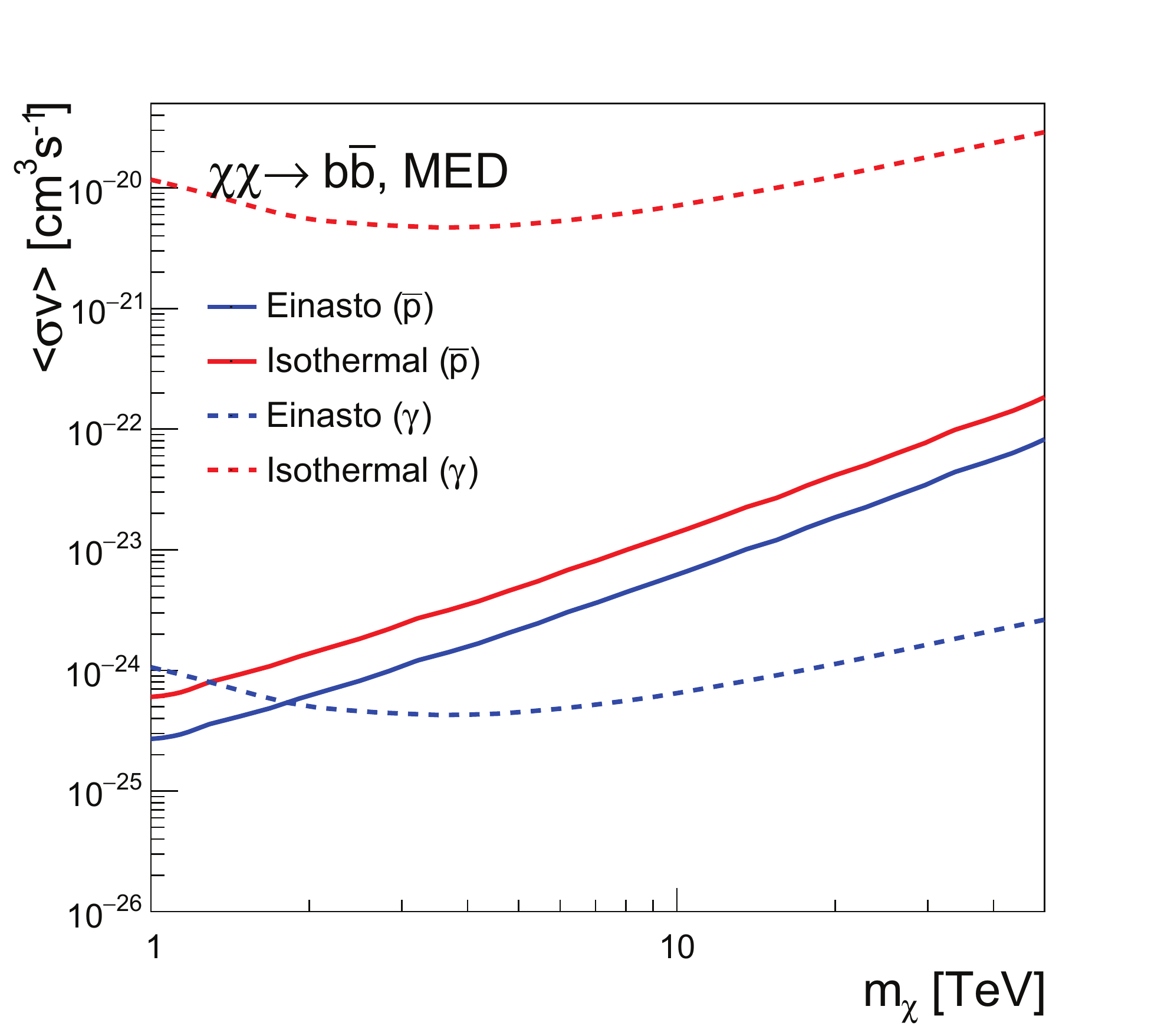}
\includegraphics[width=0.32\textwidth]{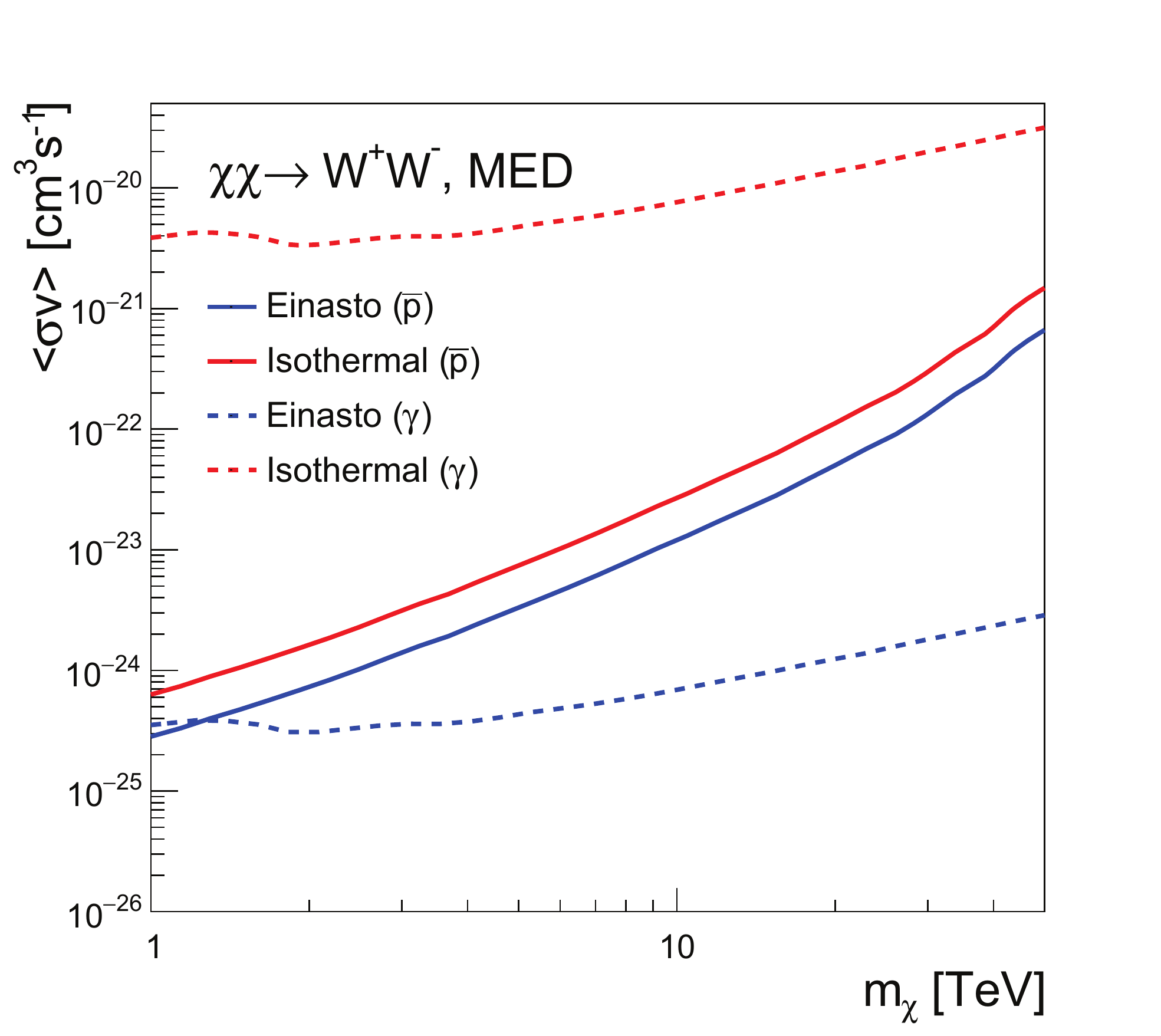}
\caption{A comparison between the upper limits from the AMS-02 and HAWC $\bar{p}/p$ data
and the HESS 254 hours GC $\gamma$-ray data. The solid lines stand for constrains from the $\bar{p}/p$ data, and
the dashed lines are the $\gamma$-ray limits. The ``MED'' propagation model are used
as a benchmark model.
}
\label{compare_ams_hess}
\end{figure}

We make a comparison between the upper limits from the $\bar{p}/p$ data
and from the $\gamma-$ray data, which is presented in Fig.~\ref{compare_ams_hess}.
We use the ``MED'' propagation model as the benchmark model, and the ``Isothermal''
and ``Einasto'' profiles represent the typical ``Cored'' and ``Cuspy'' profiles respectively.
The ``Isothermal'' profile is parameterized as follows:
\beq{Isothermal}
\rho_{\mathrm{DM}}=\rho_{\odot}\frac{1+(r_{\odot}/r_{\mathrm{Iso}})^2}{1+(r/r_{\odot})^2}~,
\eeq
where $\rho_{\odot}=0.43~\mathrm{GeV/cm}^{3}$ is the local DM energy density,
$r_{\mathrm{Iso}}=3.5~\mathrm{kpc}$. The ``Einasto'' profile can be written as:
\beq{Einasto}
\rho_{\mathrm{DM}}=\rho_{\odot}\exp\left[-\left(\frac{2}{\alpha}\right)\left(\frac{r^{\alpha}-
r^{\alpha}_{\odot}}{r^{\alpha}_{\mathrm{Ein}}}\right)\right]~,
\eeq
where $\alpha=0.17$ and $r_{\mathrm{Ein}}=20$ kpc.
As we can see, in all annihilation channels, for ``Isothermal'' profile, the upper
limits from $\bar{p}/p$ data are much more stringent than the ones from $\gamma-$ray data,
while for ``Einasto'' profile, the $\gamma-$ray data gives stricter limits than
$\bar{p}/p$ for DM mass larger than 2 TeV.

It is worth mention that the Fermi-LAT experiment~\cite{Atwood:2009ez} also collected a large amount
of $\gamma-$ray data near the GC, and the relatively large region of interest
can reduce the large gaps between different DM profiles. However, the energy range
of Fermi-LAT observations (from 20 MeV to more than 300 GeV) are much lower than HESS, and thus
provides a weaker limitation on large DM mass. For example, the analysis in Ref.~\cite{TheFermi-LAT:2017vmf}
show that, for a steep profile like ``NFW'', HESS gives stronger constraints than Fermi-LAT at
$E\gtrsim1$ TeV. For a flat profile like ``Isothermal'', the Fermi-LAT limits derived in
Ref.~\cite{Ackermann:2012rg} are slightly weaker than the $\bar{p}/p$ constraints at $E\approx 10$ TeV.
By these facts, to make the most conservative conclusion,
we use the $\bar{p}/p$ limits to calculate the maximal $\Dbar$ and $\Hebar$ fluxes
in ``Isothermal'' profile, and use the HESS $\gamma-$ray limits for ``Einasto'' profile in the following sections.

\section{The flux of $\overline{\mathrm{D}}$ and $\overline{\mathrm{He}}$ for large DM mass}\label{sec:flux}
\subsection{DM direct annihilation}
With the DM annihilation cross section upper limits at hand,
we can derive the maximal $\Dbar$ and $\Hebar$ fluxes for
different annihilation channels, propagation models, and
DM profiles. As shown in Fig.~\ref{Dbar_compare}, by constraining
the DM annihilation cross section with the $\bar{p}/p$ data,
the uncertainties from propagation models are small, thus
we present our results in the ``MED'' propagation model,
and other models do not affect our conclusions. For illustration purpose, the particle
mass of the heavy DM are set to be $m_{\chi}=10$ and 50 TeV,
which are smaller than the unitarity bound for a self-conjugate DM~\cite{PhysRevD.100.043029}.
However, it is worth mention that our analysis is largely model independent, we do not assume
the DM have thermal origins.

Currently, the AMS-02 detector has the strongest detection ability for both $\Dbar$
and $\Hebar$. For $\Dbar$ flux, the AMS-02 detection sensitivity is given in Ref.~\cite{Aramaki:2015pii},
while the sensitivity for $\Hebar$ are only released in terms
of $\Hebar/\mathrm{He}$ ratio in Ref.~\cite{Kounine:2010js}. To study
the detection prospects of AMS-02, we present our results in terms of $\Dbar$ fluxes and
$\Hebar/\mathrm{He}$ ratios.

\begin{figure}
\includegraphics[width=0.31\textwidth]{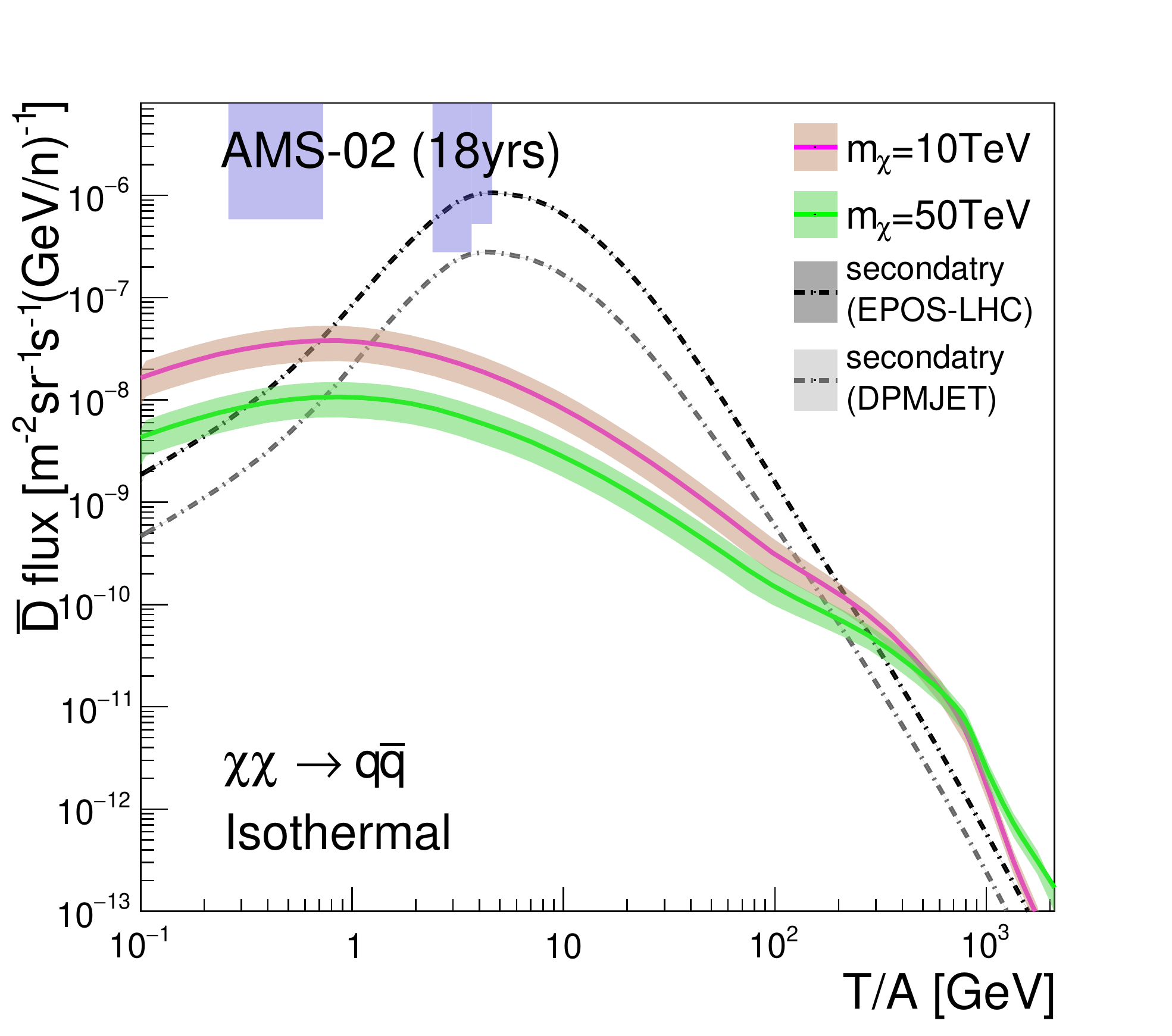}
\includegraphics[width=0.31\textwidth]{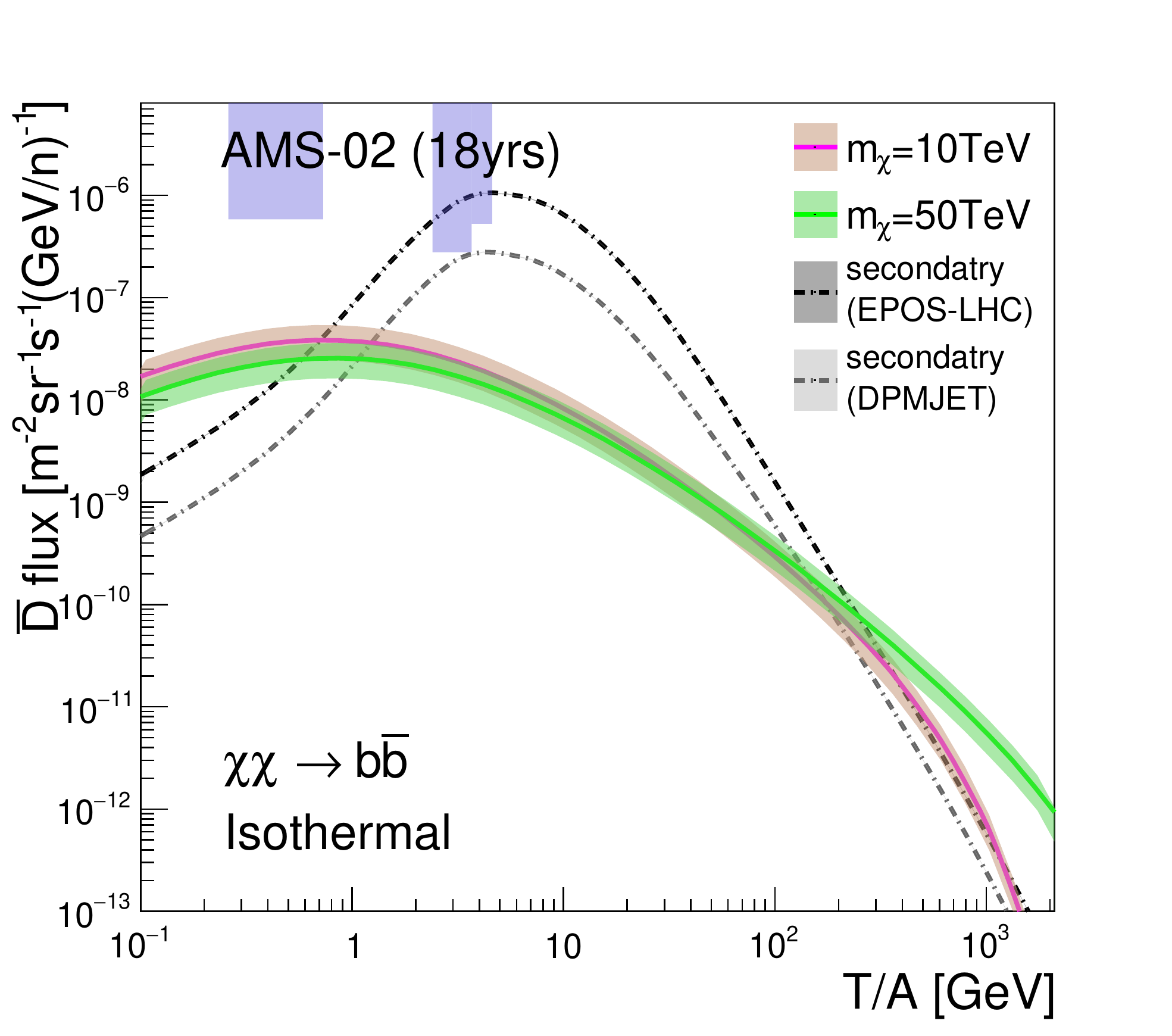}
\includegraphics[width=0.31\textwidth]{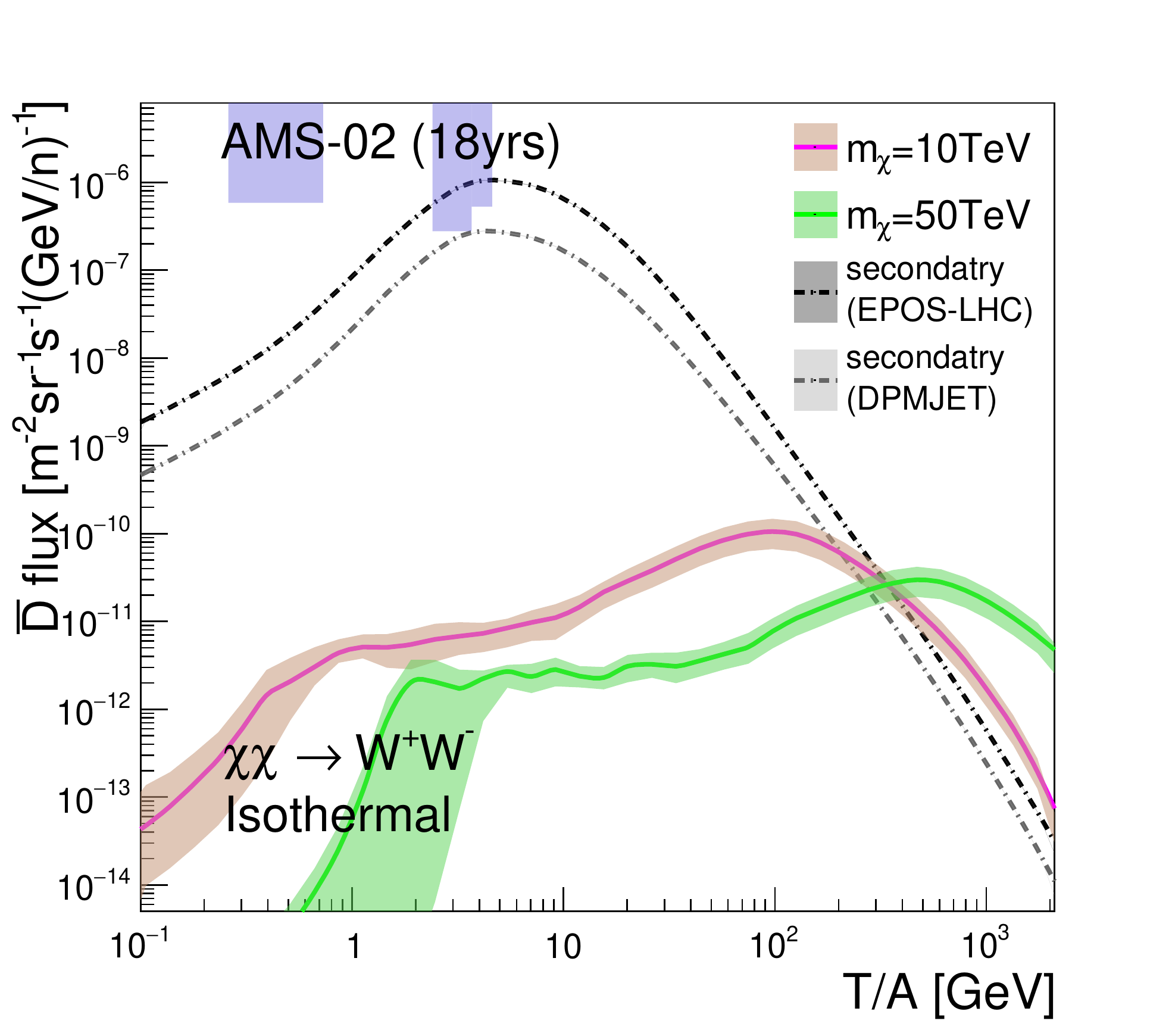}\\
\includegraphics[width=0.31\textwidth]{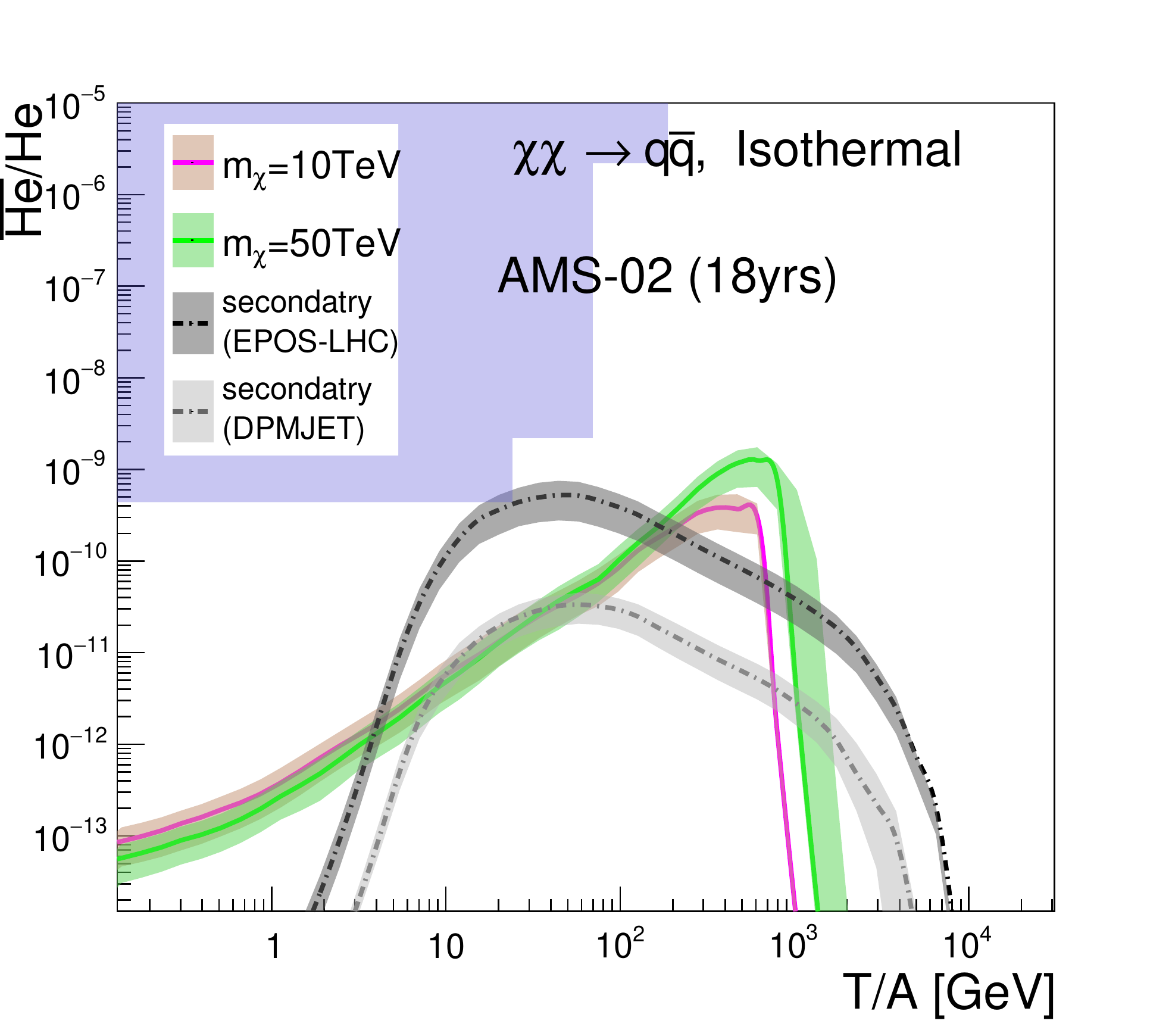}
\includegraphics[width=0.31\textwidth]{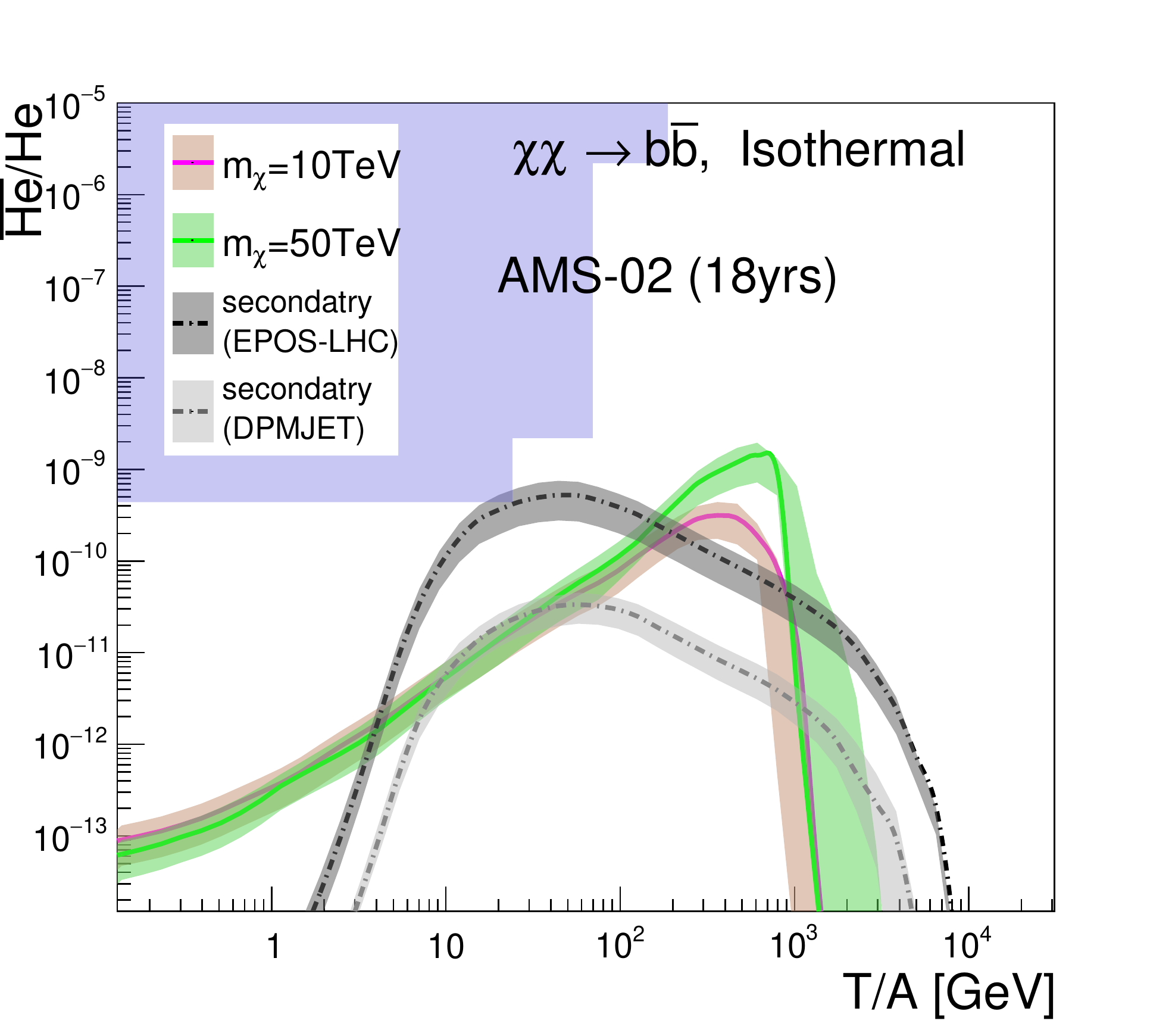}
\includegraphics[width=0.31\textwidth]{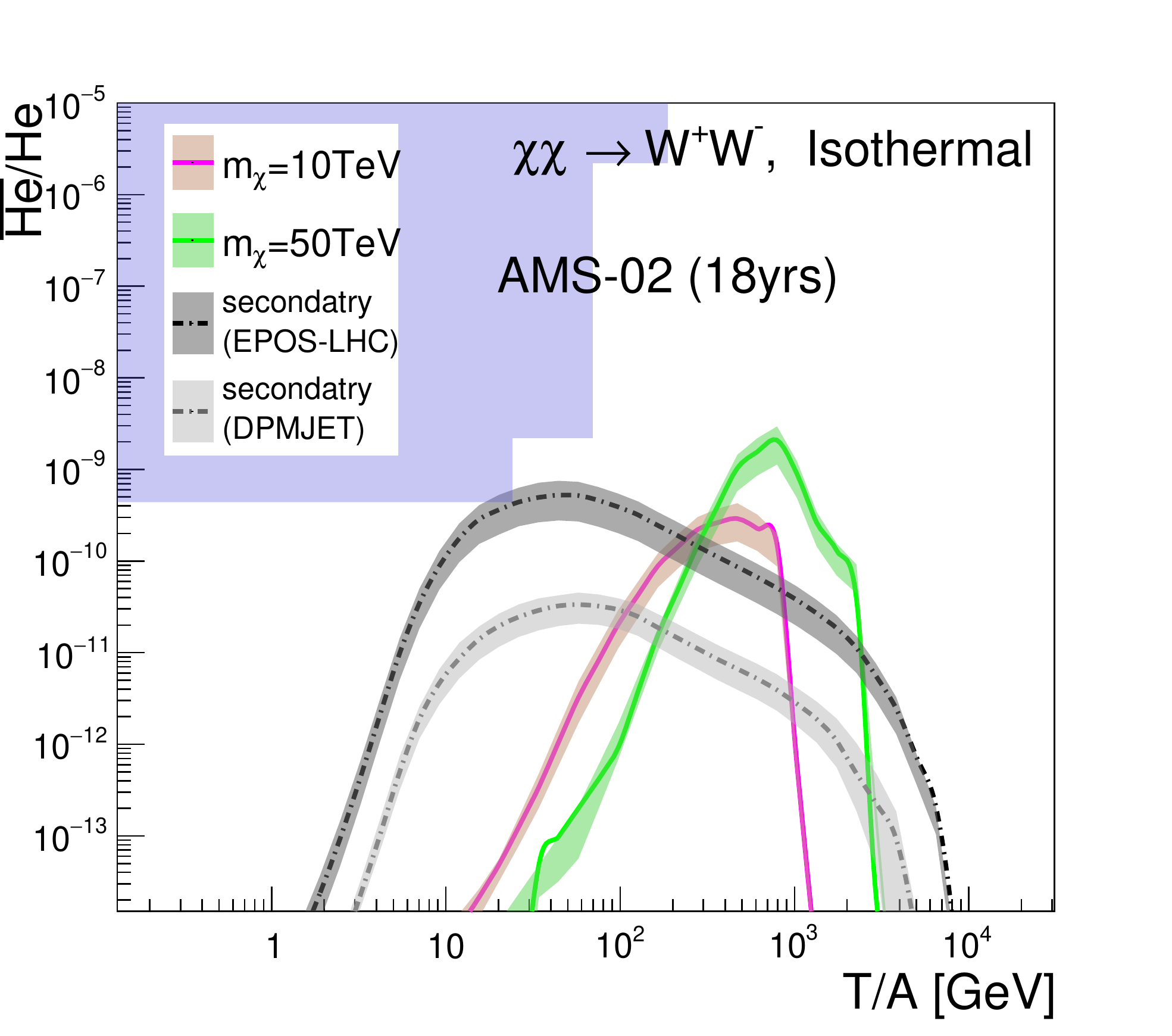}
\caption{The $\Dbar$ fluxes (top figures) and the $\Hebar/\mathrm{He}$ ratios
(bottom figures), for the ``Isothermal'' profile with large DM mass and ``MED'' propagation model.
The DM annihilation cross section is constrained by the AMS-02 and HAWC $\bar{p}/p$ data.
The blue shade represents the 18-year AMS-02 detection sensitivity.
}
\label{flux_AMS}
\end{figure}

The results for ``Isothermal'' profile with $m_{\chi}=10$ and 50 TeV are presented in
Fig.~\ref{flux_AMS}, with the DM annihilation cross section constrained by the AMS-02 and HAWC $\bar{p}/p$ data.
The top three figures show the results about $\Dbar$ fluxes for different annihilation channels, while the bottom three
figures present the $\Hebar/\mathrm{He}$ ratio results.
The blue shades represent the prospective
AMS-02 detection sensitivity after 18 years of data collection, and
the error bands show the uncertainties
from coalescence momenta. Note that for $\Dbar$, the error bands for the secondary background
are thinner than the line width.
We can see that for $\Dbar$, the DM contributions exceed the secondary backgrounds
in the energy region $T/A\gtrsim300$ GeV. For $b\bar{b}$ and $W^+W^-$
channels and $m_{\chi}=50$ TeV, the excess can be as large as one
order of magnitude. Similarly, for $\Hebar$, the excess exist at the kinetic energy
around 800 GeV per nucleon, and the primary $\Hebar$ fluxes originated by the annihilation of DM can be 20 times
larger than the secondary background with $m_{\chi}=50$ TeV.
Despite the fluxes of these anti-nuclei are small at high kinetic
energies, and are far below the AMS-02 sensitivities, these excesses can be
promising windows for future detections.

\begin{figure}
\includegraphics[width=0.31\textwidth]{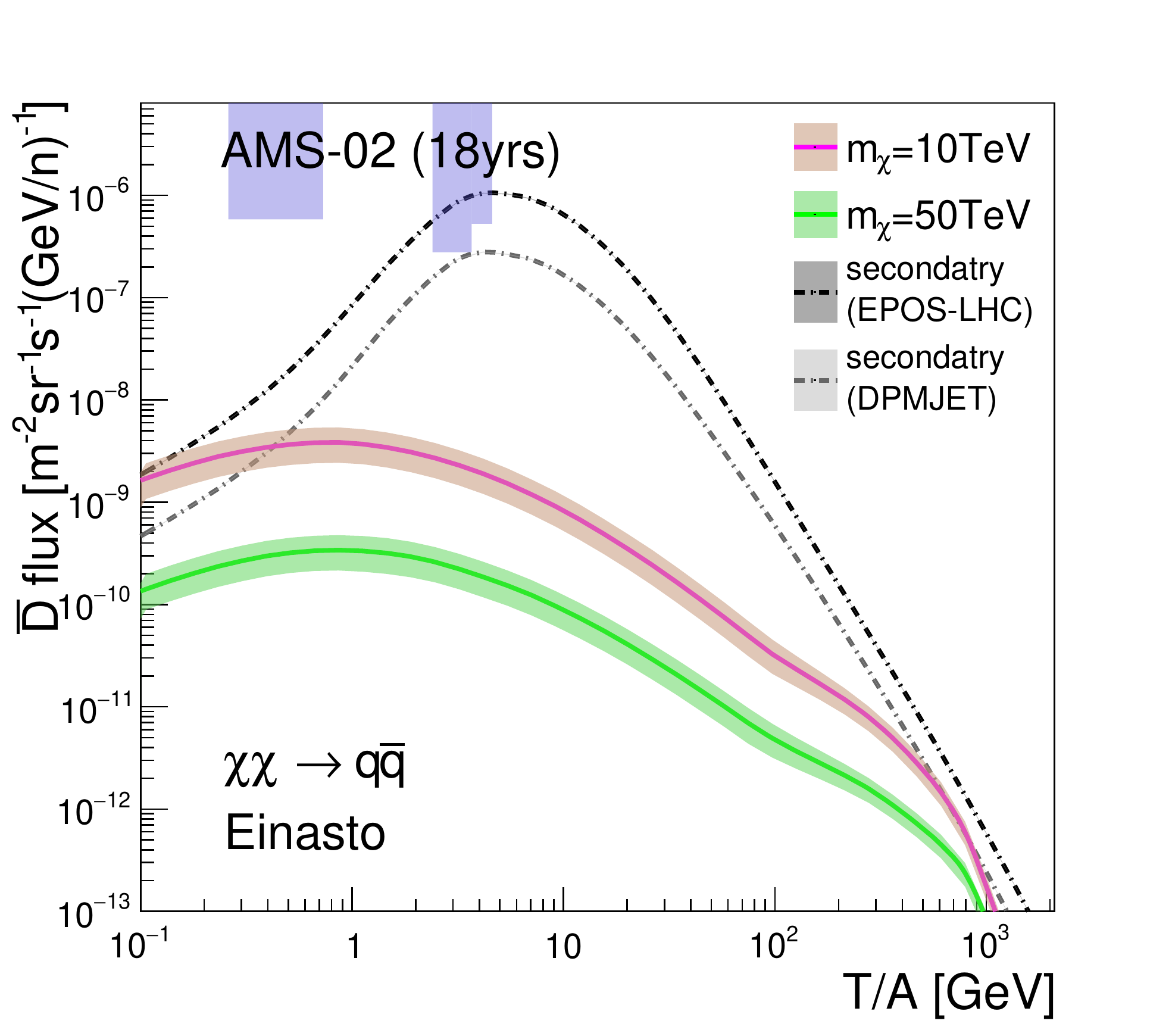}
\includegraphics[width=0.31\textwidth]{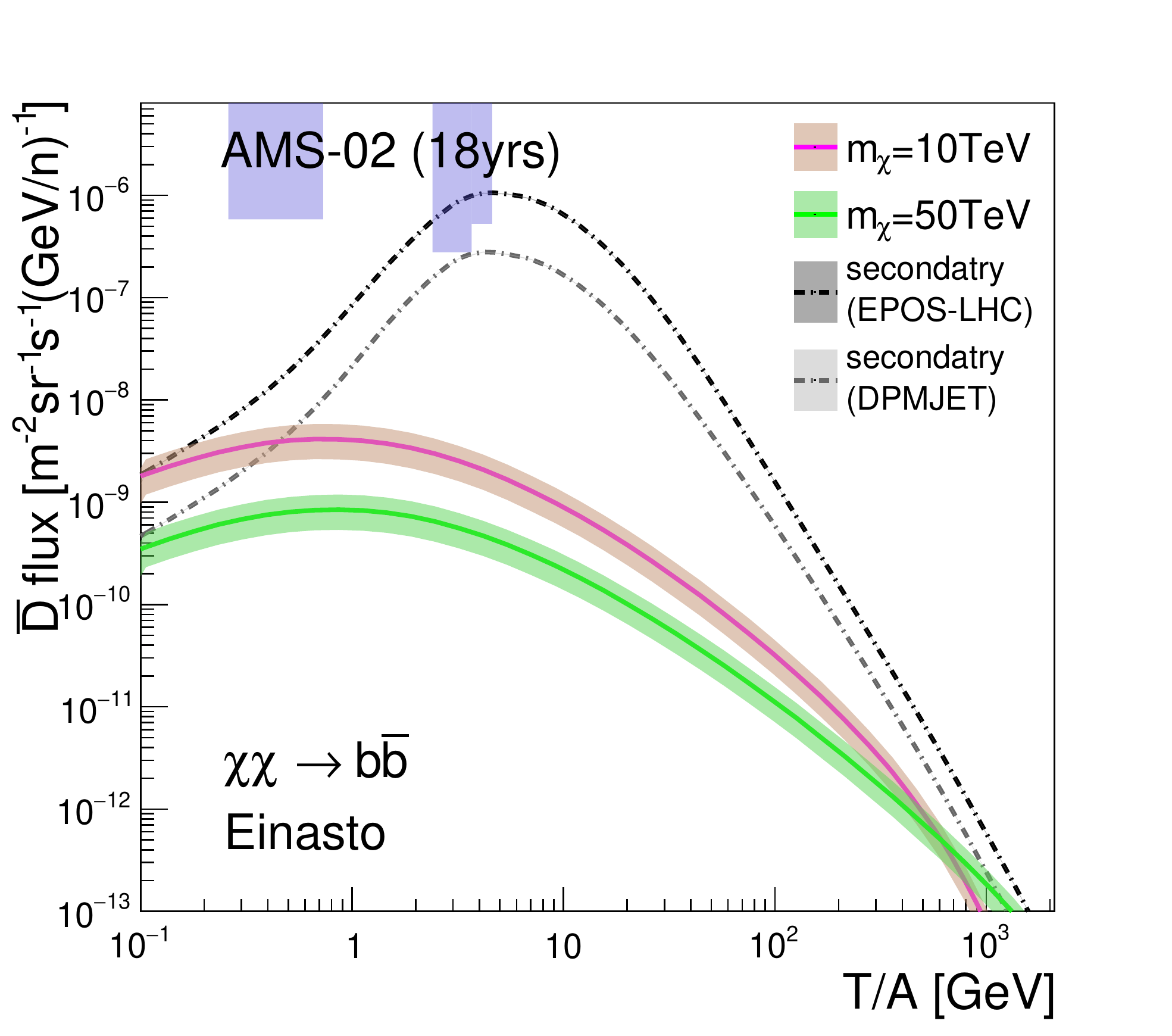}
\includegraphics[width=0.31\textwidth]{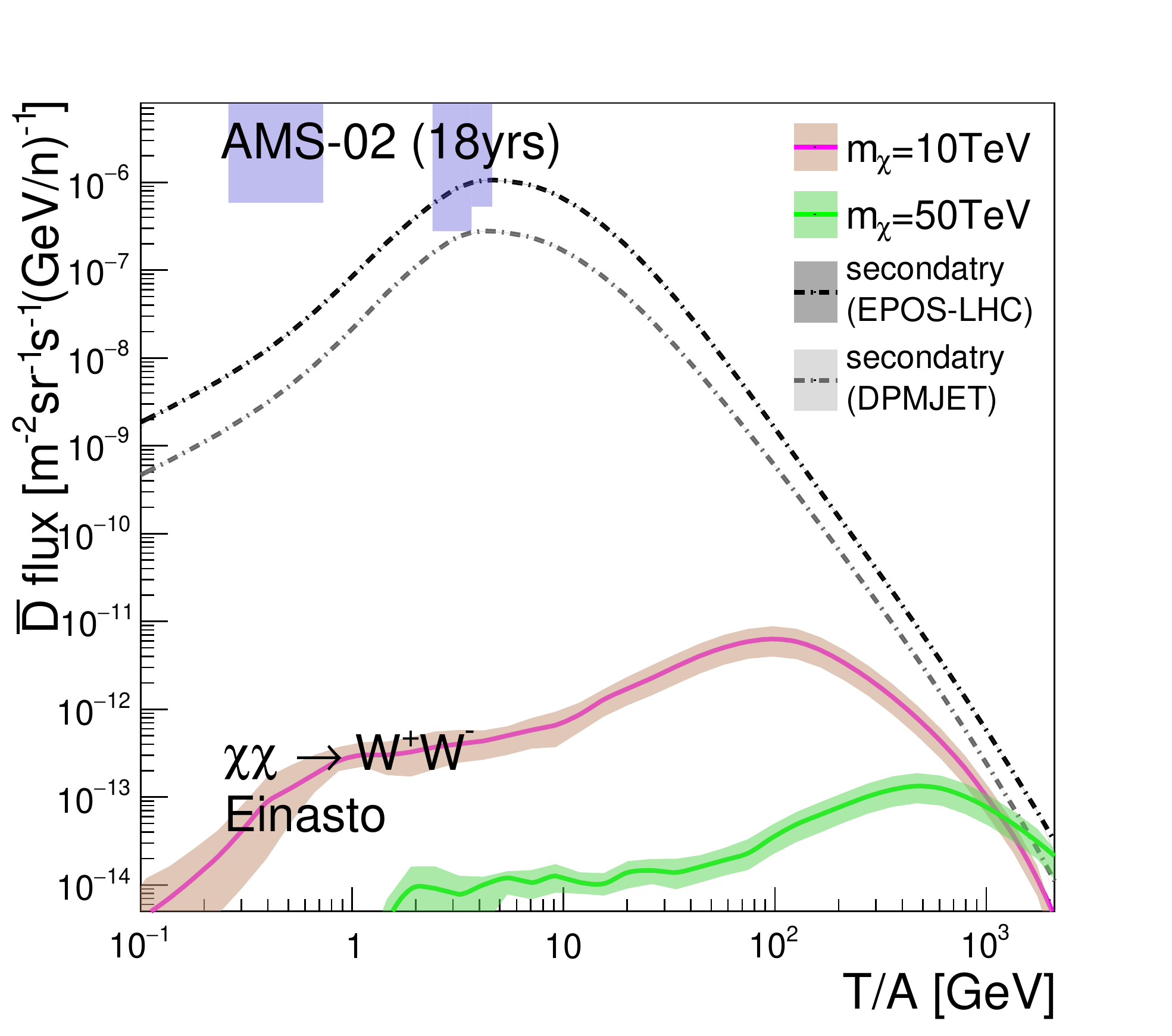}\\
\includegraphics[width=0.31\textwidth]{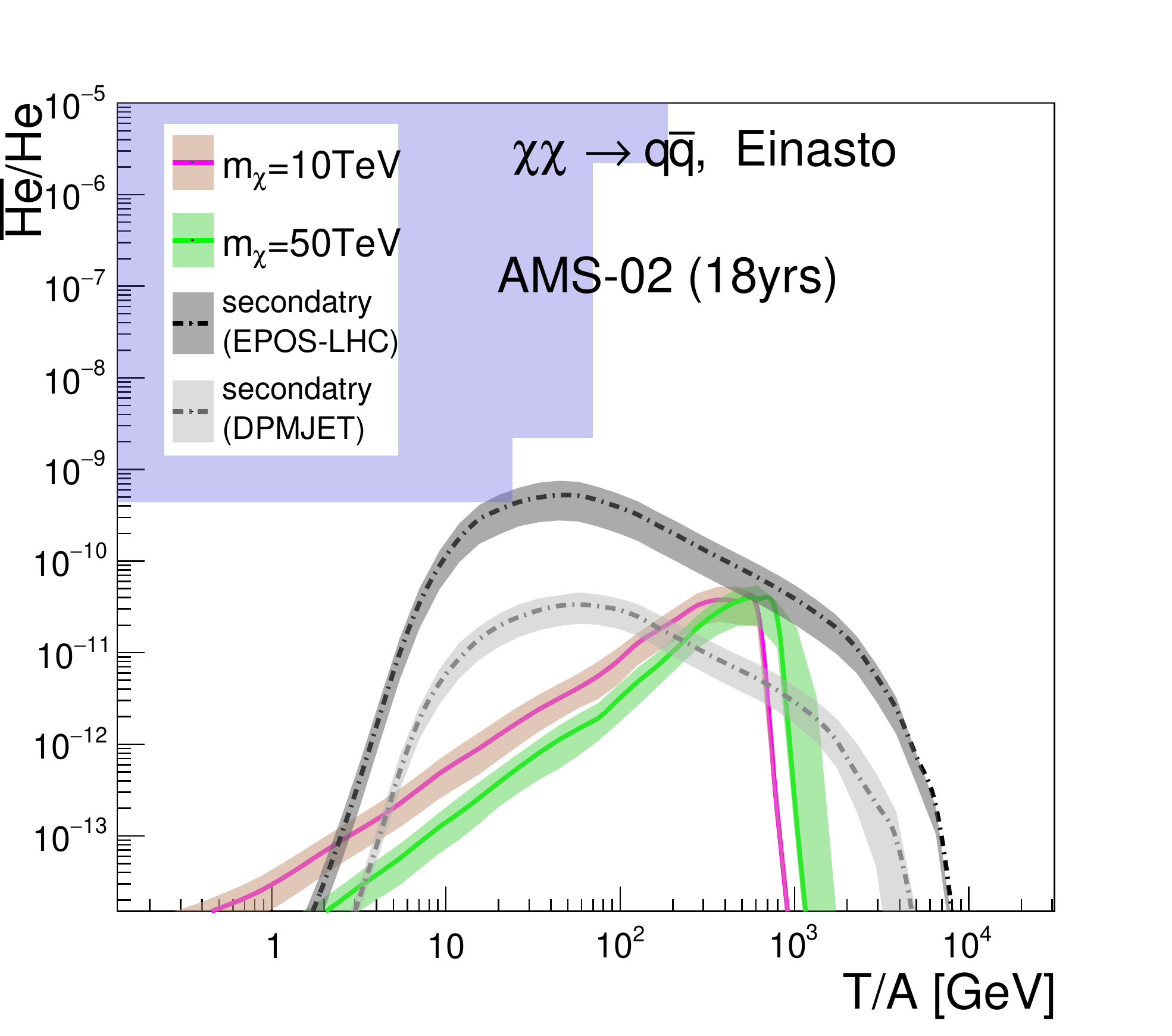}
\includegraphics[width=0.31\textwidth]{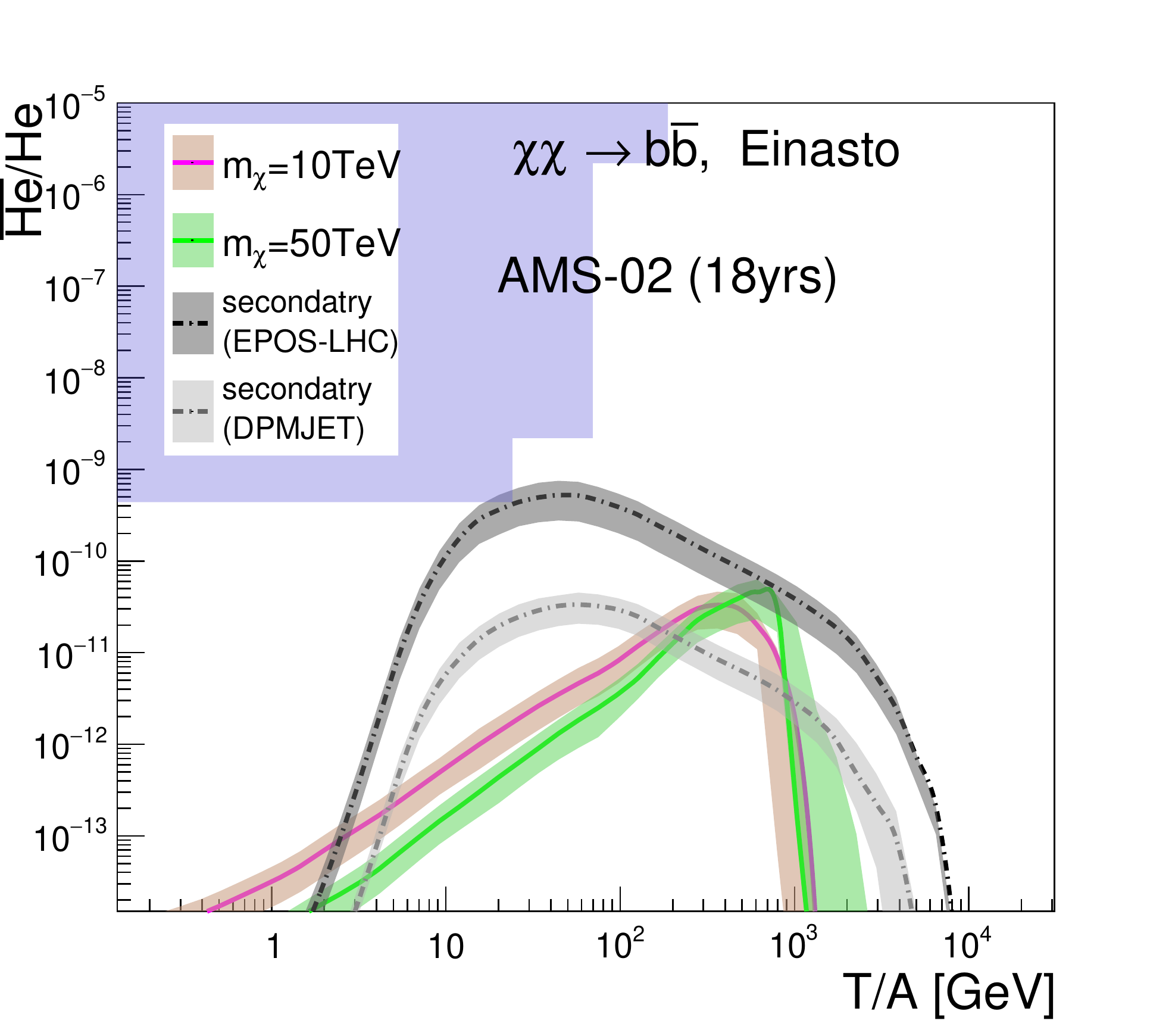}
\includegraphics[width=0.31\textwidth]{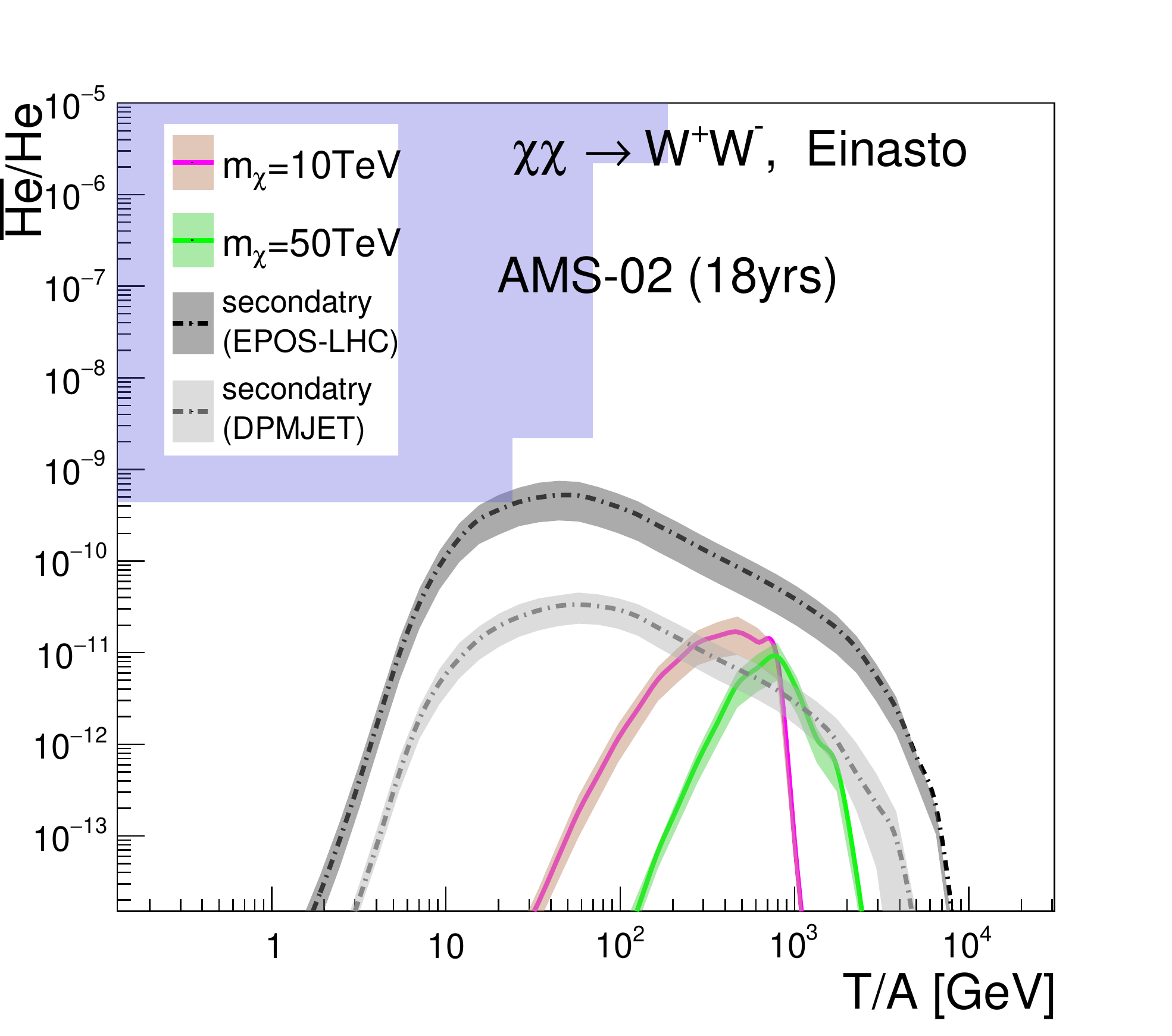}
\caption{The $\Dbar$ fluxes (top figures) and the $\Hebar/\mathrm{He}$ ratios
(bottom figures), for the ``Einasto'' profile with large DM mass and ``MED'' propagation model.
The DM annihilation cross section is constrained by the HESS 10-year GC $\gamma-$ray data.
The blue shade represents the 18-year AMS-02 detection sensitivity.
}
\label{flux_HESS}
\end{figure}

The results for ``Einasto'' profile are shown in Fig.~\ref{flux_HESS},
with the DM annihilation cross sections are constrained by the HESS 10-year GC $\gamma-$ray data.
For $\Dbar$, the DM
contributions are below the secondary background in all energy regions and annihilation channels,
and thus the high window closes.
However, for $\Hebar$, the conclusion depends
on the choice of MC generators. The DM contributions are lower than the secondary background
given by \EPOSLHC, but can still exceed the {\DPMJET} background.

\subsection{DM annihilation through mediators}
We also consider the process $\chi\chi \rightarrow \phi\phi \rightarrow f\bar{f}f\bar{f}$,
that two DM particles first annihilate into a couple of mediators, and then
the mediators decay into standard model final states. If the mass of the
mediator are much smaller than the DM particle, the mediator
wound be highly boosted, thus the $\Dbar$ and $\Hebar$ produced in
this process are expected to assemble in high energy regions,
which provides a high signal-to-background ratio for the high energy window.

\begin{figure}
\includegraphics[width=0.32\textwidth]{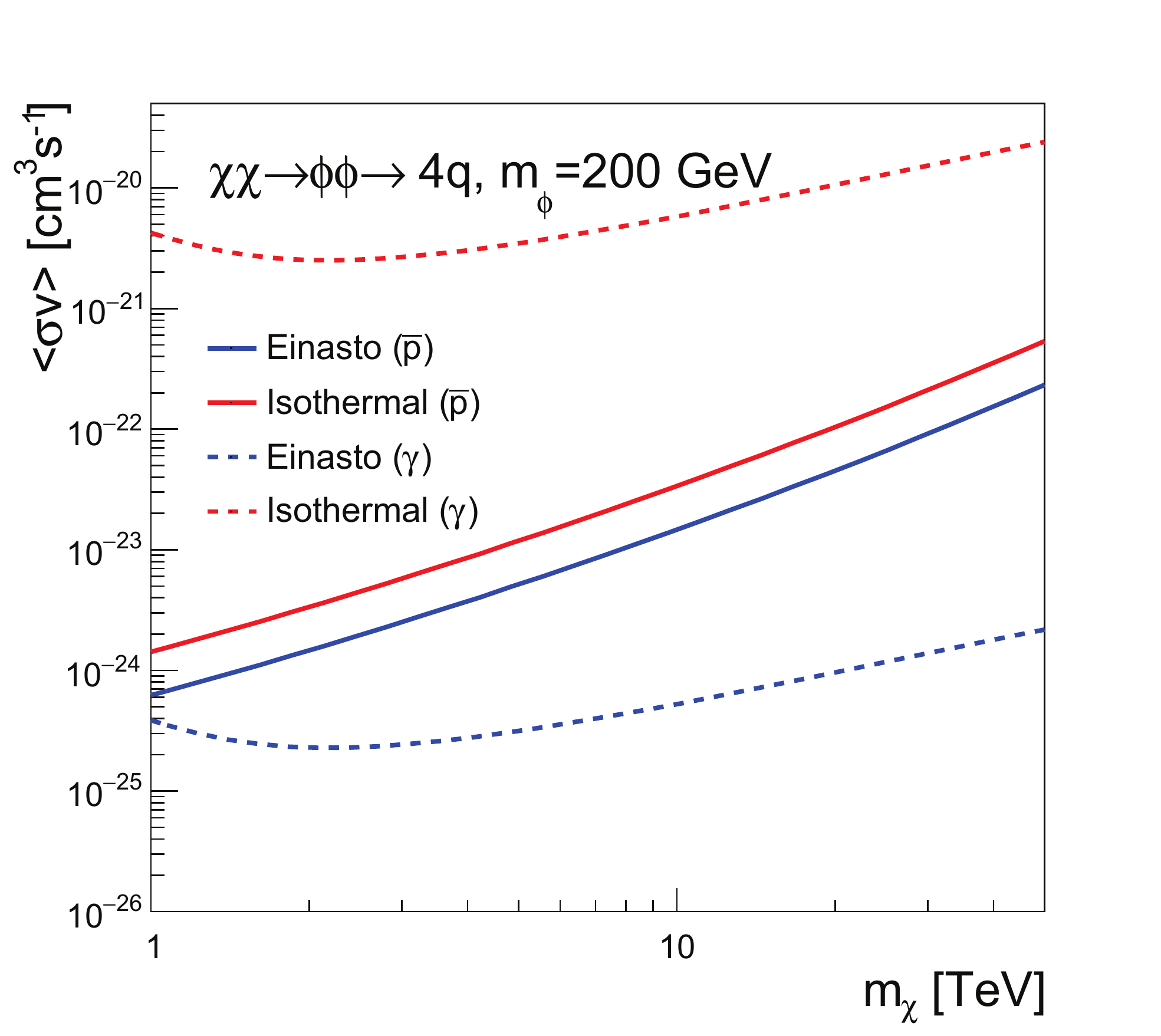}
\includegraphics[width=0.32\textwidth]{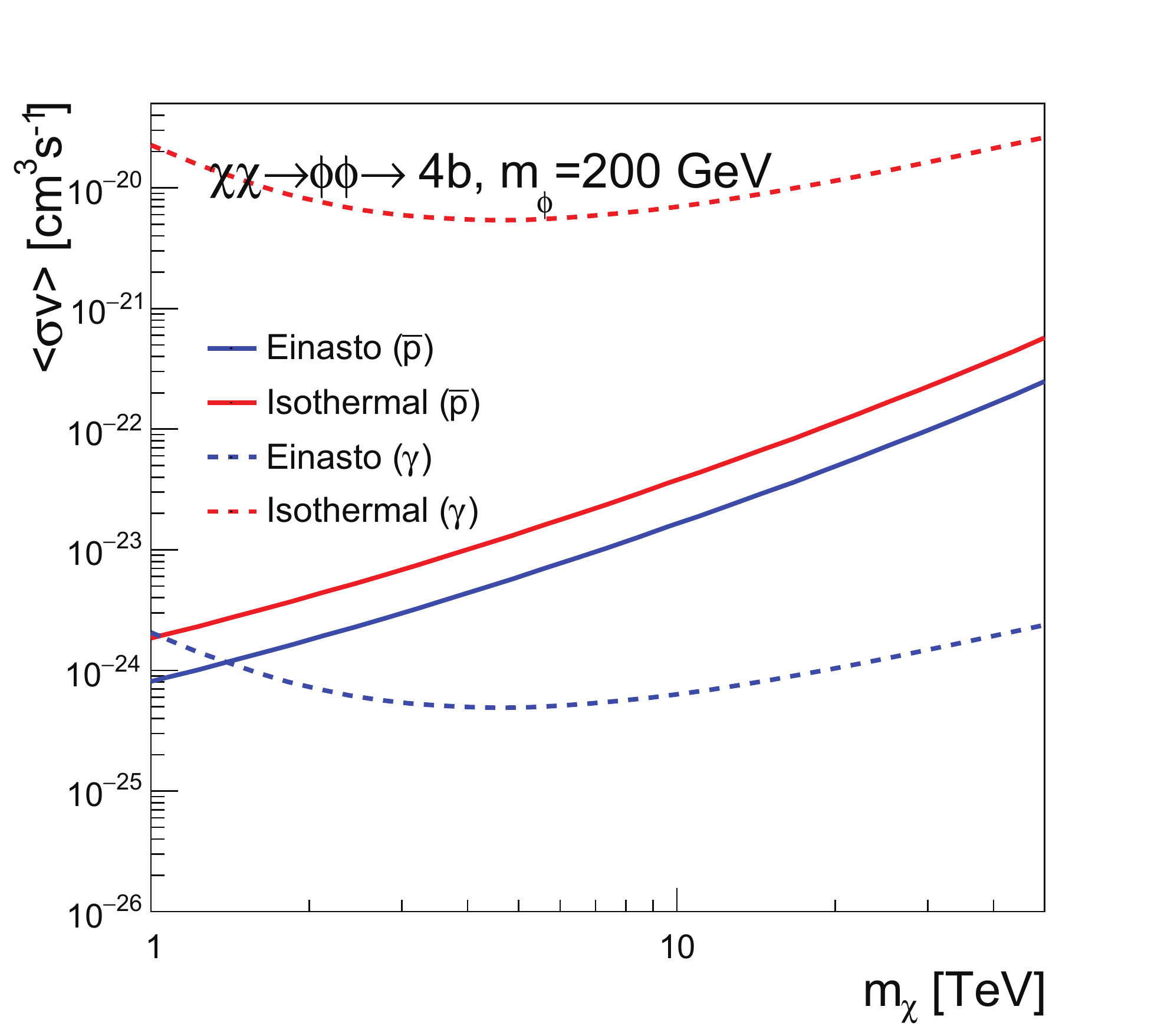}
\includegraphics[width=0.32\textwidth]{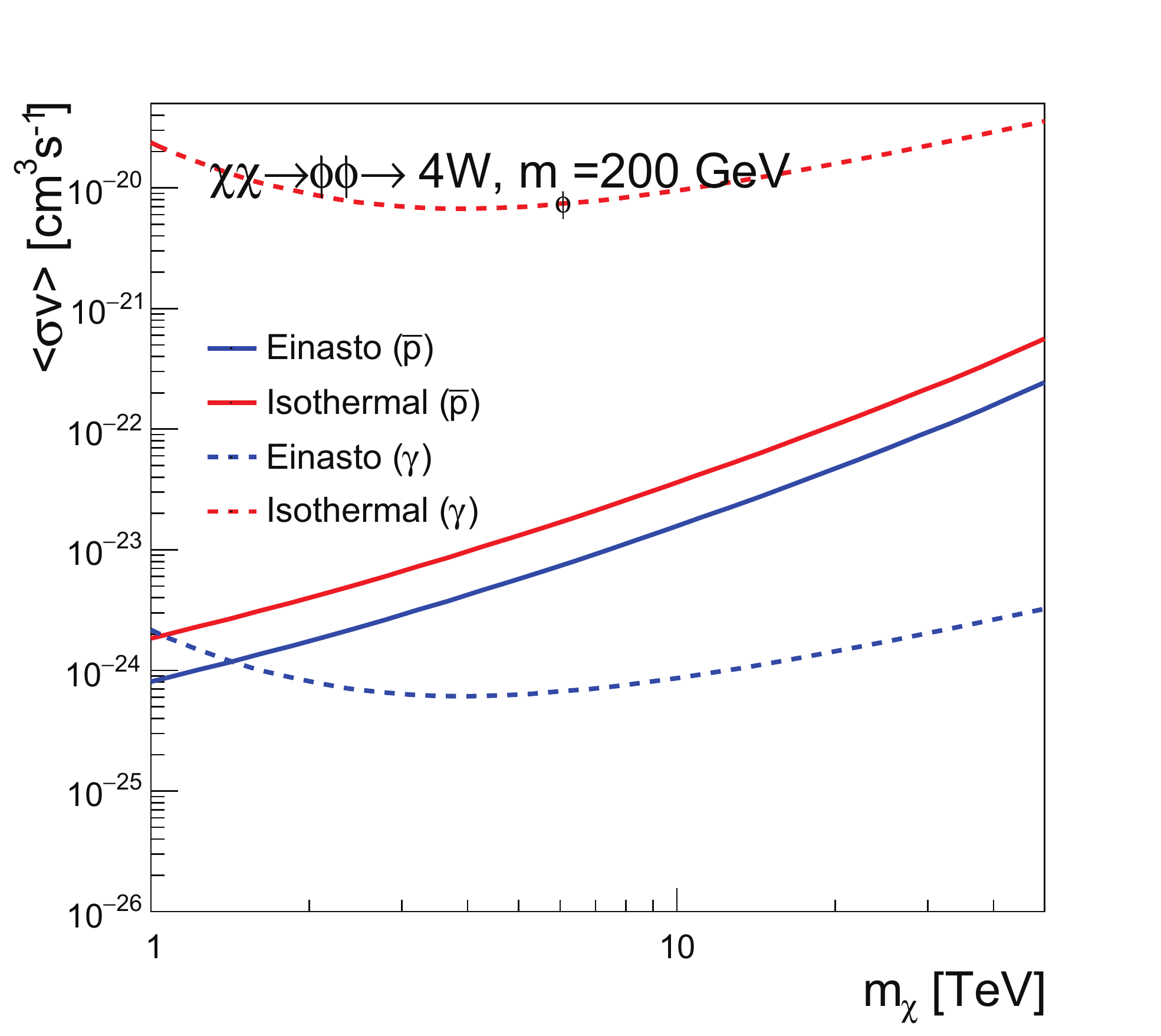}
\caption{The 95$\%$ CL upper limit of DM annihilation cross sections for different mediator decay channels. The
solid lines are derived by using the AMS-02 $\bar{p}/p$ data and the HAWC $\bar{p}/p$ upper limit,
while the dashed lines represents the limits from the HESS GC $\gamma-$ray data.
The energy spectrum of the secondary $\bar{p}$ are calculated by {\tt EPOS-LHC}, and the ``MED'' propagation model are used.
}
\label{medi_limit}
\end{figure}

By following the steps described in Sec.~\ref{sec:limit}, we obtain
the $95\%$ CL upper limit of cross sections for the annihilation
process with mediators, the results for mediator mass $m_{\phi}=200$ GeV are presented in Fig.~\ref{medi_limit}.
Similar to the results for direct annihilations, for ``Isothermal'' profile,
the most stringent constraints are from $\bar{p}/p$ data, while for
``Einasto'' profile, the $\gamma-$ray limitation are stricter.
Again, we use the AMS-02 and HAWC $\bar{p}/p$ data to constrain the ``Isothermal'' profile,
and the ``Einasto'' profile is restricted by the HESS GC $\gamma-$ray data.

\begin{figure}
\includegraphics[width=0.31\textwidth]{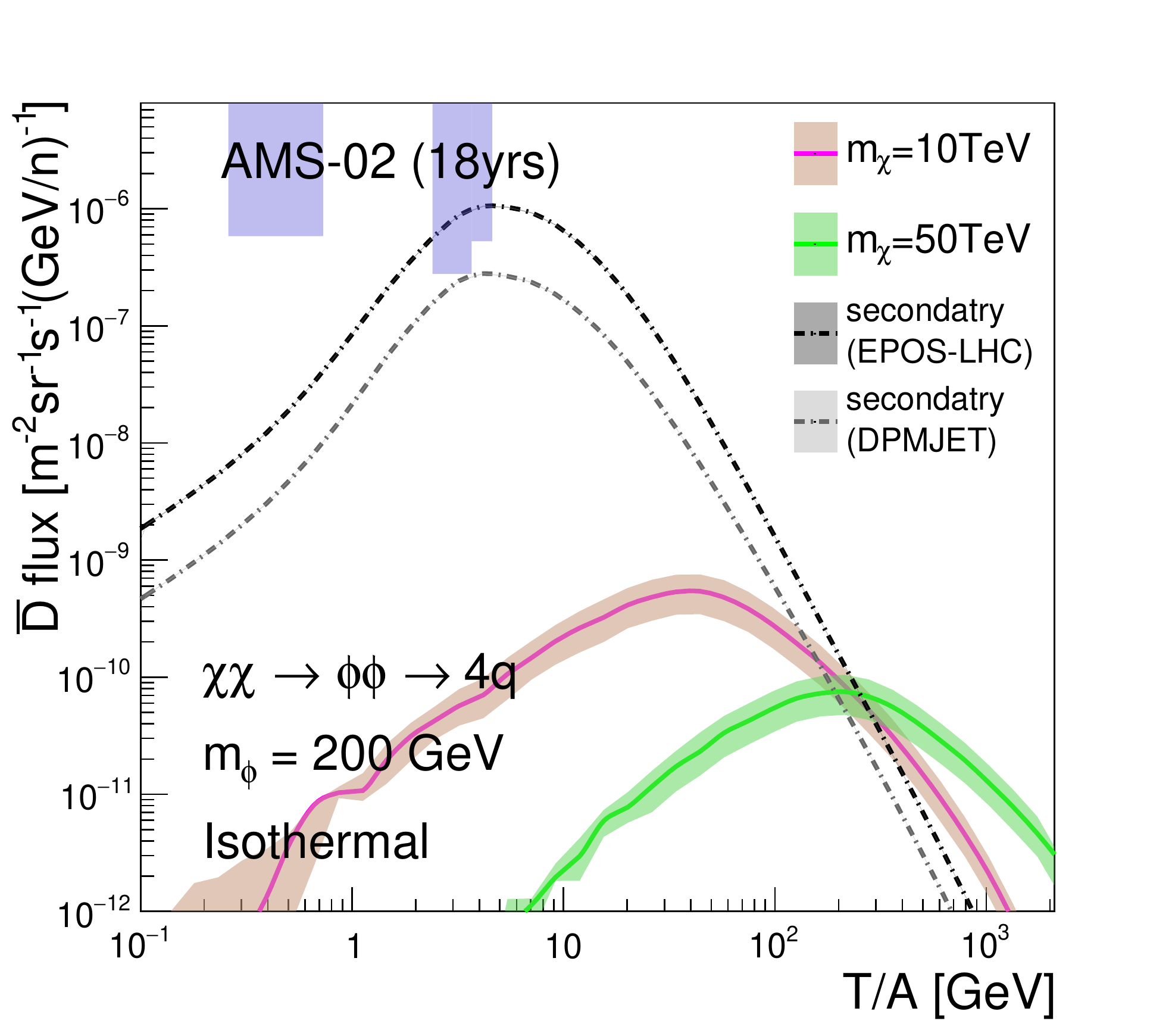}
\includegraphics[width=0.31\textwidth]{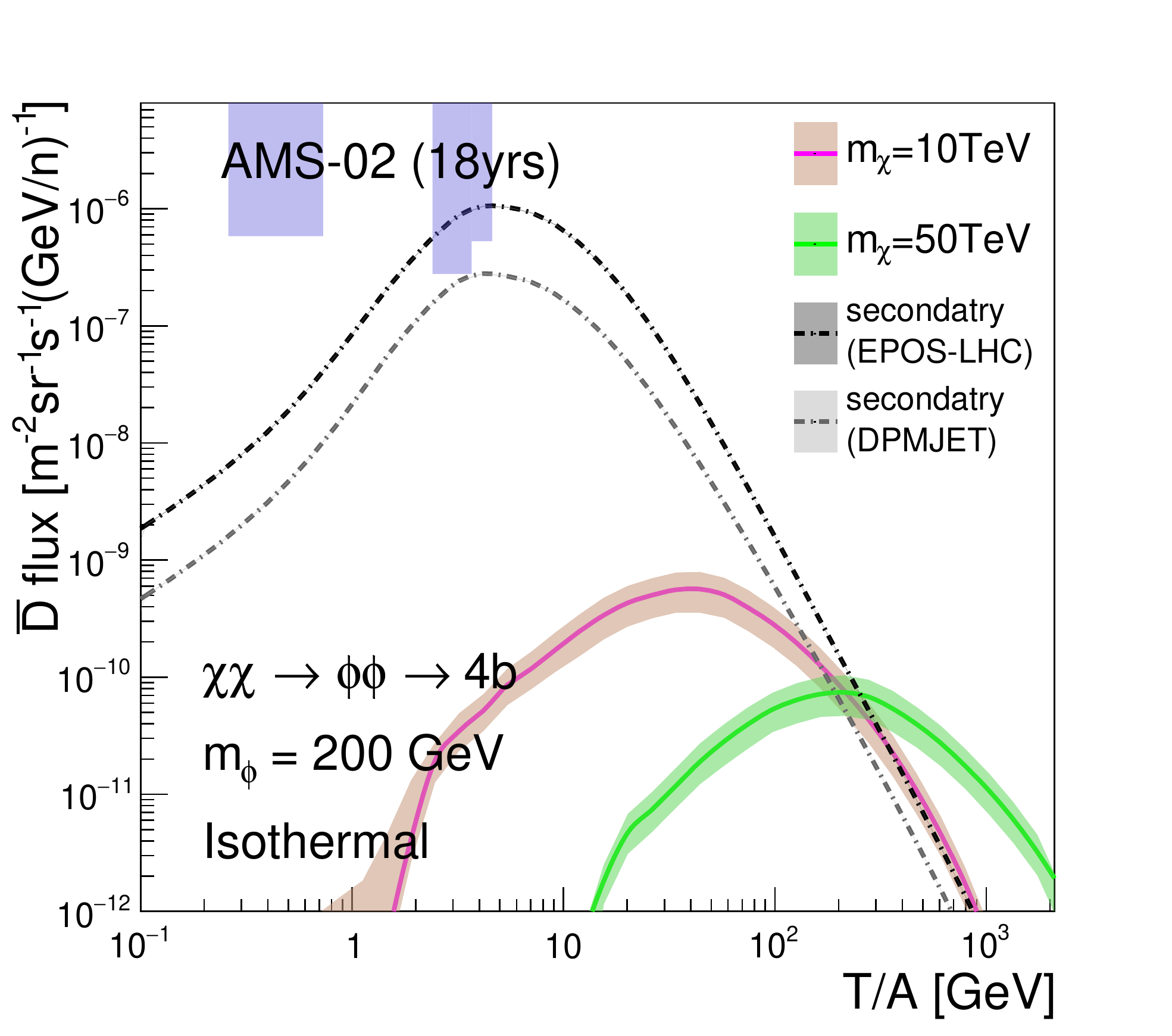}
\includegraphics[width=0.31\textwidth]{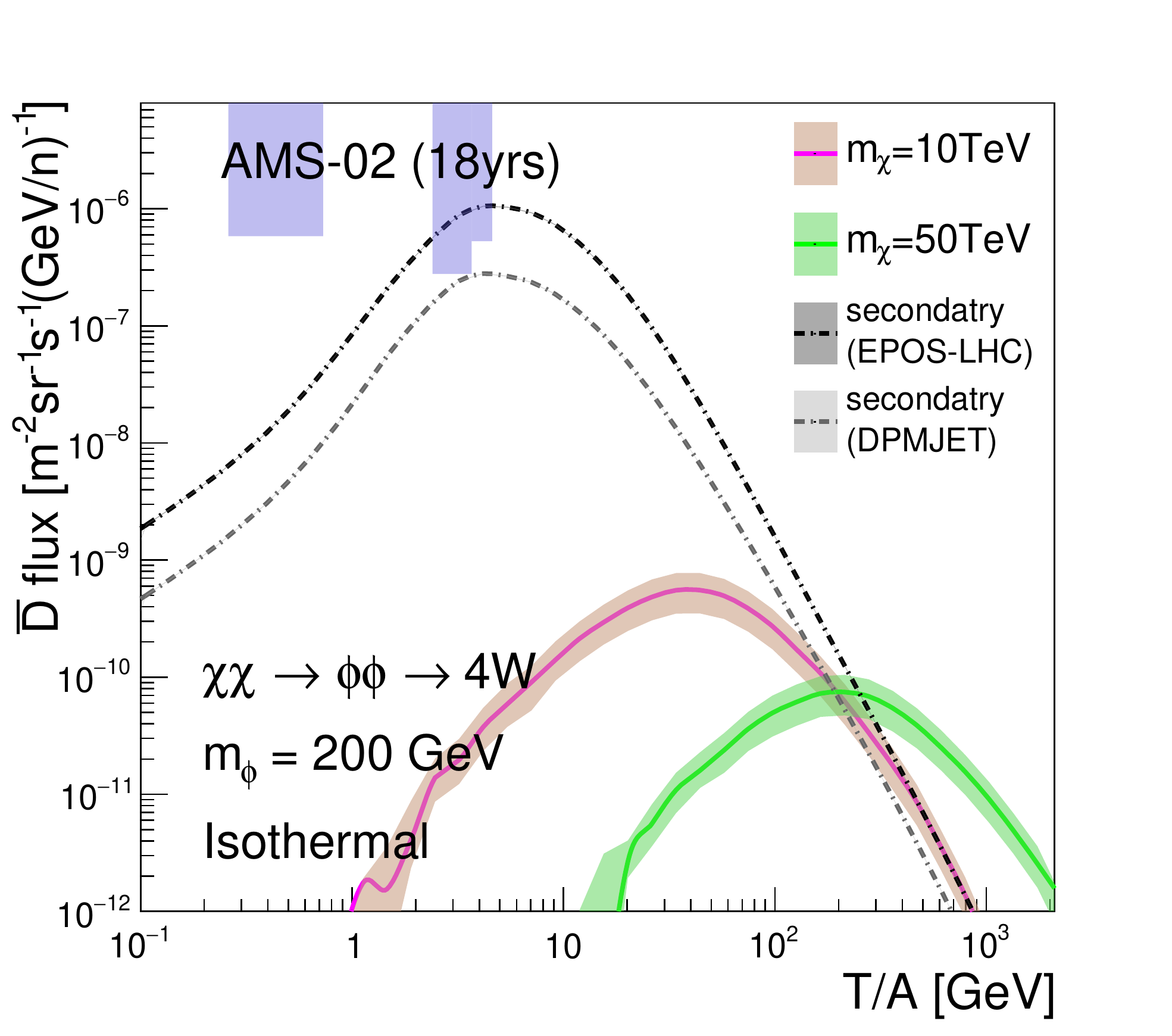}\\
\includegraphics[width=0.31\textwidth]{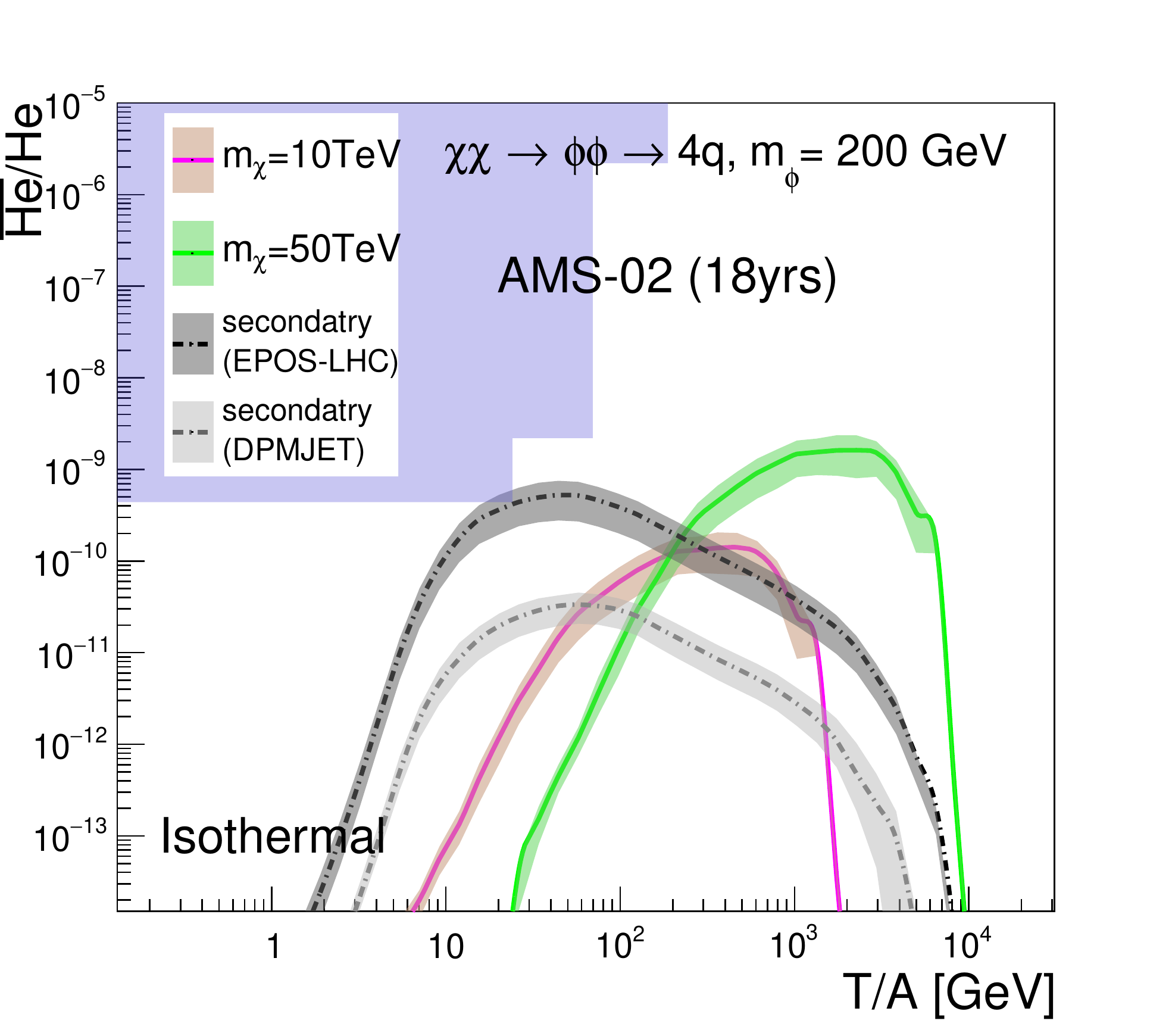}
\includegraphics[width=0.31\textwidth]{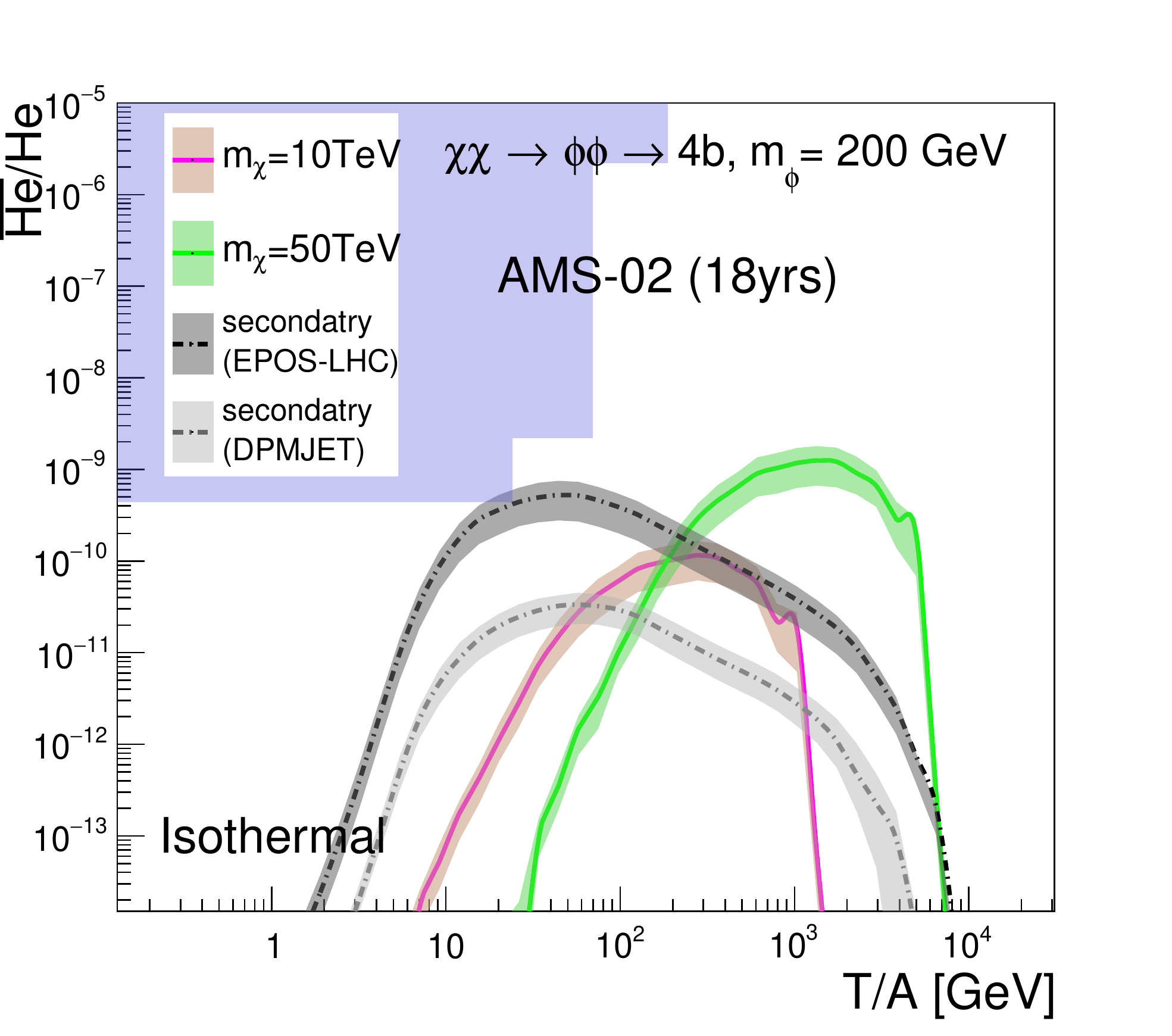}
\includegraphics[width=0.31\textwidth]{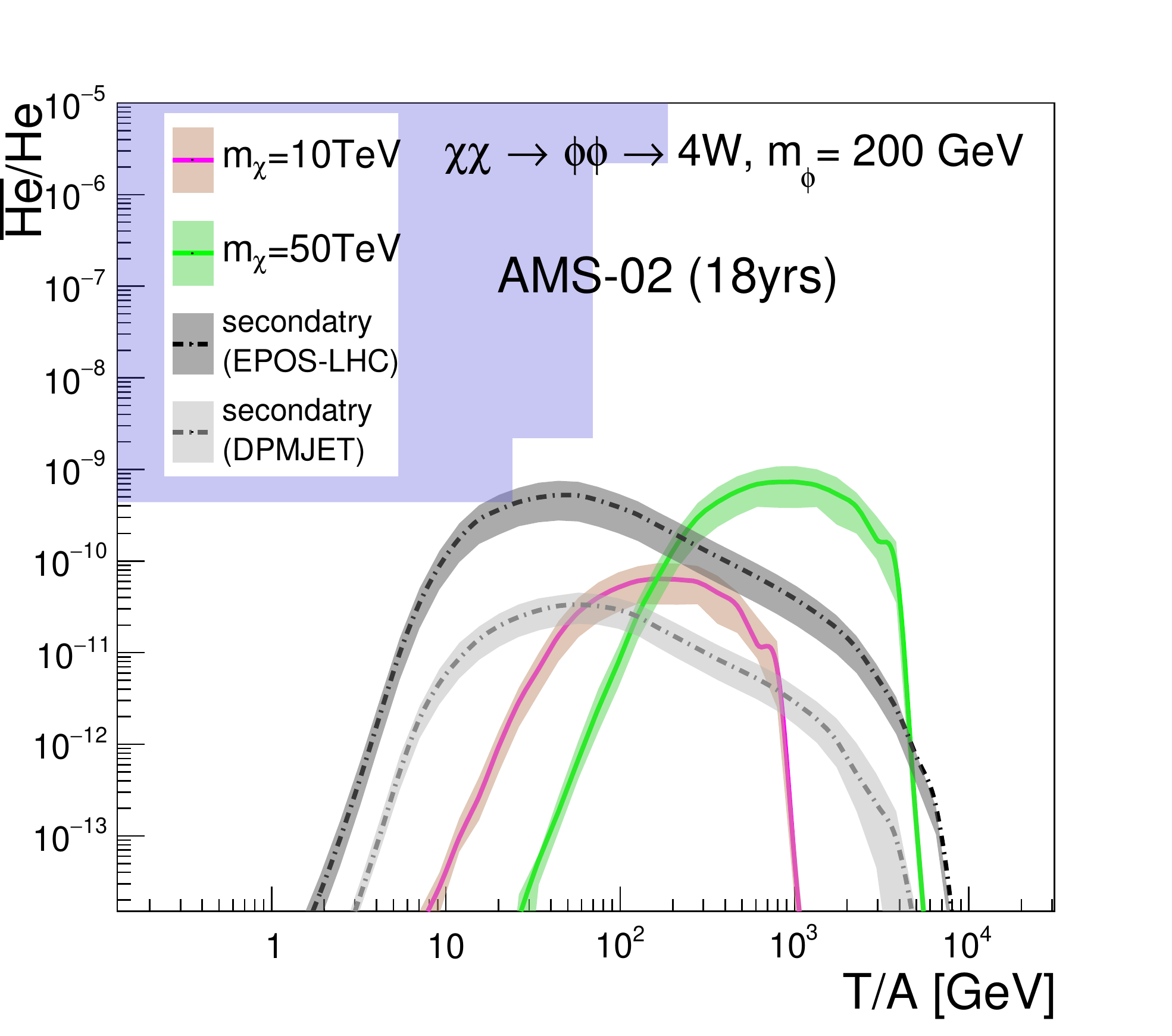}
\caption{The $\Dbar$ fluxes (top figures) and the $\Hebar/\mathrm{He}$ ratios
(bottom figures), for the ``Isothermal'' profile and different mediator decay channels,
with the ``MED'' propagation model are used.
The DM annihilation cross section is constrained by the AMS-02 and HAWC $\bar{p}/p$ data.
}
\label{medi_AMS}
\end{figure}

\begin{figure}
\includegraphics[width=0.31\textwidth]{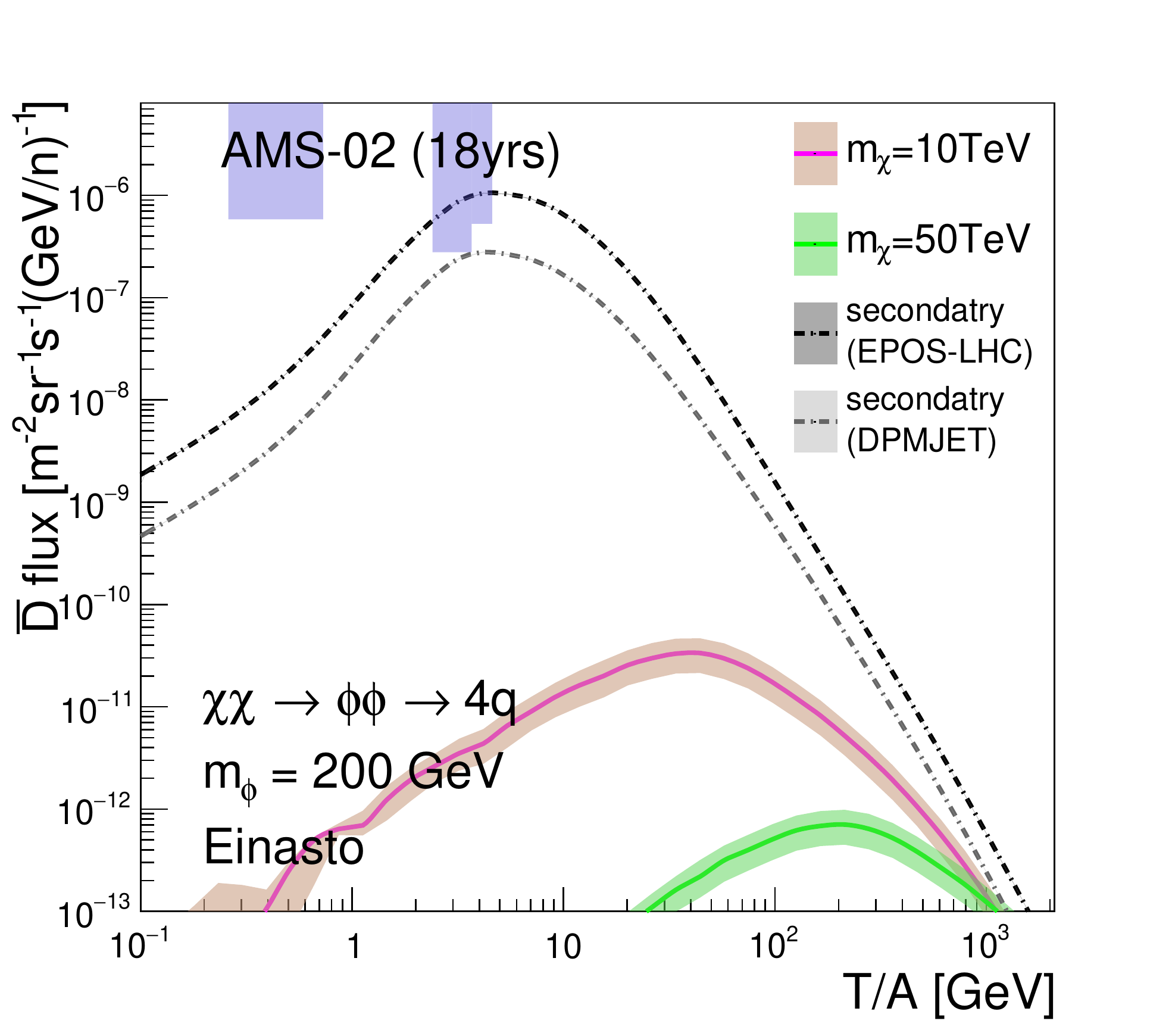}
\includegraphics[width=0.31\textwidth]{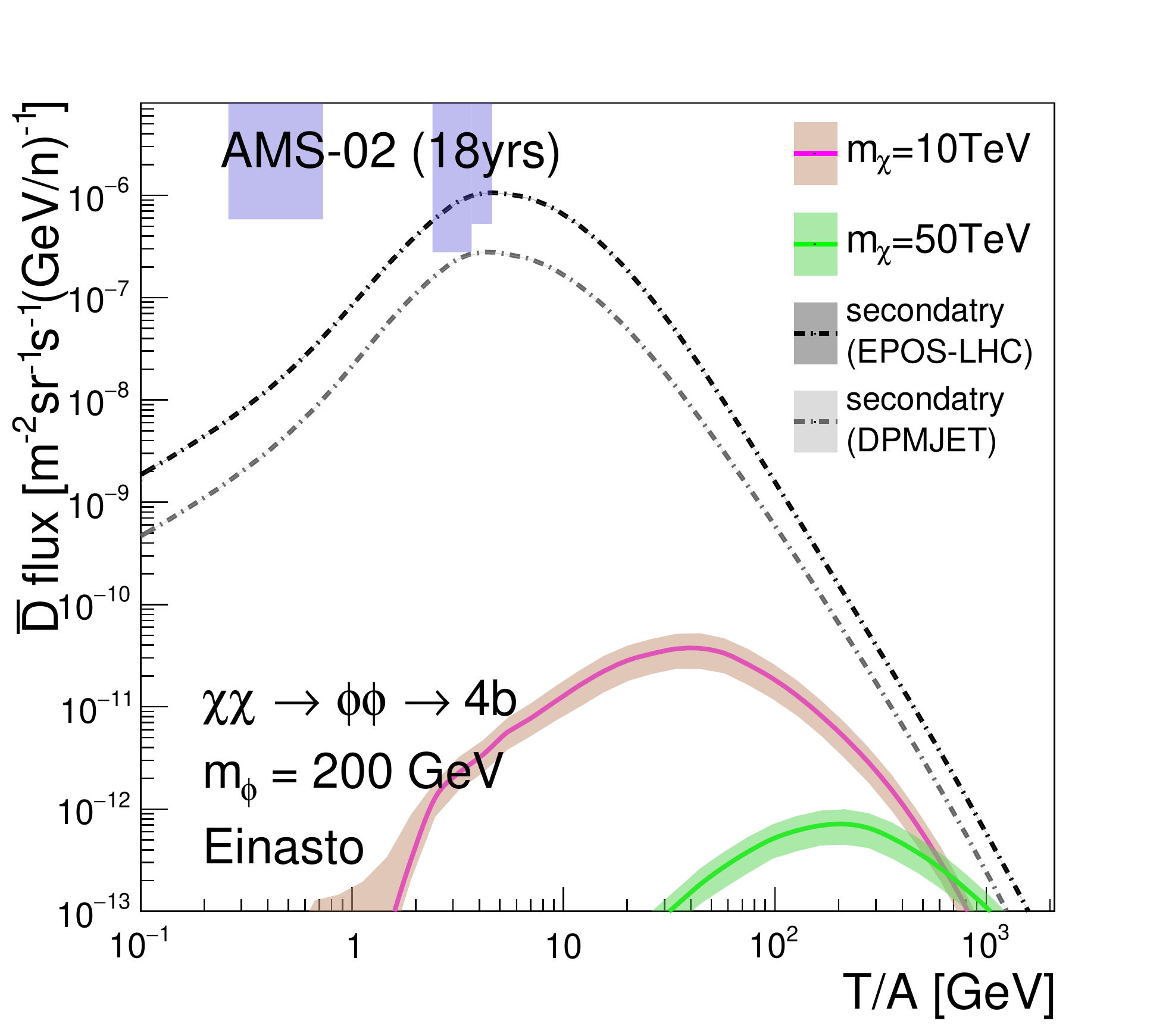}
\includegraphics[width=0.31\textwidth]{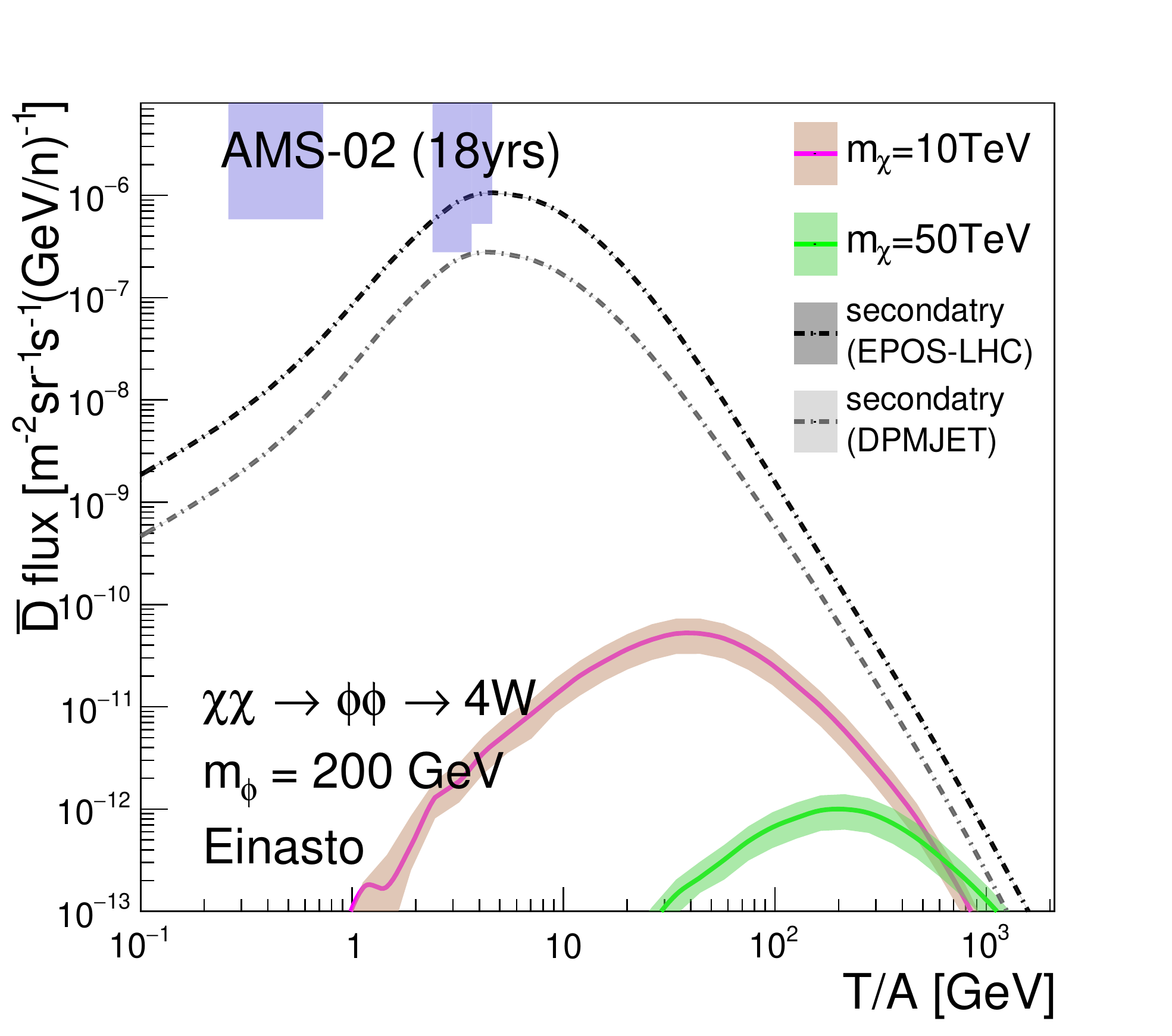}\\
\includegraphics[width=0.31\textwidth]{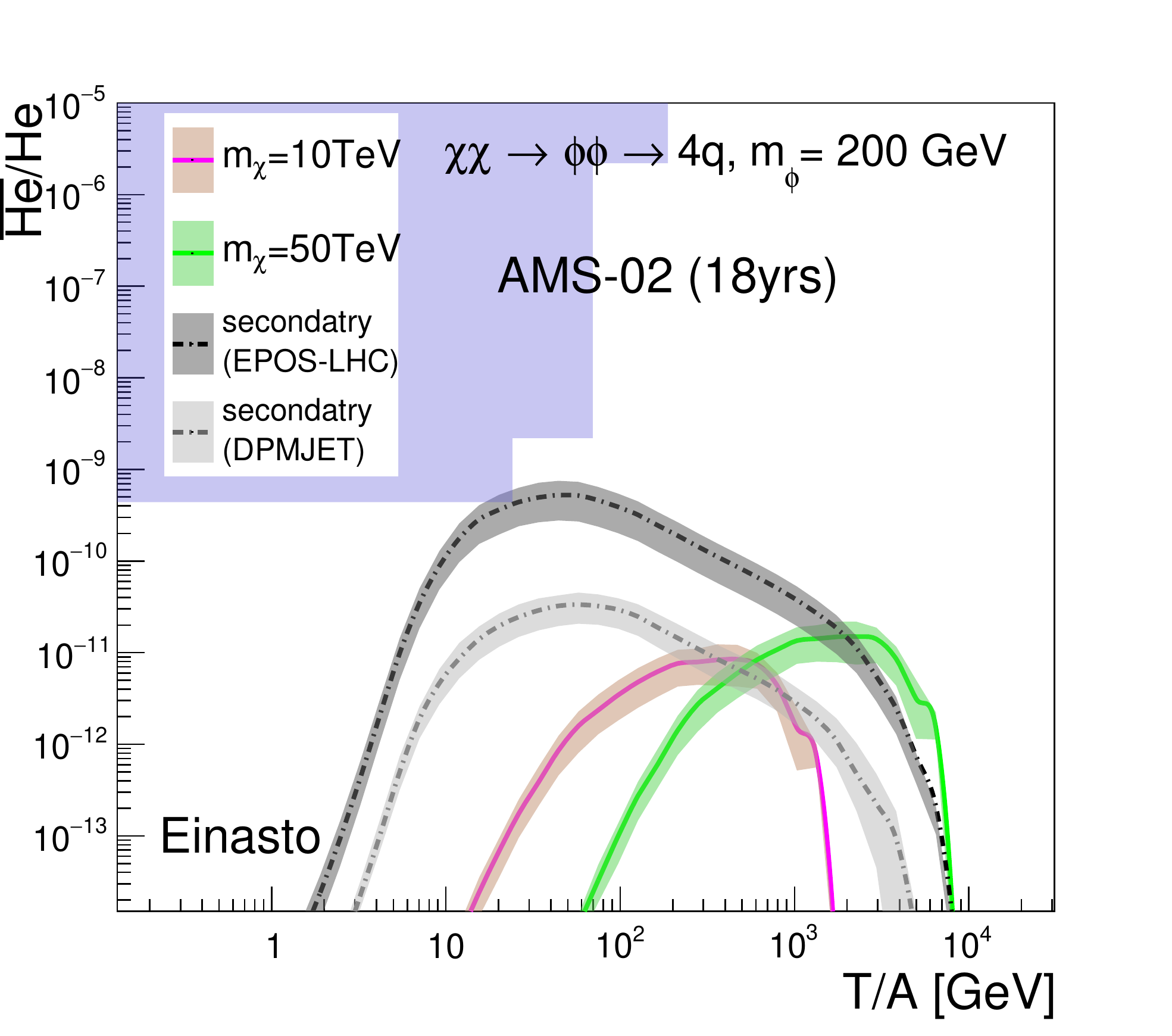}
\includegraphics[width=0.31\textwidth]{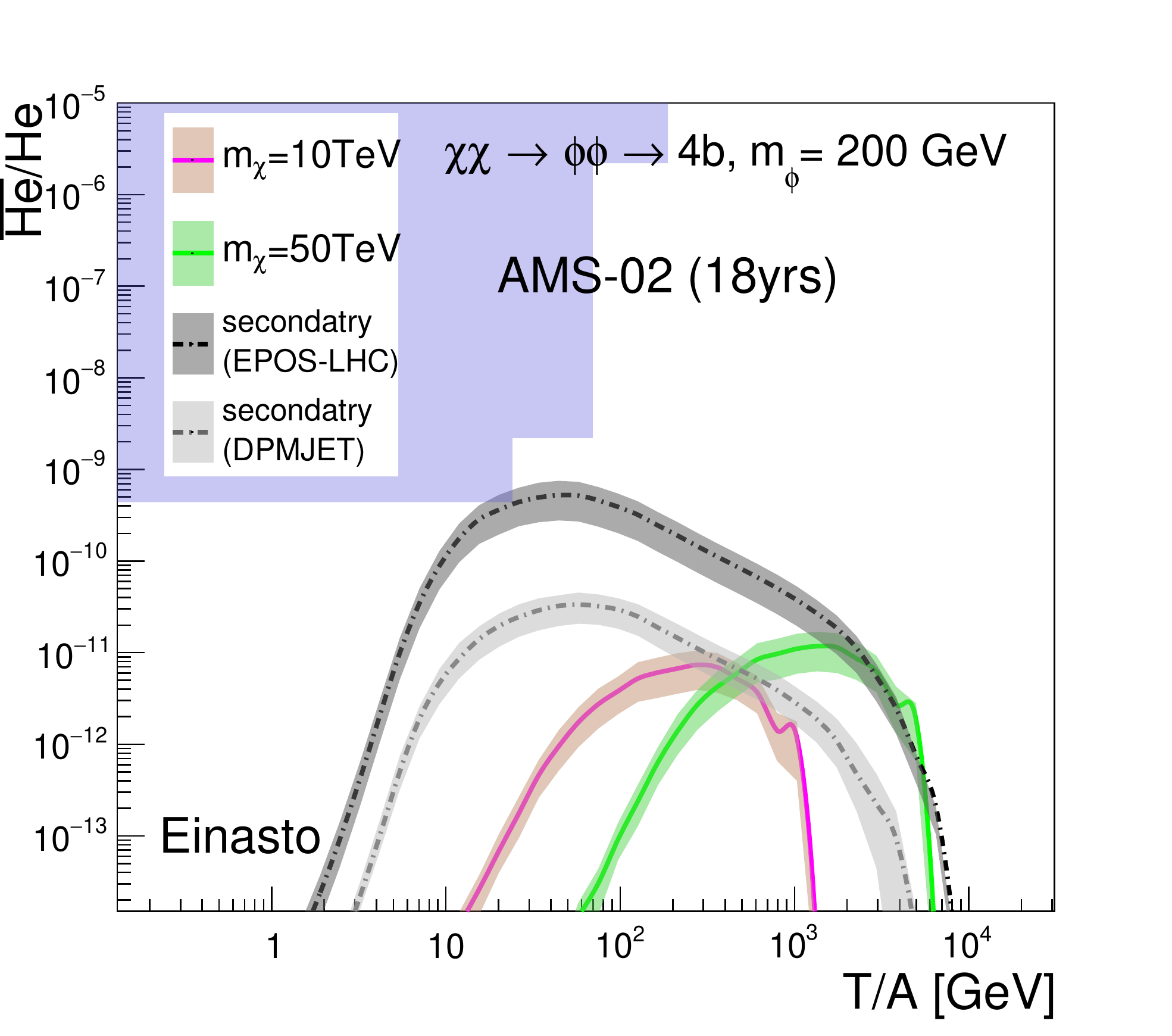}
\includegraphics[width=0.31\textwidth]{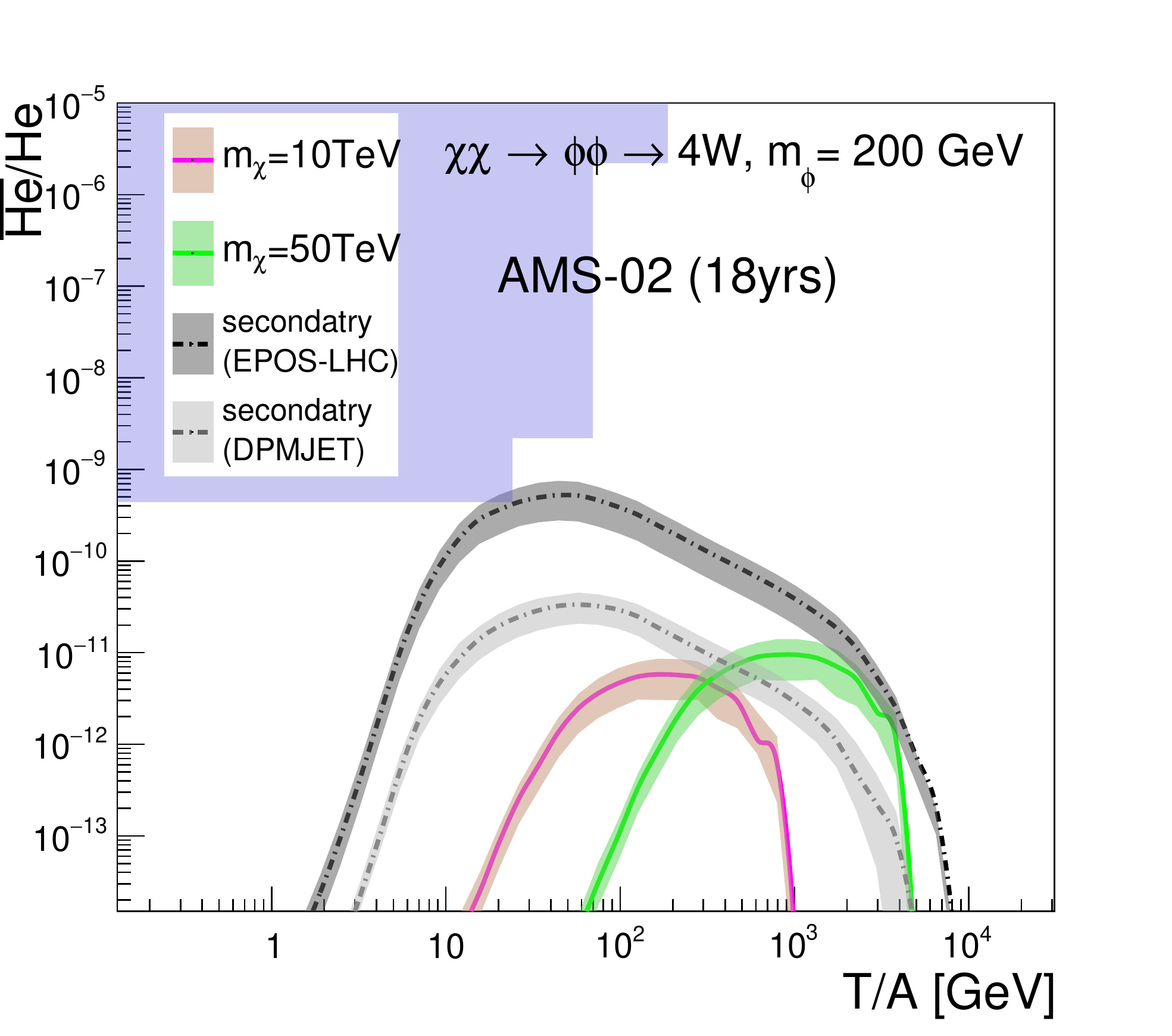}
\caption{The $\Dbar$ fluxes (top figures) and the $\Hebar/\mathrm{He}$ ratios
(bottom figures), for the ``Einasto'' profile and different mediator decay channels,
with the ``MED'' propagation model are used.
The DM annihilation cross section is constrained by the HESS 10-year GC $\gamma-$ray data.
}
\label{medi_HESS}
\end{figure}

The results for ``Isothermal'' profile and ``Einasto'' profile
with $m_{\phi}=200$ GeV are presented in Fig.~\ref{medi_AMS} and Fig.~\ref{medi_HESS}
respectively. As expected, the DM contributions are boosted
to high energy regions, and we get similar conclusions as in direct annihilation channels.
For ``Isothermal'' profile with $m_{\chi}=50$ TeV, the high energy window opens for both $\Dbar$ and
$\Hebar$ in all decay channels, especially for
$\Hebar$, the excesses can reach two order of magnitude in $q\bar{q}$ and
$b\bar{b}$ channels. While for $m_{\chi}=10$ TeV, the DM contributions for
$\Dbar$ and $\Hebar$ are comparable to the secondary backgrounds.

However, as shown in Fig.~\ref{medi_HESS}, for ``Einasto'' profile, the excesses in high energy regions
disappear for $\Dbar$ for all DM masses and mediator decay channels. For $\Hebar$, the contributions from DM with $m_{\chi}=50$ TeV can be larger
than the background calculated by using {\DPMJET}. But for {\EPOSLHC}, the only exceed appears in $q\bar{q}$
decay channel with $m_{\chi}=50$ TeV.

\begin{figure}
\includegraphics[width=0.31\textwidth]{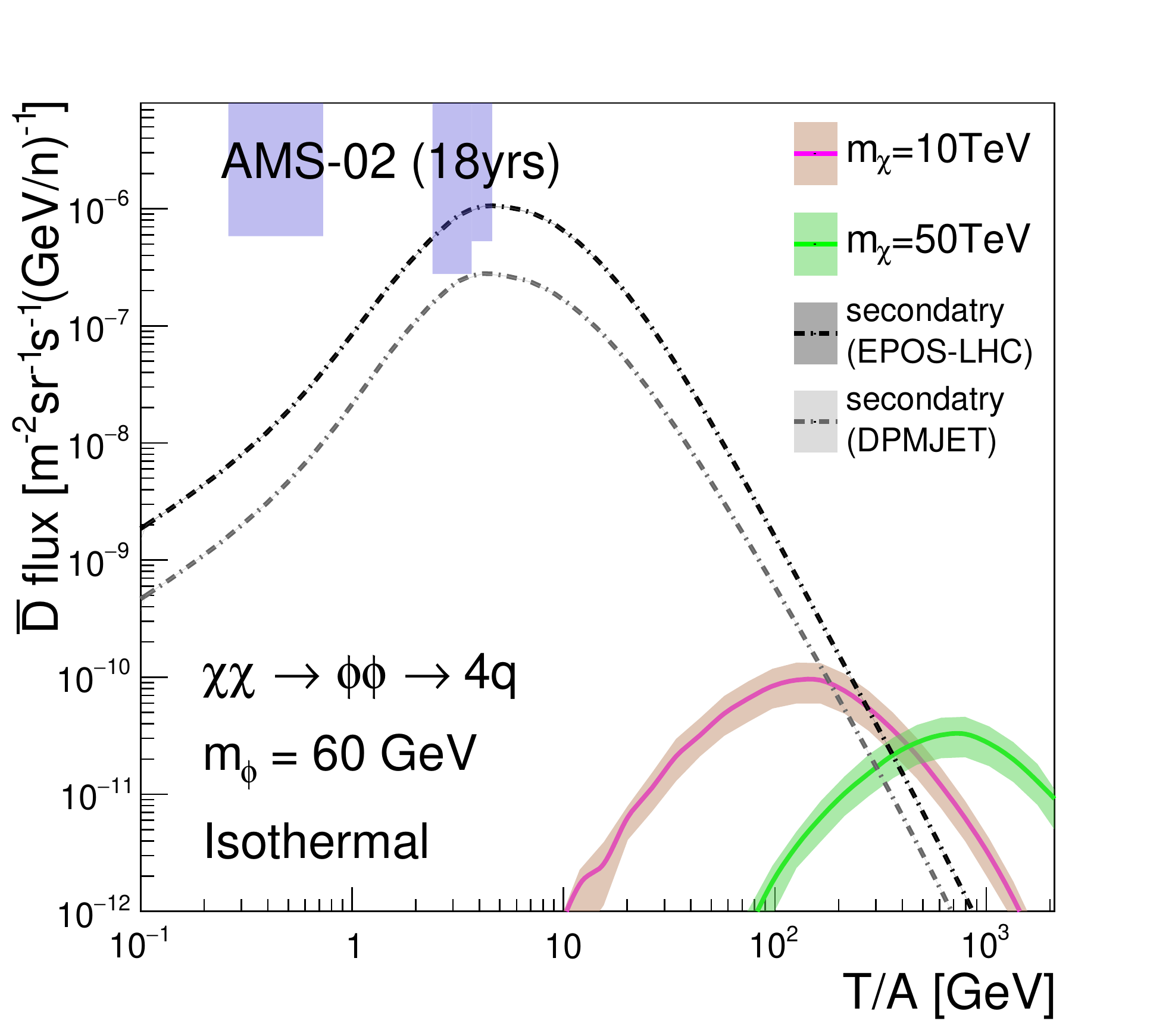}
\includegraphics[width=0.31\textwidth]{Dbar_qqbar_med200_AMS}
\includegraphics[width=0.31\textwidth]{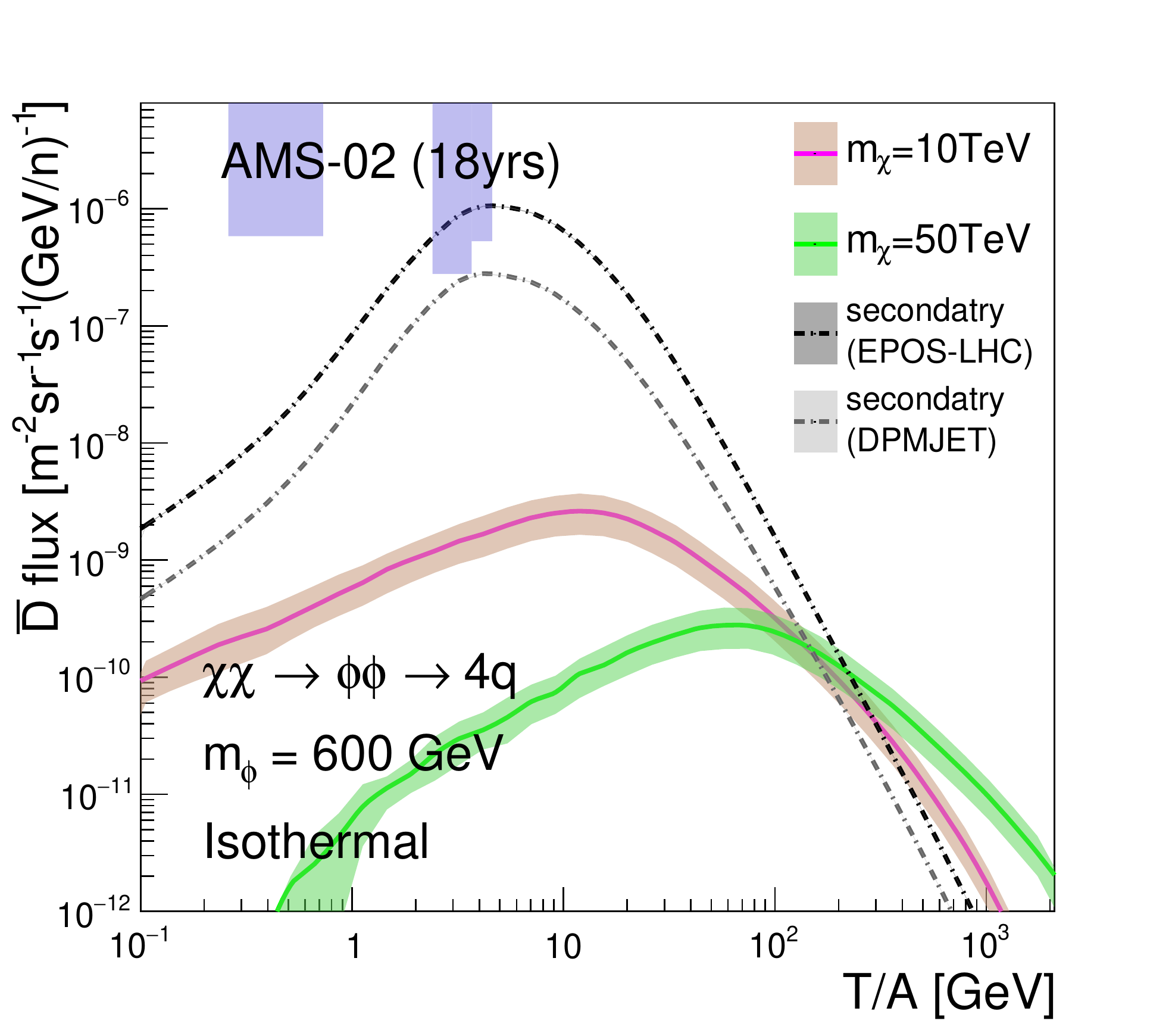}\\
\includegraphics[width=0.31\textwidth]{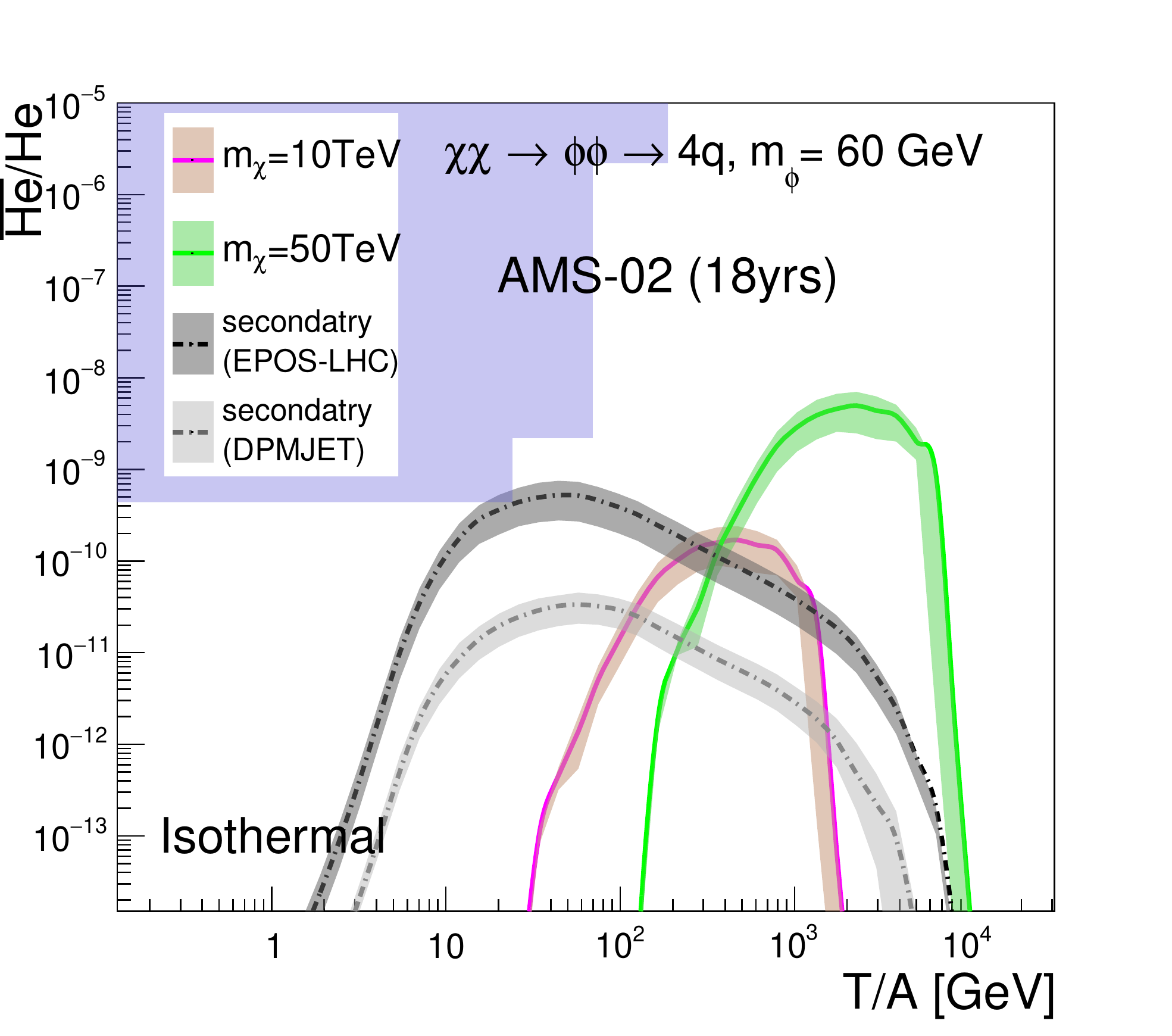}
\includegraphics[width=0.31\textwidth]{Hebar_qq_med200_AMS}
\includegraphics[width=0.31\textwidth]{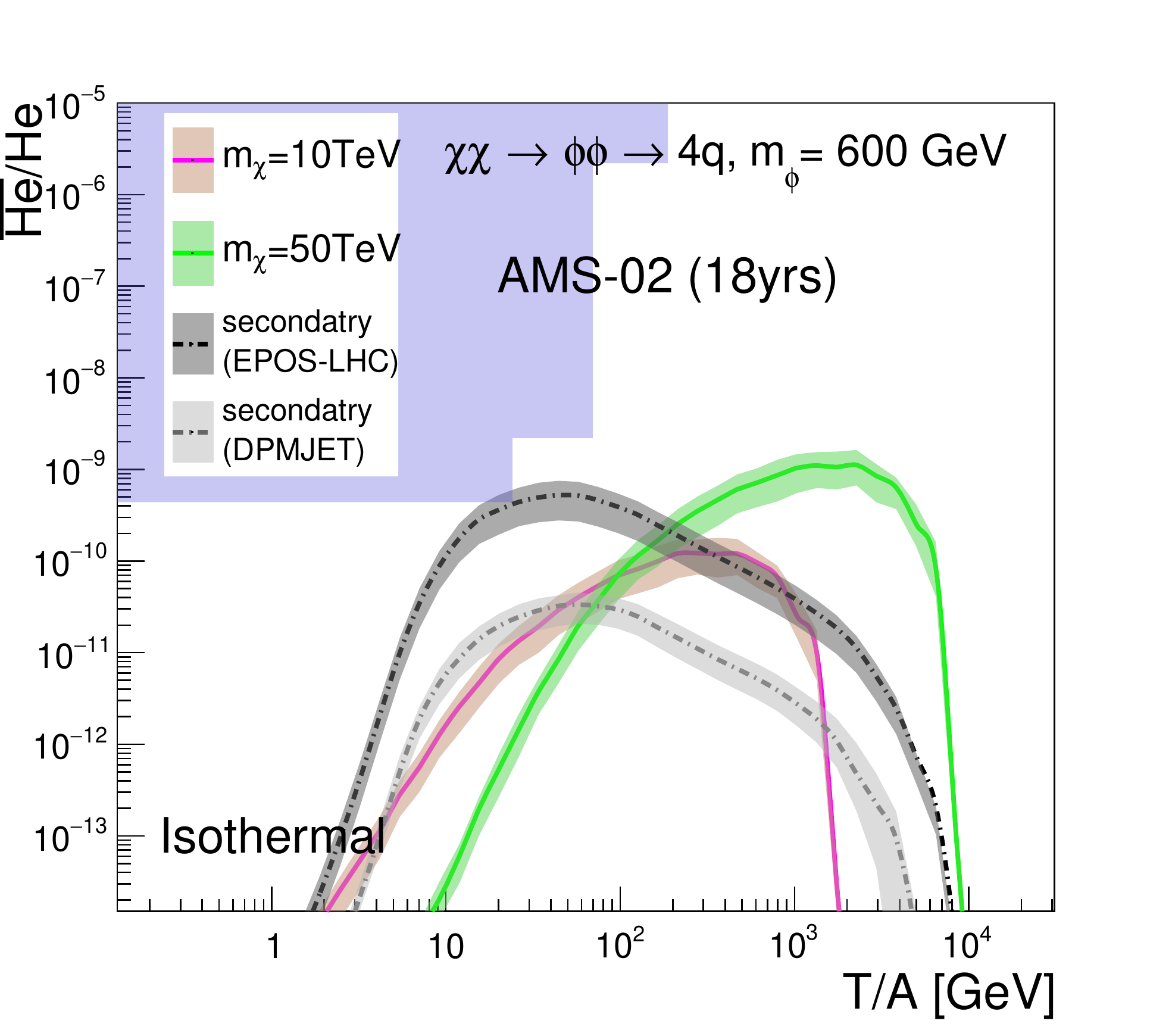}
\caption{The $\Dbar$ fluxes (top figures) and the $\Hebar/\mathrm{He}$ ratios
(bottom figures), for the ``Isothermal'' profile and $\chi\chi \rightarrow \phi\phi \rightarrow 4q$ channels,
with different mediator masses and the ``MED'' propagation model.
The DM annihilation cross section is constrained by the AMS-02 and HAWC $\bar{p}/p$ data.
}
\label{medi_masses_AMS}
\end{figure}

For other mediator masses, we get the same conclusion. Take the $\chi\chi \rightarrow \phi\phi \rightarrow 4q$
channel for an example, we calculate the $\Dbar$ fluxes and $\Hebar$ ratios for mediator mass
$m_{\phi}=60$, 200 and 600 GeV, and present the results for ``Isothermal'' profile in Fig.~\ref{medi_masses_AMS}.
It can be seen that with the growth of
the mediator mass, the DM contributions in low energy regions increase significantly. However, in high energy
regions, which we are interested in, the variations of the results are small.

\section{Conclusions}\label{sec:conclusion}
In summary, we explored the possibility of probing
DM by high energy CR anti-nuclei. We used the MC generators
\PYTHIA, \EPOSLHC~and \DPMJET~and the coalescence model
to calculate the spectra of anti-nuclei, with the coalescence
momenta of $\Dbar$ and $\Hebar$ were derived by fitting the
data from ALICE, ALEPH and CERN ISR experiments. The propagation
of charged CR particles are calculated by the \GALPROP~code,
with the inelastic interaction cross sections between the primary
CR and interstellar gases given by MC generators.
We used the HESS GC $\gamma-$ray data to constrain the DM
annihilation cross sections for the DM profiles with large
gradient at GC, while the flat profiles were limited by the
AMS-02 and HAWC $\bar{p}/p$ data.

Our results showed that, for a ``Cored'' type DM density
profile like the ``Isothermal'' profile, the high energy window opened
for both $\Dbar$ and $\Hebar$ in all channels.
However, for a ``Cuspy'' type profile like the ``Einasto'' profile,
the $\Dbar$ contributions from DM annihilations were below the secondary background
in both DM direct annihilation and annihilation through mediator decay channels.
As for $\Hebar$, the conclusion depended on the choice of MC generators,
the $\Hebar$ flux originated from DM annihilations could exceed the secondary background for {\DPMJET},
while the excess disappeared for {\EPOSLHC}.

Signals in high energy regions can effectively avoid the uncertainties
from the solar activities. Although the fluxes
of $\Dbar$ and $\Hebar$ in high energy regions were far
below the sensitivity of the current experiments like AMS-02 and GAPS,
the high energy window could be a promising probe of DM for the next generation experiments.
We believe that with the fast development of detector technologies,
people would finally be able to detect the DM through the high energy window.

\section*{Acknowledgments}
This work is supported in part by the
National Key R\&D Program of China No.~2017YFA0402204
and by the National Natural Science Foundation of China (NSFC)
No.~11825506, No.~11821505, No.~U1738209, No.~11851303 and No.~11947302.

\bibliographystyle{arxivref}
\bibliography{DbarHebarref}
 \end{document}